\documentclass[a4paper,fleqn,sort&compress]{cas-sc}
\usepackage[authoryear]{natbib}
\usepackage[nodots]{numcompress}
\usepackage{appendix}

\usepackage{float} 
\usepackage{subcaption} 
\usepackage{graphicx}
\usepackage{caption,subcaption}
\usepackage{lineno}

\begin{document}
\def\floatpagepagefraction{1}
\def\textpagefraction{.001}


\shorttitle{Wave scattering of ice floes}
\shortauthors{C Zhang et~al.}

\title [mode = title]{\textcolor{black}{A black-box-model-enhanced} interaction method for water-wave scattering by large group of arbitrary-shaped ice floes in Arctic route planning \tnotemark[1]}                      

\tnotetext[1]{The manuscript was submitted to Cold Regions Science and Technology on March 22, 2025; First revision notification was received on March 18, 2026; First revision was submitted on April 7, 2026.}

\author[1]{Chongwei Zhang}
[
      orcid=0000-0002-0062-4548]
\credit{Conceptualisation, Methodology,  Data curation, Investigation, Formal analysis, Visualisation, Validation, Software, Funding acquisition, Resources, Project administration, Writing - Original Draft, Writing - Review \& Editing}
\affiliation[1]{organization={State Key Laboratory of Coastal and Offshore Engineering, Dalian University of Technology},
    city={Dalian},
    postcode={116024}, 
    country={P.R. China}}
\cortext[cor1]{Corresponding author. E-mail: chongweizhang@dlut.edu.cn}
\cormark[1]    
\author[1]{Hongli Yang}
\credit{Methodology,  Data curation, Investigation, Formal analysis, Visualisation, Validation, Software, Writing - Original Draft, Writing - Review \& Editing}
\author[2]{Peng Wu}
\credit{Formal analysis, Writing - Review $\&$ Editing}
\affiliation[2]{organization={Department of Mechanical Engineering, University College London},
    city={London},
    postcode={WC1E 6BT}, 
    country={United Kingdom}}
\author[1]{Peng Lu}
\credit{Data curation, Visualisation, Writing - Review \& Editing}
\author[1]{Dezhi Ning}
\credit{Funding acquisition, Resources, Supervision, Project administration, Writing - Review \& Editing}

\begin{abstract}
This study develops an enhanced interaction (EI) method for efficient prediction of the water-wave field among a large group of ice floes in Arctic route planning.
A novel black-box model, termed the wave component detection (WCD) method, is proposed for constructing the diffraction transfer matrix (DTM) within the framework of interaction theory.
The DTM, which is conventionally mathematically intractable for three-dimensional ice floes with arbitrarily complex geometry, can now be determined using this readily implementable and universally applicable approach.
Without loss of generality, four ice-floe shapes are taken as example models to demonstrate the capability of the EI method.
Operation rules are recommended for the practical implementation of the EI method.
The error range of the EI method is identified in scenarios with multiple ice floes of different sizes and distances.
The super-high efficiency of the EI method is demonstrated in cases involving an ultra-large group of ice floes.
It takes less than 1.5 hours to calculate wave amplitudes at 160,000 locations in the wave field of 1,800 ice floes (based on 1,440,000 boundary elements) on an ordinary personal computer with a 2017-released CPU.
Based on the wave field predicted by the EI method, users can take advantage of the wave-sheltering effect of the ice floes to optimize routes.
For demonstration, the dynamic programming strategy is used to recommend optimized navigation routes among 1561 ice floes of mixed shapes.
The average wave amplitude the ship encounters can be reduced to about half of the incident wave amplitude.
\end{abstract}

\maketitle

\section{Introduction}
\color{black}
\subsection{Water-wave scattering problem for Arctic route planning}
\color{black}
As the global temperature keeps rising, the seasonal ice retreat in the Arctic Ocean has expanded greatly over the past few decades. 
The increased open water among ice floes enables Arctic shipping between Asia and Europe. 
The Arctic routes can cut the navigational distance between Europe and Asia by around 40\%, benefiting fuel savings and carbon emission reduction \citep{SCHOYEN2011977}. 
On the other hand, as the seasonal open water expands, the wave amplitude in the Arctic Ocean is also increasing.
The mean significant wave height may exceed 3 meters at 70° N latitudes during summer \citep{Waseda2018}. 
In the central Beaufort Sea, wind waves with 5-meter heights have been measured \citep{Thomson2024}.
Compared to conventional open-water shipping, the wave environment along Arctic routes is much more complex due to the presence of numerous ice fragments.
Without proper path planning in ice-floe fields, shipping companies may not save on fuel costs along Arctic routes. 
For example, ships must slow down to adapt to unpredictable ocean wave conditions and avoid ice floes.

\begin{figure}[!htp]
\centering
{
  \begin{minipage}{0.45\linewidth}
  \centering
  \includegraphics[width=1.0\linewidth]{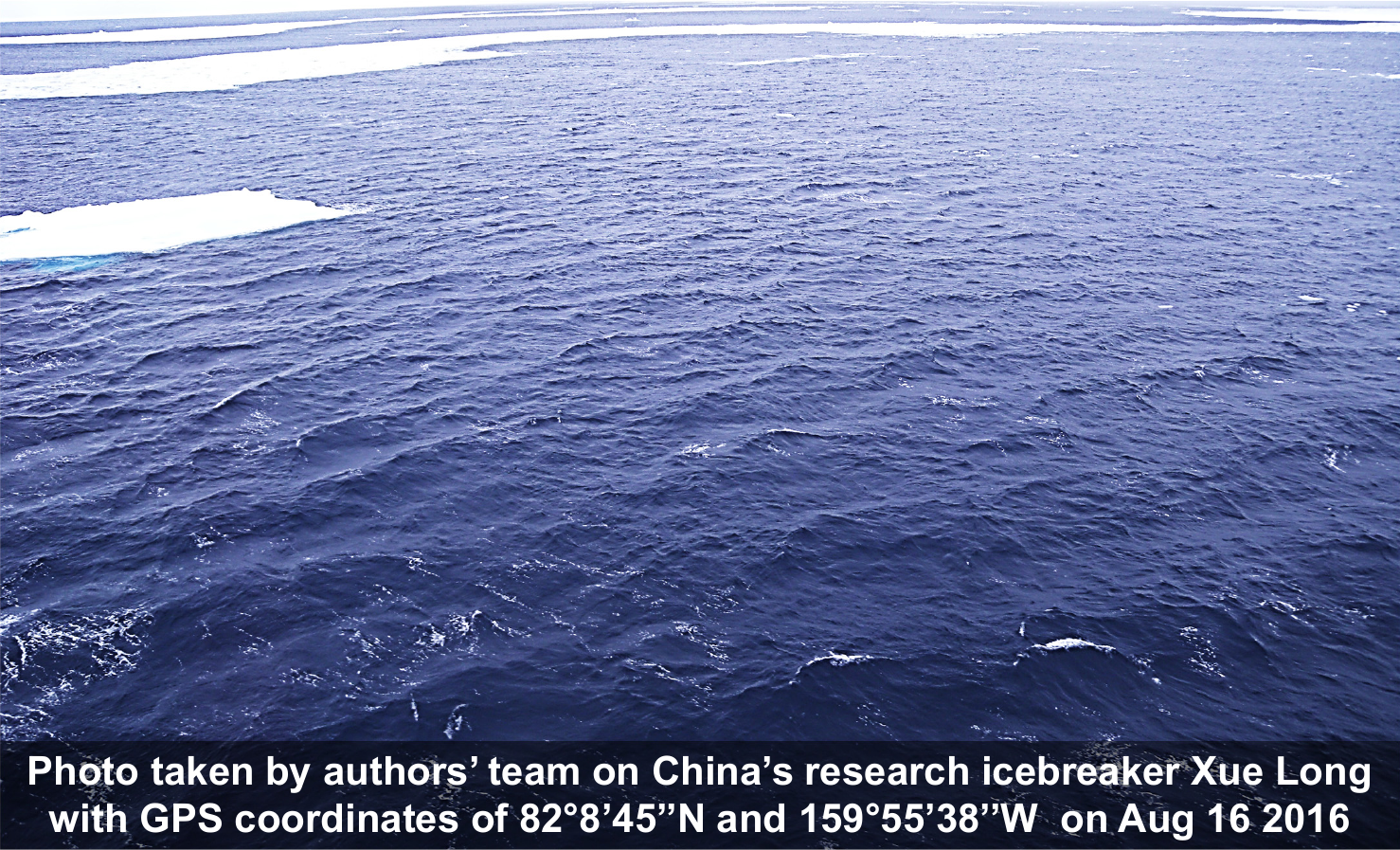}
  \subcaption{}
  \end{minipage}
}
{
  \begin{minipage}{0.45\linewidth}
  \centering
  \includegraphics[width=1.0\linewidth]{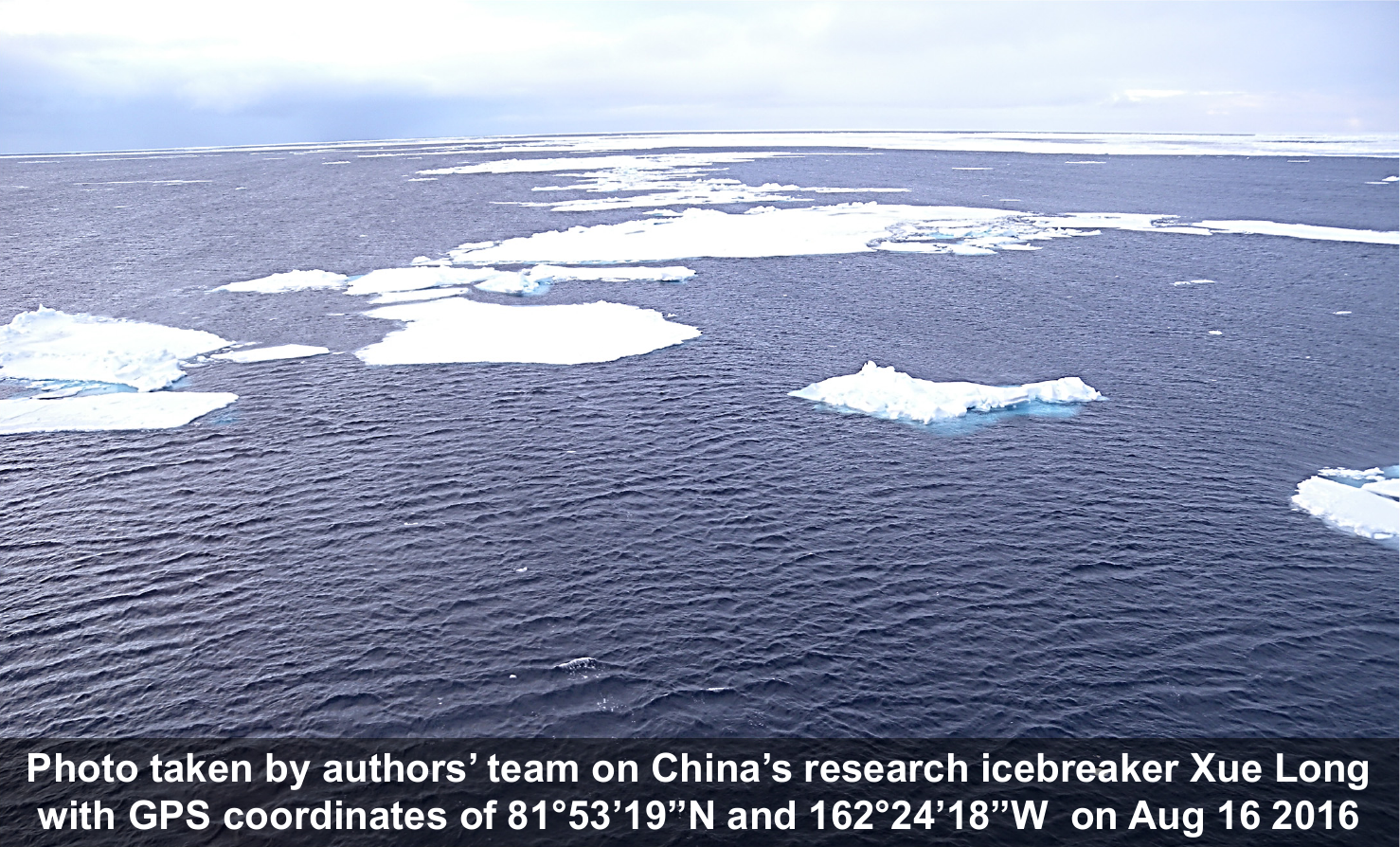}
  \subcaption{}
  \end{minipage}
}
\\
{
  \begin{minipage}{0.45\linewidth}
  \centering
  \includegraphics[width=1.0\linewidth]{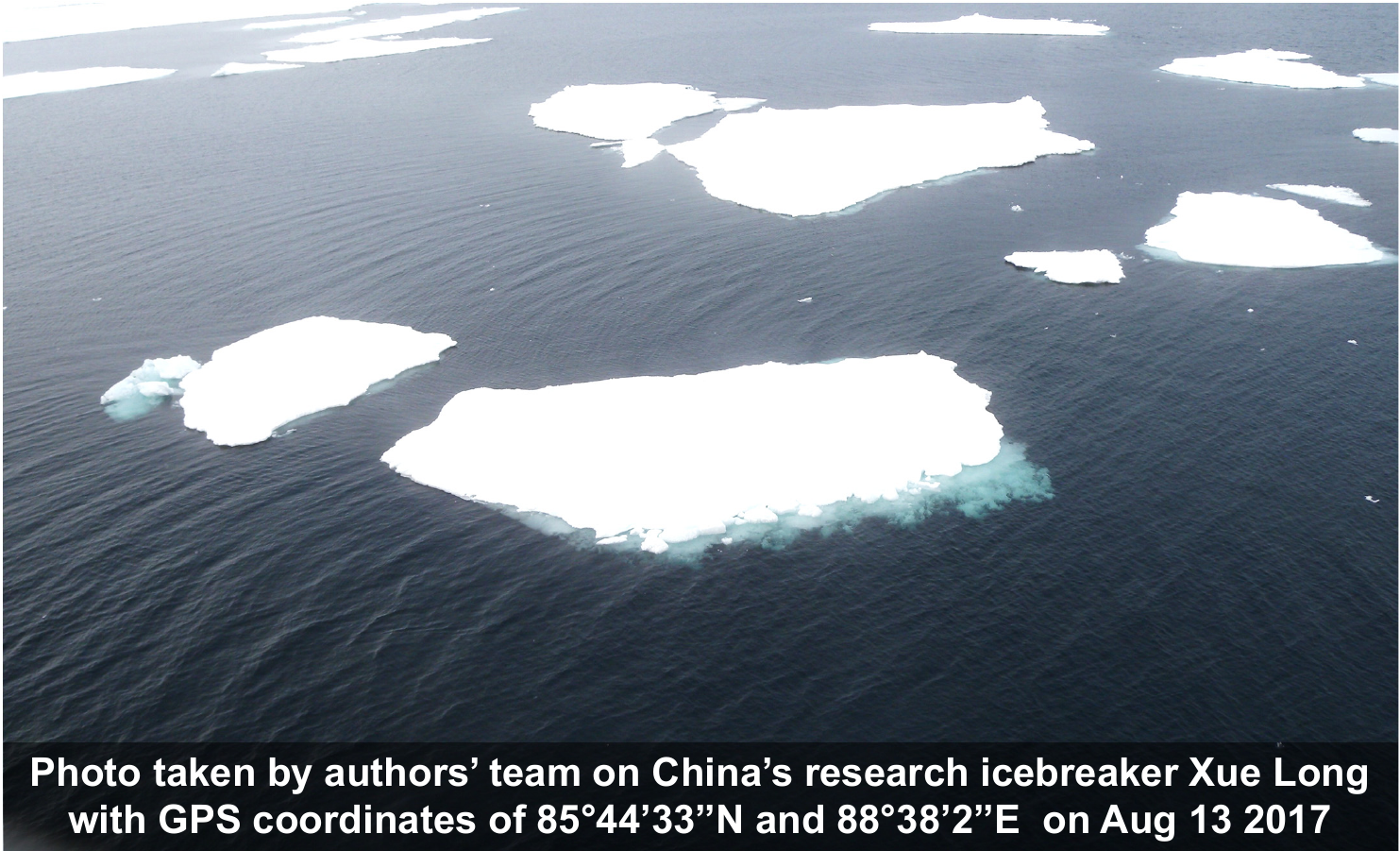}
  \subcaption{}
  \end{minipage}
}
{
  \begin{minipage}{0.45\linewidth}
  \centering
  \includegraphics[width=1.0\linewidth]{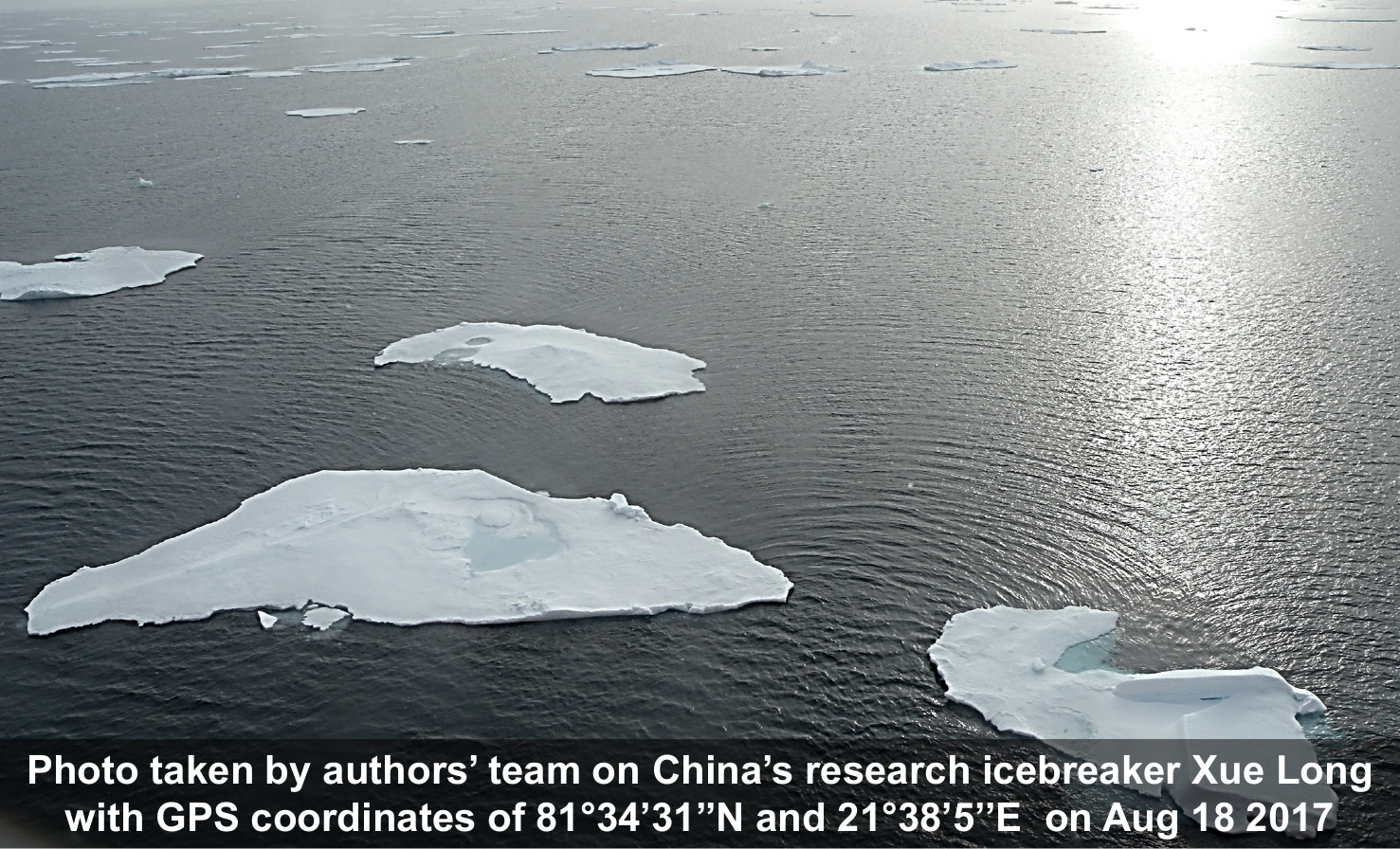}
  \subcaption{}
  \end{minipage}
}
\caption{
Photos taken by the authors' team aboard the Chinese research icebreaker Xue Long in 2016 and 2017: wave scattering field around ice floes along Arctic shipping routes, with corresponding GPS coordinates of (a)  82°8’45’’N and 159°55’38’’W; (b)  81°53’19’’N and 162°24’18’’W; (c) 85°44’33’’N and 88°38’2’’E; and (d) 81°34’31’’N and 21°38’5’’E}
\label{fig:wave_scattering_photo}
\end{figure}

The wave environment is most complex in the marginal ice zone (MIZ). 
In this zone, a large number of individual ice floes, varying in size and shape, are spread across the ocean surface.
Either wind waves formed over ice-free water or incoming swells from the open sea can penetrate a significant distance (10 to 20 km) into the ice floe field \citep{Squire2018}.
As a genuine piece of evidence, Fig. \ref{fig:wave_scattering_photo} presents some in-situ photos of the wave field around ice floes along Arctic shipping routes. 
The photos were taken by the authors' team on China's research icebreaker Xue Long in the summers of 2016 and 2017. 
From these photos, one can clearly see the redistribution of wave amplitude in the waterways between ice floes.
If the sizes of ice floes are non-negligible compared to the dominant wavelength, the propagating waves scatter strongly in the vast ice-floe fields.
The oscillations of water bodies confined in waterways between ice floes can be enlarged or weakened after wave scatterings.
An enlarged wave amplitude can cause wave-induced resistance to a ship and complicate its manoeuvring operations, harming fuel consumption and carbon reduction.
Conversely, if ships sail mainly through the small-amplitude wave field via intelligent Arctic route planning, their navigation can be both safe and economical.

For intelligent Arctic route planning, the core is to predict wave propagation and scattering in the ice-floe field efficiently and reliably.
If ice floes are sufficiently small and dense, an equivalent homogenized free surface (like a “viscous layer”) can be used to represent highly concentrated ice-floe fields. 
The ice layer can be assigned a rheology with specific viscoelastic properties, similar to those of a dense fluid \citep{Keller1998, WANG201090, Shen2022}.
However, for Arctic shipping, scattered ice floes of non-negligible sizes often attract greater concern.
Currently, wave scattering studies for large piece of ice floes are mostly based on 2D models \citep{fox1990reflection, meylan1993finite, sahoo2001scattering, linton2003reflection, meylan2015surge, mosig2019transport, zhang2023resonance}.
Some other 3D studies often focus on the wave scattering problem around one or a few ice floes \citep{meylan1996response, porter2004approximations, bennetts2010wave}.
Neither 2D nor the ``few-body'' cases are adequate for practical wave field predictions in Arctic route planning.

\color{black}
\subsection{Methodology for water-wave scattering by large groups of ice floes}\label{subsec:intro2}
\color{black}

\begin{figure}[htbp]
	\centering
	\includegraphics[width=0.8\linewidth]{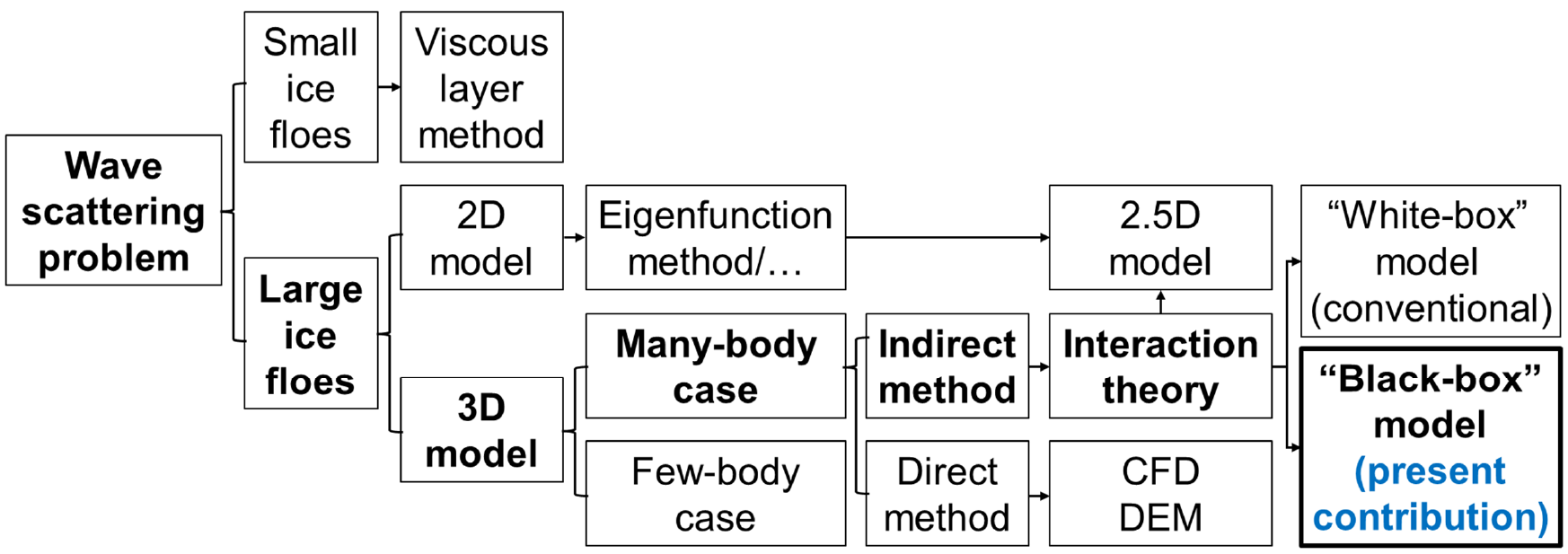}
	\caption{Knowledge structure of methodologies for ice-floe wave scattering problems}
	\label{fig:liter}
\end{figure}

For practical purposes, the 3D wave scattering around very large groups of ice floes (e.g., at least over a thousand) with arbitrary shapes should be considered.
So far, existing 3D wave-scattering methods can be briefly categorized into direct and indirect ones, as shown in Fig. \ref{fig:liter}.
For the direct 3D method, the diffraction calculation is performed for all bodies simultaneously, and the entire ensemble is treated as a unit.
Typical direct methods include various computational fluid dynamics (CFD) methods.
For example, 
\cite{orzech2018coupled} developed a coupled numerical model to investigate the physics of wave attenuation in the MIZ at small scales (of several meters). 
A multi-phase CFD model, along with a discrete element method (DEM) software package, was used to simulate ice-floe particles in waves and currents.
\cite{HUANG2020102817} applied an incorporated DEM-CFD solver to calculate a ship's ice-floe resistance. 
Ice motions, ship-ice/ice-ice collisions, and ship-generated waves were solved at the same time.
\cite{wang2023numerical} used the structured arbitrary Lagrangian–Eulerian (S-ALE) method and the numerical wave-making method based on dynamic boundary conditions to numerically study the motion response of ice floes in waves.
However, direct 3D methods are too time-consuming for use in a quick-reaction route planning system. 
This is because they must consider the entire fluid domain and all ice floes at the same time. 
When the number of ice floes and the size of the fluid domain increase, these computations become prohibitively expensive and infeasible.

The indirect 3D method, which is often based on interaction theory, only requires diffraction solutions for individual ice floes.
The interaction theory was initially proposed by 
\cite{kagemoto1986interactions} for calculating the wave hydrodynamics of a multi-body structure in finite-depth water.
The key idea is to regard the scattered wave of each body as the incident wave on all other bodies.
In other words, the total incident waves on a body are considered a summation of the diffracted waves from all other bodies along with the ambient incident field.
A transfer operator called the ``diffraction transfer matrix (DTM)'' is used to connect the coefficients of scattered waves and incident waves,  representing the diffraction property of an isolated body.
Using Graf's addition theorem, the coefficients of the scattered wave fields of all bodies are solved simultaneously.

However, in 
\cite{kagemoto1986interactions}, only the solution for identical axisymmetric bodies was described to ensure that the diffraction solution of an isolated body had to be expressed in the cylindrical eigenfunction expansion of outgoing waves. 
How to obtain the DTM for an arbitrary body was not given in \cite{kagemoto1986interactions}.
For the scenario of ice floes, the body with arbitrary geometry must be considered.
\cite{Yoshida1990ANM} represented the finite-depth free-surface Green's function in a cylindrical coordinate system, so that the diffraction solution can be expressed in terms of cylindrical eigenfunction expansion.
The panel method based on the source strength distribution function was used to solve the wave diffraction problem. 
This extended 
\cite{kagemoto1986interactions}'s interaction theory to bodies with arbitrary geometry.
\cite{peter2004infinite} further extended the finite-depth interaction theory \citep{kagemoto1986interactions, Yoshida1990ANM} to infinite depth.
The ice floe was modelled as a thin plate rather than a rigid body, incorporated with the finite-element method (FEM) to determine the modes of vibration.
For an isolated body, each eigenfunction expansion of the scattered wave field was solved using a constant panel method.
The columns of the DTM are the coefficients of the eigenfunction expansion of the scattered wave field under incident waves with unit amplitude.

In addition to direct and indirect 3D methods, some “quasi-3D” or ``2.5D'' methods are also proposed for studying the attenuation of ocean wave energy in the MIZ.
By 2.5D, ice floes in the marginal ice zone (MIZ) are split into multiple rows or strips of infinitely long, with each strip considered as a medium as a whole.
An infinite number of floes, often identical and circular, are periodically distributed in each strip.
The reflected and transmitted waves across a strip, as calculated by the interaction theory, can be expressed as a continuous superposition of plane waves travelling in all directions within the angular range.
A transfer matrix can be obtained to represent the relationship between the incident amplitudes and scattered amplitudes for the single row.
The solution of wave transmission for the $(m+1)$th strip can be deduced from that of the $m$th strip in an iterative manner, by taking the transmitted waves behind the $m$th strip as the incident waves of the $(m+1)$th strip.

As a typical study based on the 2.5D method, 
\cite{bennetts2009wave} considered an idealized array consisting of a finite number of rows. 
Each row contained an infinite number of identical and equally spaced circular floes. 
Individual floes were represented as circular elastic plates.
The model was further extended by 
\cite{bennetts2010three} to the case in which each row contains an infinite number of modules of floes with some prescribed periodicity.
\cite{peter2010water} considered an arrangement of ice floes as follows: First, several circular ice floes were grouped into modules; Then, an infinite line array was formed periodically by these modules; Finally, many line arrays were assembled one behind another in a stack.
The scattering characteristics of the module, the line array, and the stack were identified in sequence.
In 
\cite{montiel2015reflection}, the scenario of a directional wave spectrum travelling through a strip of randomly distributed circular ice floes was considered.
The single strip approach of \cite{montiel2015reflection} was further extended to include multiple strips of circular ice floes in 
\cite{montiel2016attenuation}.
\cite{porter2019coupling} described a semi-analytical approach to determine the response of a shallow-drafted rectangular elastic plate to waves, and extended the approach to doubly-periodic arrangements of ice sheets.
The 2.5D method can deal with a huge number of ice floes without difficulty. However, since it is mainly designed to analyze the wave transmission or dissipation properties behind an ice-floe field, sufficient information about the wave surface between ice floes of arbitrary shape cannot be provided for route planning.

\color{black}
\subsection{Recent studies on water-wave scattering by ice floes via interaction theory: ``white-box'' model}\label{subsec:intro3}
\color{black}

\color{black}
Among the various approaches to modelling wave–ice interactions, interaction theory provides an efficient theoretical framework for describing wave scattering by arrays of ice floes.
A typical application procedure for interaction theory is detailed in \cite{peter2004infinite} based on the DTM, which enables efficient modelling of wave interactions among multiple floes through a modular scattering representation. 
In this framework, the scattered wave field from each floe is expressed as a superposition of outgoing cylindrical wave modes, and Graf’s addition theorem is employed to translate these wave fields between different local coordinate systems. 
Consequently, the multiple scattering problem can be reduced to a linear system of equations, rendering the method particularly suitable for large-scale simulations involving numerous ice floes. 
Wave propagation in the MIZ influenced by the presence of discrete ice floes can be predicted efficiently.

A comprehensive overview of interaction theory is provided by \cite{Squire2007}, which summarized the theoretical and experimental progress in modelling wave propagation through sea ice. 
The review also highlighted the wave attenuation effect in the MIZ, which is primarily governed by wave scattering from ice floes.
In line with this guidance, a series of subsequent interaction-theory-based studies on wave scatterings in the MIZ have focused on wave attenuation, in which the elasticity of individual ice floes is of critical importance.
Therefore, ice floes are commonly modelled as floating thin elastic plates.
For example, \cite{Kohout2008} developed an elastic plate model for ice floes to investigate wave attenuation in the MIZ, in which wave scattering by individual plates is characterized via eigenfunction expansions within the framework of linear wave theory.
A three-dimensional model for wave attenuation in the MIZ was proposed by \cite{Bennetts2010}, in which the ice field is represented as a collection of randomly distributed elastic plates.
The results indicate that attenuation rates are strongly dependent on wave period, ice thickness, and floe concentration.
\cite{Bennetts2012} established a spectral attenuation model based on multiple scattering theory for application in large-scale wave models.
\cite{Meylan2018} presented a three-dimensional time-domain formulation for wave scattering in the MIZ, which enables the directional redistribution of wave energy.
\cite{meylan2021floe} developed a floe-size-dependent scattering model in both two and three dimensions to investigate wave attenuation in the MIZ.
By comparing two- and three-dimensional formulations, the study demonstrates that scattering dominates attenuation at short wave periods, whereas additional dissipative mechanisms are required to explain the observed attenuation of long-period waves.

More recently, \cite{montiel2024kernel} proposed a scattering-kernel formulation derived from the far-field amplitude of an array of floating ice floes.
This formulation was incorporated into a radiative transfer equation to characterize the directional redistribution and transport of wave energy across large-scale ice fields.
Their analysis revealed that the floe size distribution exerts a significant influence on the scattering kernel, particularly within the weak localization regime.
\cite{mohapatra2025three} developed a three-dimensional analytical hydroelastic model for a floating elastic plate under oblique wave–current interactions. 
By employing Green’s function techniques combined with a geometrical symmetry velocity decomposition method, the model derives the dispersion relation and predicts wave reflection, wave forces, and plate deflection under varying current conditions.
Meanwhile, alternative modelling strategies have also been proposed.
For instance, \cite{yu2022new} introduced a data-driven parameterization method based on dimensional analysis, employing ice thickness as a scaling parameter. Their model reproduces the nonlinear dependence of wave dissipation on ice thickness and wave frequency, and resolves apparent inconsistencies between large-scale field observations and small-scale laboratory measurements.

\subsection{Practical maritime motivations and special considerations of present study}

\begin{figure}[htbp]
	\centering
	\includegraphics[width=0.8\linewidth]{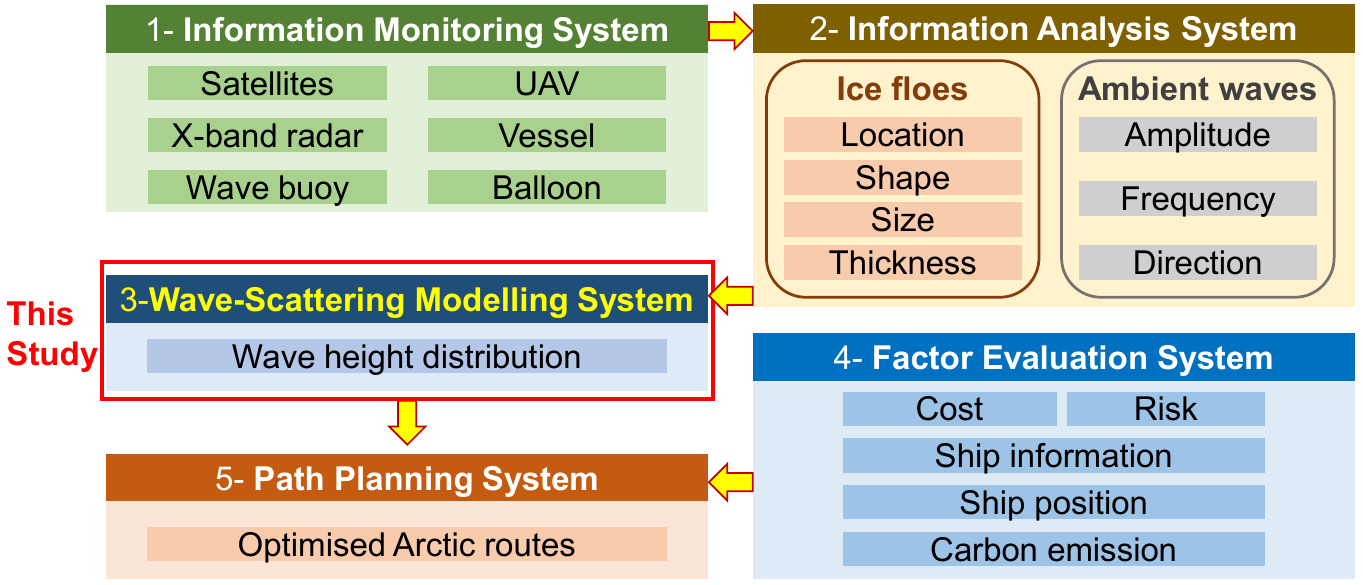}
	\caption{Project structure of Arctic route planning system for Xue Long icebreaker}
	\label{fig:arctic}
\end{figure}

This study is motivated by the practical maritime requirements of the Chinese research icebreaker Xue Long.
Maritime safety issues related to ship navigation in polar waters are particularly focused on collision avoidance in environments containing icebergs or deep-draught large ice floes.
An intelligent route planning system is demanded for ship navigation in Arctic waters.
This study forms part of the Arctic route planning system developed by the authors’ team since 2020.
As shown in Fig. \ref{fig:arctic}, the system consists of five subsystems: the information monitoring system (IMS), the information analysis system (IAS), the wave-scattering modelling system (WMS), the factor evaluation system (FES), and the path planning system (PPS).
The purpose of IMS is to collect environment information of Arctic routes through satellites, unmanned aerial vehicles (UAVs), ship-based X-band radars, wave buoys, weather balloons, and so on.
Different formats of information data are input into the IAS. 
Through IAS, the information of ice floes and ambient wave parameters is recognized.
With the ice and wave information known, the WMS is used to calculate the water-wave scattering field. 
Specifically, the distribution of wave height in waterways formed by ice floes is confirmed.
Together with a comprehensive evaluation of factors such as ship information, cost, risks, carbon emission, and so on, the PPS provides an optimized Arctic route. 
The Arctic route planning system operates dynamically, keeping up with the real-time updates of the IMS.
The mathematics behind the WMS is the focus of this study.

The interaction theory forms a promising framework for the WMS of the Arctic route planning system. 
The determination of the DTM for an individual floe model forms the core of all variants of the interaction theory.
However, the scenarios examined in the aforementioned studies presented in Subsec. \ref{subsec:intro3} focus primarily on thin ice plates.
For ice-plate models, analytical or semi-analytical solutions for deriving DTM have been well established for regular-shaped ice plates, especially circular ice plates.
Analytical solutions can be derived using eigenfunction expansions, separation of variables, the Wiener–Hopf technique, and other methods of mathematical physics.
The ``velocity potential'', as a mathematical variable that cannot be intuitively measured in the physical fluid domain, is solved to construct the DTM.
In summary, the conventional approach to obtaining DTM involves establishing a solvable boundary value problem for wave-ice interaction, typically within the framework of linear potential-flow theory, and solving the well-posed problem using advanced mathematical methods. 
Each individual element of the DTM is explicitly and accurately represented by the obtained solutions.
In this context, the conventional approach for deriving DTM constitutes a fully transparent ``white-box'' model throughout the entire process.

The ``white-box'' model is indeed mathematically elegant and accurate, provided that a precise, complete, and solvable boundary value problem can be formulated.
However, unlike the thin ice-plate models addressed in the aforementioned literature, this study focuses on large icebergs and massive ice floes with considerable draught, which is of significance to the Xue Long icebreaker.
Such ice floes often exhibit distinct arbitrary three-dimensional geometries, characterized by irregular shapes, complex sub-ice surfaces, variable thickness, and even the presence of underwater cracks.
Deriving analytical or semi-analytical solutions of the velocity potential for such irregular ice floes thus becomes extremely difficult or even infeasible.
Therefore, a more easily implemented and universally applicable approach should be developed to obtain DTM, thereby integrating the interaction theory into the Arctic route planning system.
This constitutes the core contribution of the present study.

\color{black}
\subsection{Core contribution and novelty of present study: A ``black-box'' model for DTM}

The core contribution and novelty of this study lie in the proposal of a ``black-box'' model for constructing the DTM of three-dimensional ice floes with arbitrarily complex geometry, namely the wave component detection (WCD) method.
Using this method, it is no longer necessary to accurately establish or analytically solve a well-posed boundary-value problem for wave-ice interaction.
Instead, the DTM for each ice floe is regarded as a ``black box'' that receives incident waves and emits disturbed waves.
The internal structure of the black box  (i.e., the elements of the DTM) is detected by transmitting a sufficient number of predefined distinct incident plane waves to each ice floe (as ``excitation'' to a system) and examining the diffracted wave field (as ``response'' of the system).
The diffracted wave field around each isolated ice floe is constructed via a fitting procedure based on free-surface elevations measured at a set of ``virtual wave gauges''.
Through input-output mapping of the characteristic parameter arrays for all pairs of ``excitation'' and ``response'', the internal structure of the black box (i.e., the elements of the DTM) can be determined through sufficient tests.

It should be highlighted that the diffracted wave field (i.e., the ``response'' of each ice floe) can be obtained via efficient online calculations using any suitable numerical solver, data-driven predictions based on an experimental database, or any other user-specified tools.
Therefore, the proposed ``black-box'' model provides an easily implemented and universally applicable method for constructing the DTM of three-dimensional ice floes with arbitrarily complex geometry.

Further incorporation of the WCD approach into the framework of classical interaction theory yields the ``enhanced interaction (EI) method'' of the present study.
Combined with an self-executing script that further integrates the IAS output ports and a fast automatic mesh generator, the EI method is practically applied in the WSM of the Arctic route planning system for calculating the wave field among multiple ice floes with arbitrary geometries.
This method retains the overall framework of classical interaction theory while extending to three-dimensional ice floes of arbitrary geometry, providing a more efficient, flexible, and portable solution for wave scattering predictions in complex MIZ environment.
The present EI method fits into the existing knowledge structure of methodologies for ice-floe wave scattering problems, as shown in Fig. \ref{fig:liter}.

Sec. \ref{sec:theory} describes the mathematics behind the EI method. 
The implementation details of the WCD approach and the numerical procedure of the EI method are presented in detail.
Sec. \ref{sec:resul} examines the effectiveness of the WCD approach and the EI method.
The super-efficiency performance of the EI method in handling an ultra-large group of complex-shaped ice floes is verified.
An example of optimal path planning is provided to show how the EI method is incorporated into the Arctic route planning system.
Sec. \ref{sec:discuss} provides further discussion on the advantages and limitations of the ``black-box'' model.
Conclusions are given in Sec. \ref{sec:concl}.
\color{black}

\section{Enhanced interaction method for wave scattering of ice floes}\label{sec:theory}
\color{black}
\subsection{Theoretical assumptions and problem definition}\label{subsec:define}

Based on the practical maritime operational requirements of the Chinese research icebreaker Xue Long, three background assumptions are initially proposed.
First, this study specifically accounts for large icebergs and massive ice floes of considerable thickness.
Owing to their size, spatial geometry and deep draught, these ice floes can be reasonably approximated as rigid bodies, with elastic deformation neglected.
Second, the ice floes are assumed to be fixed in space, and wave radiation effects induced by floe motion can therefore be reasonably neglected.
This assumption is justified by their large mass and inertia, which lead to relatively small wave-induced motions. 
Wave radiation arising from floe motion is weak and exerts a limited influence on the overall wave field. 
Third, short-term near-field forecasts of the wave environment for the upcoming approximately 3–6 hours are required to support navigation and maneuvering decisions.
For such short prediction windows, the horizontal drift of large ice floes is generally limited relative to the wavelength of ocean waves and can be reasonably approximated as fixed.

\begin{figure}[htbp]
	\centering
	\includegraphics[width=0.5\linewidth]{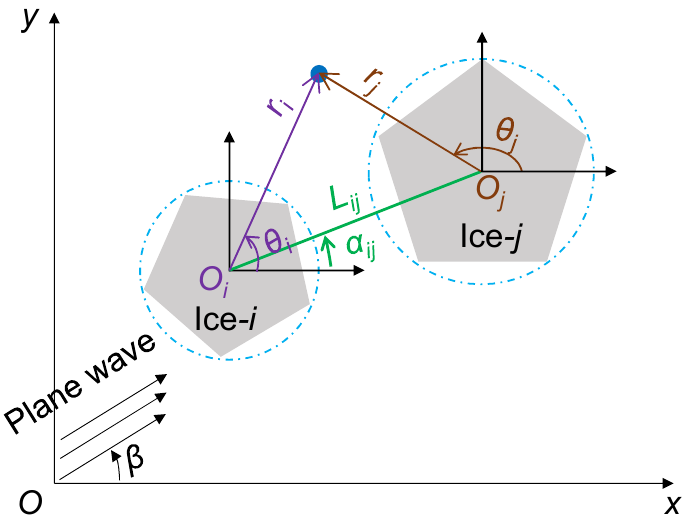}
	\caption{Definition of coordinate systems}
	\label{fig:coord_sys}
\end{figure}

\color{black}
To describe the EI method, a global Cartesian coordinate system, $O-xyz$, is established on the still water surface.
The $Oz$ axis points upward. 
Assume there are $N$ ice floes with arbitrary shapes.
The $i$-th ice floe is called “ice-$i$” for short.
For each ice floe, a local polar coordinate system $O-r\theta$ is established on the $O-xy$ plane. 
The origin is at the center of the circumcircle of the ice floe on the still water surface.
As defined in Fig. \ref{fig:coord_sys}, the origin of the local system on  ice-$i$, $O_i$, is located at $(X_i, Y_i)$ in $O-xy$. 
The polar coordinates expressed in the local system of ice-$i$ is $(r_i, \theta_i)$.

\color{black}
Assuming the fluid is incompressible, non-viscous, and flow-irrotational, the potential-flow theory can be used to describe the wave field.
The velocity potential $\Phi(x,y,z,t)$, whose gradient is the fluid velocity, is defined. 
Here, $t$ represents the time.
According to linear potential-flow theory, $\nabla^2 \Phi = 0$ holds within the fluid domain, with $\partial \Phi/\partial n = 0$ on the seabed and the mean wetted surface of ice floes, and 
\begin{equation} 
	\eta = -\frac{1}{g}\frac{\partial \Phi}{\partial t} \textrm{ and }
	\frac{\partial \Phi}{\partial z} = \frac{\partial \eta}{\partial t}
\end{equation}
on the mean free-surface position,
where $\partial\Phi/\partial n$ denotes the normal derivative of $\Phi$ on the boundary surface, $\eta(x,y,t)$ is the free-surface elevation, and $g$ is the gravitational acceleration.
\color{black} 

In steady-state wave conditions with small wave amplitude, the velocity potential and free-surface elevation can be further expressed as
\begin{equation} 
	\Phi(x,y,z,t) = \mathrm{Re} \left \{ {\phi (x,y,z)e^{\mathbb{i}\omega t}} \right \} 
	\text{ and }
	\eta(x,y,t) = \mathrm{Re} \left \{ {\bar{\eta} (x,y)e^{\mathbb{i}\omega t}} \right \},
\end{equation}
where $\mathrm{Re}$ denotes the real part of a complex quantity, $\phi$ is the complex velocity potential,  $\bar{\eta}$ is the complex elevation, $\omega$ is the angular frequency of wave motion, and $\mathbb{i}=\sqrt{-1}$.
The following relationship is satisfied
\begin{equation} \label{eq:potent_ele}
	\bar{\eta}(x,y)= \frac{\mathbb{i} \omega}{g} \phi(x,y,0).
\end{equation}

The wave field around $N$ ice floes is formed by the ambient plane wave and the scattered waves from each ice floe.
The total complex velocity potential  can be expressed as
\begin{equation} \label{eq:total_potent_field}
	\phi = \phi^{\text{A}} + \sum_{\substack{j=1}}^{N} \phi_{j}^{\text{S}},
\end{equation}
where the superscript symbol A indicates the quantity of the ambient plane wave, S indicates that of the scattered wave, and the subscript $j$ refers to ice-$j$.
Employing Eq. (\ref{eq:potent_ele}) to $\phi^{\text{A}}$ and $\phi_{j}^{\text{S}}$ in Eq. (\ref{eq:total_potent_field}) results in the corresponding decomposition of free-surface elevation profiles
\begin{equation} \label{eq:total_ele_field}
	\bar{\eta} = \bar{\eta}^{\text{A}} + \sum_{\substack{j=1}}^{N} \bar{\eta} _{j}^{\text{S}}.
\end{equation}

\subsection{Decomposition of incident waves acting on each ice floe}

\begin{figure}[htbp]
	\centering
	\includegraphics[width=0.5\linewidth]{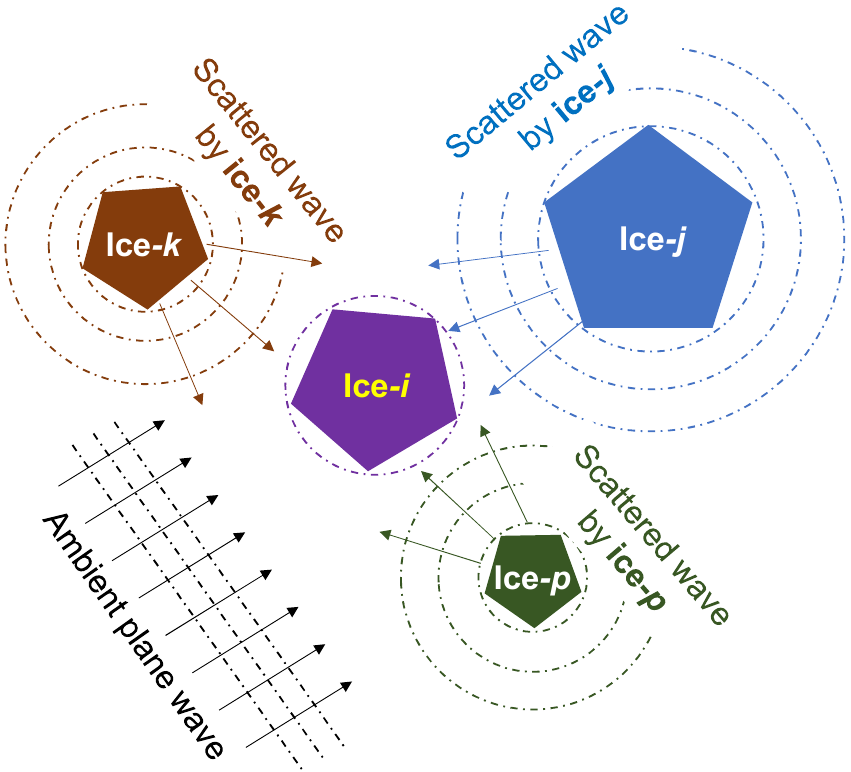}
	\caption{Demonstration of incident waves acting on ice-$i$}
	\label{incident_wave_demo}
\end{figure}

This subsection takes ice-$i$ as the object of research. 
The total incident waves acting on ice-$i$ comprise the ambient plane wave and the scattered waves from all the other ice floes, as shown in Fig. \ref{incident_wave_demo}. 
Accordingly, the complex velocity potential of the total incident wave, denoted as $\phi_i^{\text{I}}$, can be decomposed in the local system $O_i - r_i\theta_i$ as
\begin{equation} \label{eq:potent_sum}
	\phi_i^{\text{I}} = \phi_i^{\text{A}} + \sum_{\substack{j=1\\j\ne i}}^{N} \phi _{ij}^{\text{S}},
\end{equation}
where $\phi_i^{\text{A}}$ is the complex velocity potential of the ambient plane wave, and $\phi _{ij}^{\text{S}}$ denotes that of the scattered wave from ice-$j$.

The complex velocity potential of the ambient plane wave can be expanded using vertical eigenfunctions in the polar coordinate system $O_i - r_i\theta_i$
\begin{equation}\label{eq:plane_wave}
	\phi _i^{\text{A}} = \mathbb{i}\frac{g}{\omega } \sum_{m=-\infty }^{\infty } Ae^{\mathbb{i}k(X_i\cos \beta +Y_i\sin \beta )}\frac{\cosh k(z+h)}{\cosh kh} J_m(kr_i)e^{-\mathbb{i}m(\beta +\pi /2)}e^{\mathbb{i}m\theta _i},
\end{equation}
where $A$ is the complex wave amplitude, $\beta$ is the wave incidence angle, $k$ is the wave number, $h$ is the water depth, and $J_m$ is the $m$-th order Bessel function of the first kind.
The wave number $k$ is determined by the dispersion equation.
\begin{equation} 
		\omega ^2 = gk\tanh kh.
\end{equation} 
In practice, the infinite summation in Eq. (\ref{eq:plane_wave}) can be truncated to a summation of a finite number of terms.
For $m$ ranging from $-M$ to $M$, Eq. (\ref{eq:plane_wave}) can be expressed in a neater form
\begin{equation} \label{eq:plane_wave_neater}
	\phi_i ^{\text{A}} = \mathbb{i}\frac{g}{\omega} \left ( \boldsymbol{a}_i ^{\text{A}} \right ) ^{\top } \Psi_i ^{\text{A}},
\end{equation}
where $\boldsymbol{a}_i ^{\text{A}}$ is the coefficient vector of ambient plane waves, and $\Psi_i ^{\text{A}}$ is the vector of the basis functions.
The superscript T indicates the vector or matrix transpose.
The length of both $\boldsymbol{a}_i^{\text{A}}$ and $\Psi_i^{\text{A}}$ is $2M + 1$.
The $m$-th elements of $\boldsymbol{a}_i^{\text{A}}$ and $\Psi_i^{\text{A}}$ possess the following expressions
\begin{equation}\label{eq:incident_wave_coeff}
	(\boldsymbol{a}_i ^{\text{A}})_m = A e^{-\mathbb{i}k\left ( X_i\cos \beta + Y_i\sin \beta  \right ) } e^{-\mathbb{i}m\left ( \beta + \frac{\pi }{2}  \right )},
\end{equation}
and 
\begin{equation} 
	(\Psi_i ^{\text{A}} )_m = \frac{\cosh k(z+h)}{\cosh kh} J_m(kr_i)e^{\mathbb{i}m\theta _i}.
\end{equation}

The complex velocity potential of the scattered wave by ice-$j$ can be expressed in the local system $O_j - r_j\theta_j$
\begin{equation} \label{eq:scat_potent}
	\phi_{j} ^{\text{S}} = \mathbb{i}\frac{g}{\omega } (\boldsymbol{a}_j ^{\text{S}}) ^{\top } \Psi_j ^{\text{S}},
\end{equation}
where $\boldsymbol{a}_j^{\text{S}}$ is the coefficient vector of the scattered wave, and $\Psi_j^{\text{S}}$ is the associated vector of basis functions.
The length of $\boldsymbol{a}_j^{\text{S}}$ and $\Psi_j^{\text{S}}$ is the same as that of $\boldsymbol{a}_i^{\text{A}}$ and $\Psi_i^{\text{A}}$, which is $2M + 1$.
The $m$th element of $\Psi_j ^{\text{S}}$ is
\begin{equation} \label{eq:scat_basis_func}
	(\Psi_j ^{\text{S}} )_m = \frac{\cosh k(z+h)}{\cosh kh} H_m^{(2)}(kr_j)e^{\mathbb{i}m\theta _j},
\end{equation}
where $H_m^{(2)}$ is the $m$-th order Hankel function of the second kind.

According to Graf's addition theorem \citep{Abramowitz1964}, an $m$-th order Hankel function of the second kind in $O_j - r_j\theta_j$ can be expressed using Bessel functions of the first kind in $O_i - r_i\theta_i$
\begin{equation} 
	H_m^{(2)}(kr_j)e^{\mathbb{i}m\theta _j} = \sum_{n = -\infty }^{\infty } H_{m-n}^{(2)}(kL_{ij})e^{\mathbb{i}(m-n)\alpha _{ij}}J_n(kr_i)e^{\mathbb{i}n\theta _i},
\end{equation}
where $L_{ij}$ is the distance between $O_i$ and $O_j$, and $\alpha_{ij}$ denotes the direction angle of $O_i$ with respect to $O_j$.
Here, $(r_i,\theta_i)$ and $(r_j,\theta_j)$ represent the same point in different local coordinate systems.
The $m$-th element of the basis function $\Psi_j ^{\text{S}}$ in Eq. (\ref{eq:scat_basis_func}) can be represented in the local system $O_i-r_i\theta_i$ as
\begin{equation} 
	(\Psi_j ^{\text{S}} )_m = \frac{\cosh k(z+h)}{\cosh kh} \sum_{n = -\infty }^{\infty } H_{m-n}^{(2)}(kL_{ij})e^{\mathbb{i}(m-n)\alpha _{ij}}J_n(kr_i)e^{\mathbb{i}n\theta _i}
= \sum_{n = -\infty }^{\infty } H_{m-n}^{(2)}(kL_{ij})e^{\mathbb{i}(m-n)\alpha _{ij}}(\Psi_i ^{\text{A}} )_n.
\end{equation}
The basis function vectors satisfy the following relationship
\begin{equation}\label{eq:scat_amb_trans}
\Psi_j^{\text{S}} = [T_{ij}]\Psi_i^{\text{A}},
\end{equation}
where $[T_{ij}]$ is a transfer matrix of basis functions of size $(2M+1)\times(2M+1)$.
The element of $[T_{ij}]$ in row-$m$ and column-$n$ is 
\begin{equation} \label{eq:trans_matrix_basis}
	[T_{ij}]_{mn} = H_{m-n}^{(2)}(kL_{ij})e^{i(m-n)\alpha _{ij}}.
\end{equation}
By substituting Eq. (\ref{eq:scat_amb_trans}) into Eq. (\ref{eq:scat_potent}), the complex velocity potential $\phi_{j}^{\text{S}}$ can be rewritten as
\begin{equation} \label{eq:scat_potent_Oi}
	\phi_{ij} ^{\text{S}} = \mathbb{i}\frac{g}{\omega } (\boldsymbol{a}_j ^{\text{S}}) ^{\top }[T_{ij}]\Psi_i^{\text{A}},	
\end{equation}
where $\phi_{j}^{\text{S}}$ is replaced by $\phi_{ij}^{\text{S}}$ to indicate an expression in $O_i - r_i\theta_i$.

By substituting Eqs. (\ref{eq:plane_wave_neater}) and (\ref{eq:scat_potent_Oi}) into Eq. (\ref{eq:potent_sum}), the complex velocity potential of the total incident waves acting on ice-$i$ gets a new form in $O_i - r_i\theta_i$
\begin{equation} \label{eq:total_incident_potent}
	\phi _i^{\text{I}} = \mathbb{i}\frac{g}{\omega } (\boldsymbol{a}_i^{\text{A} })^{\top }\Psi_i^{\text{A}} + \mathbb{i}\frac{g}{\omega }\sum_{\substack{j=1\\j\ne i}}^{N}(\boldsymbol{a}_j^{\text{S} })^{\top }\Psi_j^{\text{S}}
=  \mathbb{i}\frac{g}{\omega }\left [ (\boldsymbol{a}_i^{\text{A}})^{\top } +  \sum_{\substack{j=1\\j\ne i} }^{N} (\boldsymbol{a}_j^{\text{S}})^{\top } [T_{ij}] \right ] \Psi_i^{\text{A} }  
=\mathbb{i}\frac{g}{\omega } (\boldsymbol{a}_i^{\text{I}})^{\top } \Psi_i^{\text{A} }.
\end{equation}

\subsection{Diffraction transfer matrix and interaction theory} 

Analogous to Eq. (\ref{eq:scat_potent}), the complex velocity potential of the scattered wave by ice-$i$ can be written as
\begin{equation} \label{eq:scat_potent_ice_i}
	\phi_{i} ^{\text{S}} = \mathbb{i}\frac{g}{\omega } (\boldsymbol{a}_i ^{\text{S}}) ^{\top } \Psi_i ^{\text{S}}.
\end{equation}
Eqs. (\ref{eq:total_incident_potent}) and (\ref{eq:scat_potent_ice_i}) stand for the complex velocity potential of the total incident wave and the scatter wave of an arbitrary ice-$i$.
Since $\boldsymbol{a}_i^{\text{A}}$, $\Psi_j^{\text{A}}$ and $\Psi_j^{\text{S}}$ are known in advance, the coefficient vectors $\boldsymbol{a}_i^{\text{I}}$ and $\boldsymbol{a}_i^{\text{S}}$ completely determine the incident and scattered wave fields around ice-$i$ respectively.
A unique diffraction transfer matrix (DTM), $[\boldsymbol{D}_i]$, exists, which can transform the coefficient vectors from $\boldsymbol{a}_i^{\text{I}}$ to $\boldsymbol{a}_i^{\text{S}}$.
Specifically, the following relationship is satisfied
\begin{equation} 
	\boldsymbol{a}_i^{\text{S}} 
	= [\boldsymbol{D}_i] 
	\left [ 
	\boldsymbol{a}_i ^{\text{A}} 
	+  \sum_{\substack{j=1\\j\ne i} }^{N} [T_{ij}]^{\top }\boldsymbol{a}_j^{\text{S}}
	\right ].
\end{equation}
The size of $[\boldsymbol{D}_i]$ is $(2M+1)\times(2M+1)$.

Going through all ice floes from ice-1 to ice-$N$, the coefficient vectors $a_i^{\text{S}}$ of each ice floe satisfy the following relationship
\begin{equation} \label{eq:lin_eq_system}
	\begin{bmatrix}
		\boldsymbol{a}_1^{\text{S}} \\
		\boldsymbol{a}_2^{\text{S}} \\
		\vdots \\
		\boldsymbol{a}_N^{\text{S}} 
	\end{bmatrix} = 
	\begin{bmatrix}
		[\boldsymbol{D}_1]& \boldsymbol{0} & \dots  & \boldsymbol{0}\\
		\boldsymbol{0}  & [\boldsymbol{D}_2] &  & \vdots \\
		\vdots  &  & \ddots  & \boldsymbol{0}\\
		\boldsymbol{0}  & \dots  & \boldsymbol{0} & [\boldsymbol{D}_N]
	\end{bmatrix} 
	\left (  
	\begin{bmatrix}
		\boldsymbol{a}_1^{\text{A}}\\
		\boldsymbol{a}_2^{\text{A}} \\
		\vdots  \\
		\boldsymbol{a}_N^{\text{A}}
	\end{bmatrix} +
	\begin{bmatrix}
		\boldsymbol{0} & [T_{12}]^{\top } & \dots  & [T_{1N}]^{\top }\\
		[T_{21}]^{\top }  & \boldsymbol{0} &  & \vdots \\
		\vdots  &  & \ddots & [T_{(N-1)N}]^{\top } \\
		[T_{N1}]^{\top }  & \dots & [T_{N(N-1)}]^{\top } & \boldsymbol{0}
	\end{bmatrix} 
	\begin{bmatrix}
		\boldsymbol{a}_1^{\text{S}} \\
		\boldsymbol{a}_2^{\text{S}} \\
		\vdots \\
		\boldsymbol{a}_N^{\text{S}} 
	\end{bmatrix}
	\right ).
\end{equation}
As long as the DTM $[\boldsymbol{D}_i]$, and the relative locations of ice floes (which determine $[\boldsymbol{T}_{ij}]$) are known for each ice floe, the complex velocity potential of the scattered wave by any ice floe can be found by solving the system of linear equations in Eq. (\ref{eq:lin_eq_system}).

Once $\boldsymbol{a}_i^{\text{S}}$ is known, the complex velocity potential of the scattered wave by each ice floe can be found using Eq. (\ref{eq:scat_potent}). 
The total complex velocity potential can be calculated with Eq. (\ref{eq:total_potent_field}). 
The distribution of the free-surface elevation can be computed using Eq. (\ref{eq:total_ele_field}).

\subsection{Wave component detection approach: ``Black-box'' model for DTM}

\color{black}

For ice floes with arbitrary three-dimensional geometries, analytical solutions required to construct the DTM are generally unavailable.
The wave component detection (WCD) approach is a more easily implemented and universally applicable approach to obtain DTM.
The overall idea of this approach is first outlined as follows.
A set of ``virtual wave gauges (VWGs)'', serving as wave-amplitude monitors, is arranged around an isolated ice floe.
Using the WCD approach, the DTM is essentially regarded as a ``black box'' that embodies the inherent wave diffraction characteristics corresponding to the hydrodynamic geometry of each individual ice floe.
To detect the internal structure of the black box, predetermined incident plane waves are transmitted to each ice floe from different directions (as ``excitation'' to the system).
The diffracted wave fields around the ice floe are recorded by the VWGs and reconstructed via surface fitting (as ``response'' of the system).
Through input–output mapping of each ``excitation'' to the corresponding ``response'', the internal structure of the black box can be identified following sufficient tests.
VWGs are arranged regularly in both circumferential and radial directions around each individual ice floe to capture the spatial distribution characteristics of the wave field.
It should be noted that this wave diffraction problem can be simulated using any efficient numerical solver.
It is also possible to generate the response wave field using intelligent data-driven methods based on experimental, numerical or analytical databases.

Mathematical details for identifying elements of the DTM by comparing incident wave parameters with those obtained from VWGs' measurements are presented below. 
Details of the VWGs' distribution and data processing procedures are provided to ensure reproducibility of the proposed method.
\color{black}

The DTM of each ice floe, i.e. $[\boldsymbol{D}_i]$, is key to Eq. (\ref{eq:lin_eq_system}) in the interaction theory.
For any ice floe, the total incident waves and scattered waves are connected through the DTM.
The matrix reflects the inherent wave diffraction characteristics of an isolated ice floe, which is independent of the input of incident waves.
This study proposes the WCD approach for establishing the DTM. 

\begin{figure}[htbp]
	\centering
	\includegraphics[width=0.5\linewidth]{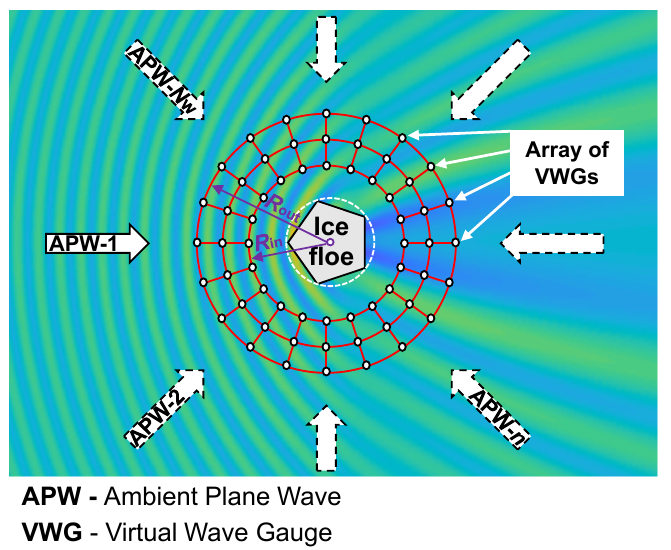}
	\caption{Ambient plane waves and array of virtual wave gauges (VWGs) around an isolate ice floe model in wave component detection approach}
	\label{fig:wave_gauge}
\end{figure}

As shown in Fig. \ref{fig:wave_gauge}, an array of VWGs is deployed around an isolated ice floe.
For convenience, the VWGs can be regularly deployed in an annular zone surrounding the ice floe. 
The inner radius of this zone is denoted by $R_{\rm{in}}$, and the outer radius is $R_{\rm{out}}$. 
There are $N_{\rm{cir}}$ VWGs along the circumferential direction and $N_{\rm{rad}}$ in the radial direction. 
The total number of VWGs is $N_{\rm{p}} = N_{\rm{cir}} \times N_{\rm{rad}}$.
The position of the $j$-th VWG has coordinates $(r_j,\theta_j)$ in the local system of the ice floe.
A series of unit-amplitude ambient plane waves from $N_{\rm{w}}$ directions are shot at each isolated ice floe to test its wave diffraction characteristics.
The incidence angle of the $n$-th wave is $\beta_n = 2n\pi/N_{\rm{w}}$. 

Taking ice-$i$ as an example, the coefficient vector of the ambient plane wave with incidence angle $\beta_n$ is denoted as $\boldsymbol{a}_i^{\text{A},\beta_n}$ (determined by Eq. (\ref{eq:incident_wave_coeff})).
Any frequency-domain hydrodynamic solver (e.g. WAMIT, HydroSTAR, Nemoh, HAMS, WAFDUT) can be used to calculate the complex free-surface elevation $\bar{\eta}$ at the locations of the VWGs. 
The complex elevation of the scattered wave at the $j$-th VWG can be obtained by subtracting $\bar{\eta}^{\rm{A}}_i(r_j,\theta_j)$ from $\bar{\eta}_i(r_j,\theta_j)$, resulting in $\bar{\eta}^{\rm{S}}_i(r_j,\theta_j)$.
Combining Eqs. (\ref{eq:scat_potent_ice_i}) with (\ref{eq:potent_ele}) results in the expression
\begin{equation} \label{eq:scat_ele}
	\bar{\eta}_i^{\text{S},\beta_n}(r_j,\theta_j) = -(\boldsymbol{a}_i^{\text{S},\beta_n}) ^{\top } \Psi_i ^{\text{S0}}(r_j,\theta_j)
\textrm{, with } j=1,2,\cdots,N_{\rm{p}},
\end{equation}
where the $\Psi ^{\text{S0}}$ is a vector with a length of $2M+1$.
The $m$-th element of $\Psi ^{\text{S0}}$ is
\begin{equation} 
	(\Psi^{\text{S0}})_m = H_m^{(2)}(k r_j)e^{im\theta_j}.
\end{equation}
Here, $\beta_n$ is appended to the superscript of $\bar{\eta}_i^{\text{S}}$ and $\boldsymbol{a}_i^{\text{S}}$ to mark the direction of the incident wave.
If $N_{\rm{p}} = 2M + 1$, the linear system of equations represented by Eq. (\ref{eq:scat_ele}) can be solved to find the elements of the coefficient vector $\boldsymbol{a}_i^{\text{S},\beta_n}$. 
Otherwise, if $N_{\rm{p}} > 2M + 1$, the system is overdetermined. 
In this case, a least-squares solution can be found for the unknown vector $\boldsymbol{a}_i^{\text{S},\beta_n}$.

The coefficient vectors of ambient plane waves and scattered waves are connected through the DTM as follows
\begin{equation} \label{eq:DTM_ice_i}
\left[\boldsymbol{a}_i^{\text{S},\beta_1}, \boldsymbol{a}_i^{\text{S},\beta_2}, \cdots, \boldsymbol{a}_i^{\text{S},\beta_{N_{w}}}\right]_{(2M+1) \times N_{\rm{w}}}
=[\boldsymbol{D}_i]_{(2M+1) \times (2M+1) }
\left[\boldsymbol{a}_i^{\text{A},\beta_1}, \boldsymbol{a}_i^{\text{A},\beta_2}, \cdots, \boldsymbol{a}_i^{\text{A},\beta_{N_{w}}}\right]_{(2M+1) \times N_{\rm{w}}}.
\end{equation}
Transposing both sides of Eq. (\ref{eq:DTM_ice_i}) leads to 
\begin{equation} \label{eq:DTM_ice_i2}
\left[\boldsymbol{a}_i^{\text{S},\beta_1}, \boldsymbol{a}_i^{\text{S},\beta_2}, \cdots, \boldsymbol{a}_i^{\text{S},\beta_{N_{w}}}\right]^{\top}_{ N_{\rm{w}}\times(2M+1)}
=\left[\boldsymbol{a}_i^{\text{A},\beta_1}, \boldsymbol{a}_i^{\text{A},\beta_2}, \cdots, \boldsymbol{a}_i^{\text{A},\beta_{N_{w}}}\right]^{\top}_{ N_{\rm{w}}\times(2M+1)}
[\boldsymbol{D}_i]^{\top}_{(2M+1) \times (2M+1) }.
\end{equation}
Each column of $[\boldsymbol{D}_i]^{\top}$ can be solved independently.
Similar to Eq. (\ref{eq:scat_ele}), $N_{\rm{w}}$ should be larger than or equal to $2M + 1$. 
When $N_{\rm{w}} = 2M + 1$, the linear system of equations for any column of $[\boldsymbol{D}_i]^{\top}$ can be solved directly. 
If $N_{\rm{w}} > 2M + 1$, the system for each column of $[\boldsymbol{D}_i]^{\top}$ is overdetermined. 
In this case, a least-squares solution can be found for each column of $[\boldsymbol{D}_i]^{\top}$.
Applying the same procedure to the remaining ice floes from ice-1 to $N$, the DTM $[\boldsymbol{D}_i]$ for each ice floe in Eq.(\ref{eq:lin_eq_system}) can be solved.

\subsection{Numerical procedure of enhanced interaction method}

\begin{figure}[htbp] 
	\centering
	\includegraphics[width=0.85\linewidth]{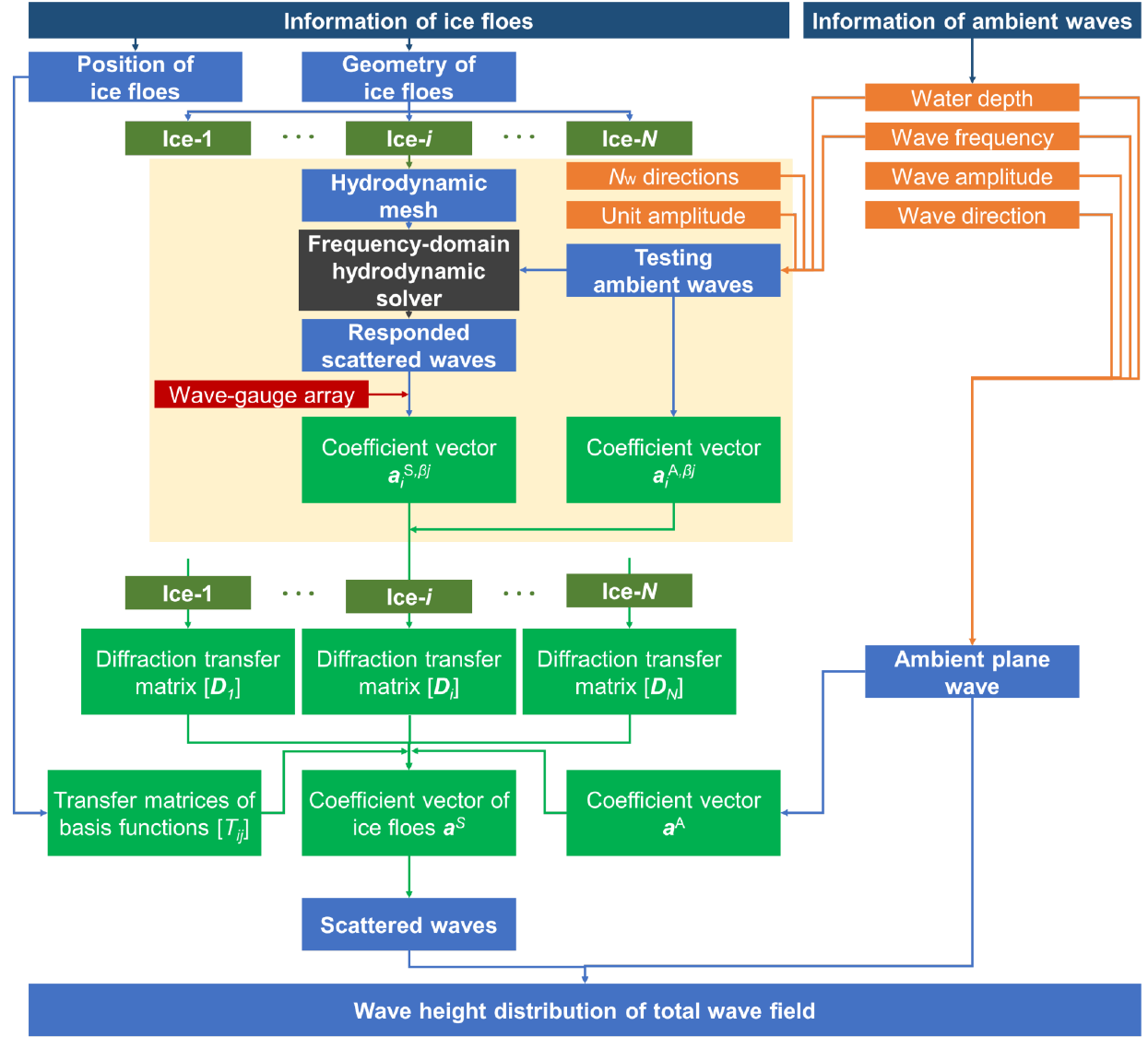}
	\caption{Structure of wave-scattering modelling system}
	\label{fig:diagram}
\end{figure}

The present EI method is formed by introducing the proposed WCD approach to the classical interaction theory continuously developed by 
\cite{kagemoto1986interactions}, 
\cite{peter2004infinite}, and others.
The numerical procedure for using the enhanced interaction method is shown in Fig. \ref{fig:diagram}, which also reflects the brief inner structure of the WMS module in Fig. \ref{fig:arctic}.
The IAS module of the Arctic route planning system provides ambient wave and ice floe information to the WMS module. 
The WMS module then outputs the wave height distribution of the total wave field around the ice floes for the PPS module.

The wave information mainly includes the frequency, amplitude, and direction of ambient water waves before they enter the MIZ, as well as the water depth.
Since this study mainly considers swell-type ambient waves, the frequency, amplitude, and direction of water waves can be relatively easily measured and approximated in open water.
Once the information of the ambient plane waves is determined, the coefficient vector $\boldsymbol{a}_i^{\text{A}}$ can be found through Eq. (\ref{eq:incident_wave_coeff}).
Ice floe information falls into two categories: the position information of all ice floes and the geometry information (such as shape and draught) of each ice floe.

Using the position information, the transfer matrix of basis functions $[T_{i,j}]$ for every two ice floes (for example, taking ice-$i$ and ice-$j$) can be calculated through Eq. (\ref{eq:trans_matrix_basis}).
Once the geometry information of each ice floe is known, an in-house hydrodynamic mesh generator is activated. 
Boundary elements are generated automatically on the mean wet surface of the ice floe.
With the same water depth and wave frequency, a series of unit-amplitude testing ambient waves from $N_{p}$ directions are prepared.
The coefficient vector of each testing ambient wave, $\boldsymbol{a}_i^{\text{A},\beta_n}$, is determined via Eq. (\ref{eq:incident_wave_coeff}).
A frequency-domain hydrodynamic solver is called to calculate the scattered waves for each ice floe in response to the action of each testing wave.
The complex free-surface elevations, $\bar{\eta}_i^{\text{S}, \beta_n}$, at the preset wave-gauge positions can be determined.
The coefficient vector $\boldsymbol{a}_i^{\text{S}, \beta_n}$ is found by solving Eq. (\ref{eq:scat_ele}).
Given that $\boldsymbol{a}_i^{\text{A},\beta_n}$ and $\boldsymbol{a}_i^{\text{S}, \beta_n}$ are known, the DTM $[\boldsymbol{D}_i]$ for each ice floe can be solved using Eq. (\ref{eq:DTM_ice_i}).

Finally, the coefficient vector of the actual scattered wave by each ice floe, $\boldsymbol{a}_i^{\text{S}}$, in the ambient plane wave can be solved through Eq. (\ref{eq:lin_eq_system}).
According to Eqs. (\ref{eq:scat_potent}) and (\ref{eq:potent_ele}), the scattered wave field generated by each ice floe is determined.
By combining all scattered waves with the ambient plane wave, Eq. (\ref{eq:total_ele_field}) can be used to generate the wave height distribution of the total wave field around the ice floes.

\section{Results and discussion}\label{sec:resul}

\begin{figure}[htbp] 
	\centering
	\includegraphics[width=0.75\linewidth]{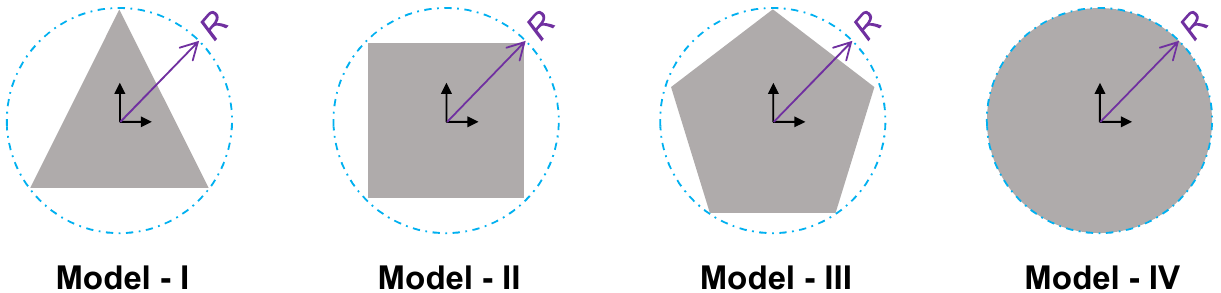}
	\caption{Projection shapes of ice floes under consideration}
	\label{fig:ice_shape}
\end{figure}

The EI method can be applied to ice floes of arbitrary shapes. 
Without losing generality, four shapes of ice floes are taken as example models to demonstrate the capability of the method.
As shown in Fig. \ref{fig:ice_shape}, the horizontal cross sections of four ice models are a regular triangle, a square, a regular pentagon, and a circle respectively. 
For convenience, these four ice models are named Model-I, Model-II, Model-III, and Model-IV. 
The circumradius of the horizontal projection of each model is $R$. 
The draught of each model is denoted by $h_{\rm{D}}$.
In this study, the geometrical parameters are set as $R = 10$ m and $h_{\rm{D}} = 0.3R$.
The water depth is $h = 10R$. 
The wave amplitude of ambient incident waves is $A = 0.1R$.
Two typical wave periods are considered: $T = 3$ s and $T = 10$ s.
The ambient incident waves with $T = 3$ s and $T = 10$ s are called the short waves and long waves, respectively.
The wavelengths of the short and long waves are 14 m and 156 m, respectively.

\begin{figure}[!htp]
\centering
{
  \begin{minipage}{0.22\linewidth}
  \centering
  \includegraphics[width=1.0\linewidth]{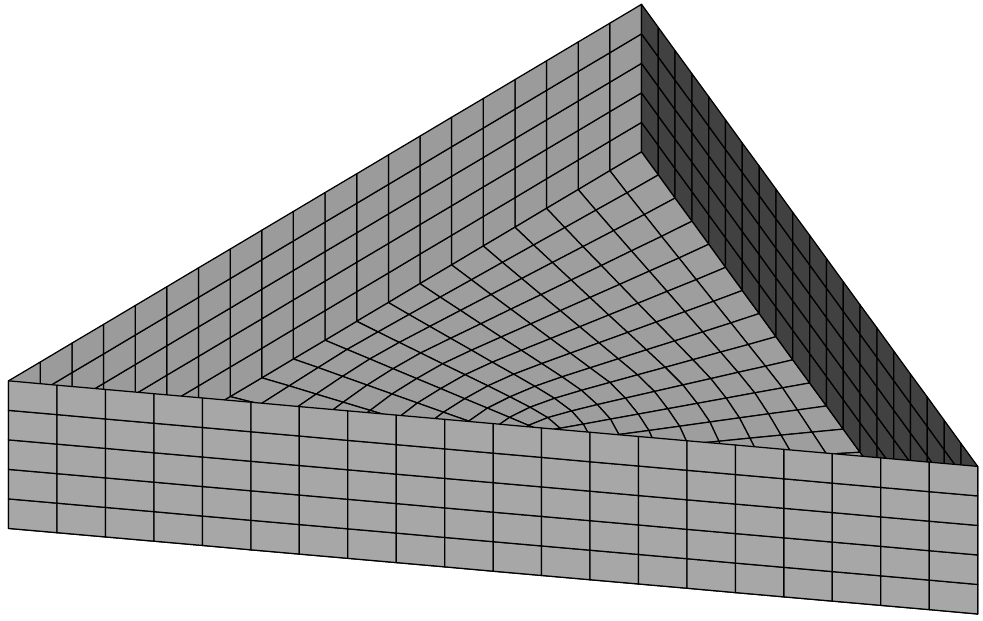}
  \subcaption{}
  \end{minipage}
}
{
  \begin{minipage}{0.22\linewidth}
  \centering
  \includegraphics[width=1.0\linewidth]{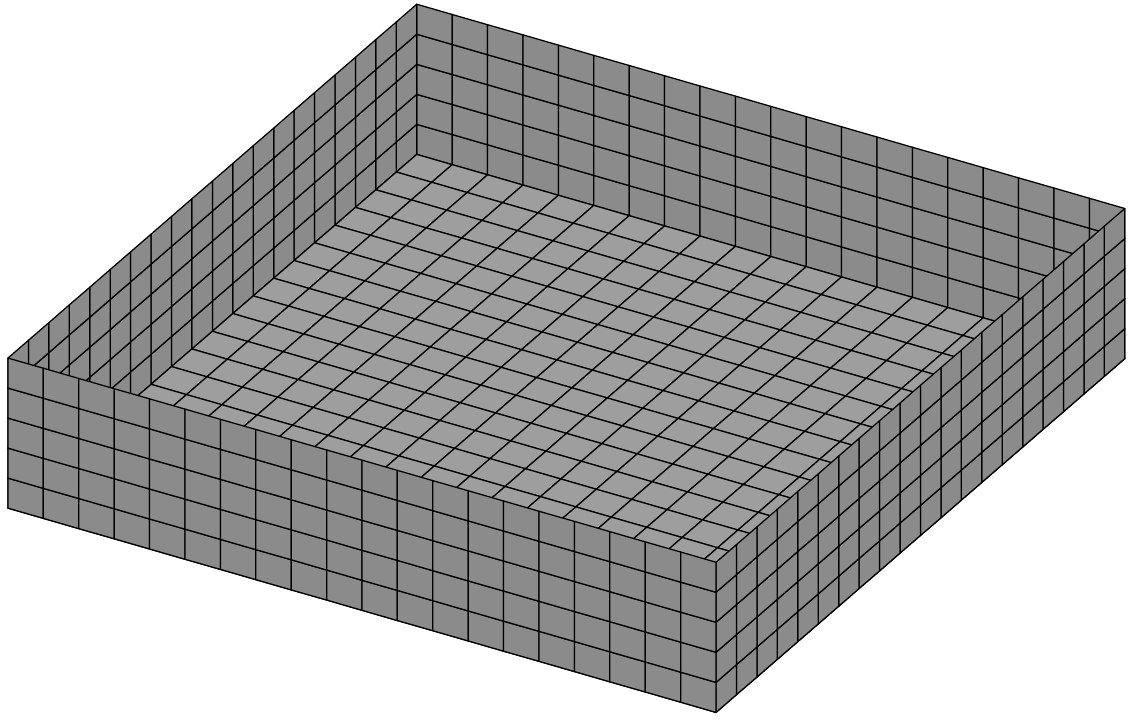}
  \subcaption{}
  \end{minipage}
}
{
  \begin{minipage}{0.22\linewidth}
  \centering
  \includegraphics[width=1.0\linewidth]{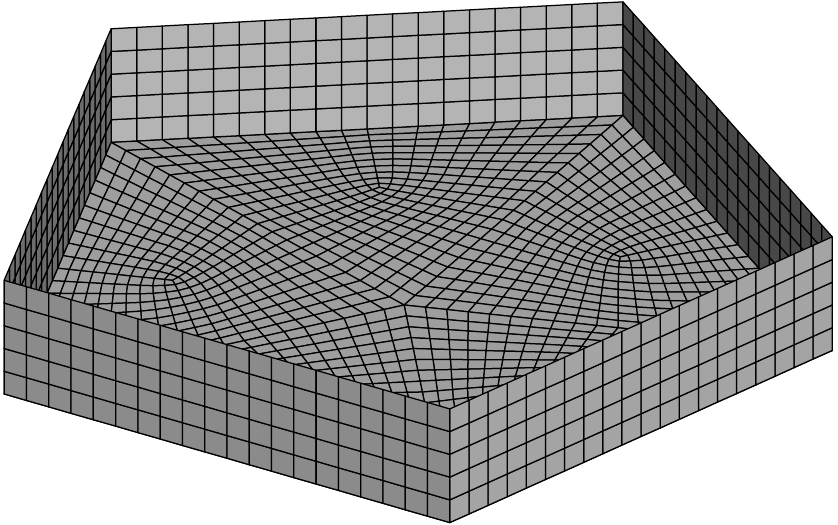}
  \subcaption{}
  \end{minipage}
}
{
  \begin{minipage}{0.22\linewidth}
  \centering
  \includegraphics[width=1.0\linewidth]{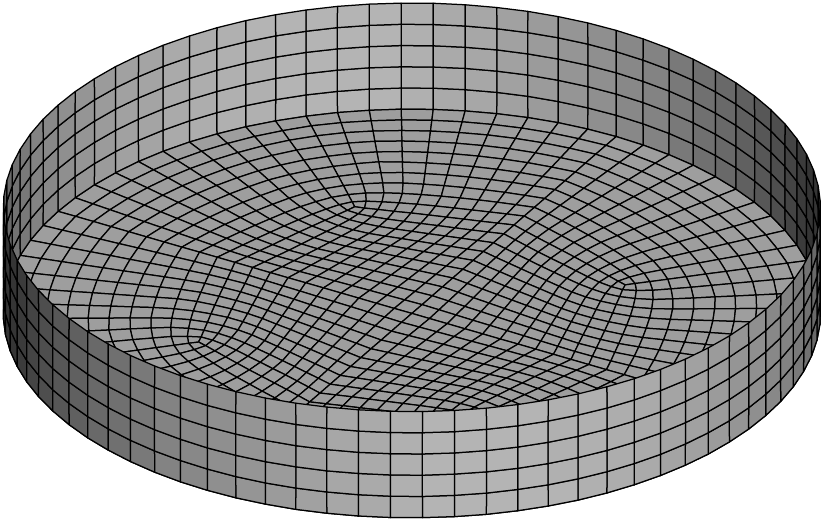}
  \subcaption{}
  \end{minipage}
}
\caption{
Boundary element distribution on wet surface of ice floe for (a) Model-I; (b) Model-II; (c) Model-III; (d) Model-IV}
\label{fig:mesh_model}
\end{figure}

This study applies the open-source code “HAMS” as the frequency-domain hydrodynamic solver in Fig. \ref{fig:diagram}.
The code is based on the boundary element method (BEM) within the framework of linear potential-flow theory.
In this section, HAMS serves two purposes.
First, it calculates the scattered waves from an isolated ice floe in response to each testing wave. 
This provides $\bar{\eta}_i^{\text{S},\beta_n}(r_j,\theta_j)$ for Eq. (\ref{eq:scat_ele}).
Second, it calculates the wave field of multiple ice floes within its computable capabilities. 
This offers a reference to validate the results of the present EI method.
To apply a BEM code, the averaged wet surface of an ice floe is discretized into boundary elements.
In this study, the boundary element distribution for Model-I, Model-II, Model-III, and Model-IV is depicted in Fig. \ref{fig:mesh_model}.
For Model-I, Model-II, and Model-III, each vertical side of the ice floe is divided into $20\times5$ identical rectangular elements.
For Model-IV, 80 elements are used along the circumference and 5 elements are used in the vertical direction.
Unstructured quadrilateral elements are used on the bottom surface of each ice floe.
Model-I, Model-II, Model-III, and Model-IV have 600, 800, 2000, and 1600 elements in total, respectively.
The numerical convergence of the current boundary element distributions has been routinely verified without any technical difficulties.
The results are not presented redundantly here.

\subsection{Examination of wave component detection approach}

The core of the present EI method is to use the WCD approach to establish the DTM for each ice floe.
The WCD approach uses the measured data from wave gauges, and reconstructs the scattered wave field around an isolated ice floe in the cylindrical coordinate system.
This subsection uses an example to show the wave reconstruction effectiveness of the WCD approach.
The ice floe of Model-II is considered.
The arrangement of the wave gauges follows
$R_{\textrm{in}}=2\lambda$, $R_{\textrm{out}}=R_{\textrm{in}}+R/2$, $N_{\textrm{cir}}=180$, and $N_{\textrm{rad}}=11$.

\begin{figure}[!htp]
\centering
{
  \begin{minipage}{0.32\linewidth}
  \centering
  \includegraphics[width=1.0\linewidth]{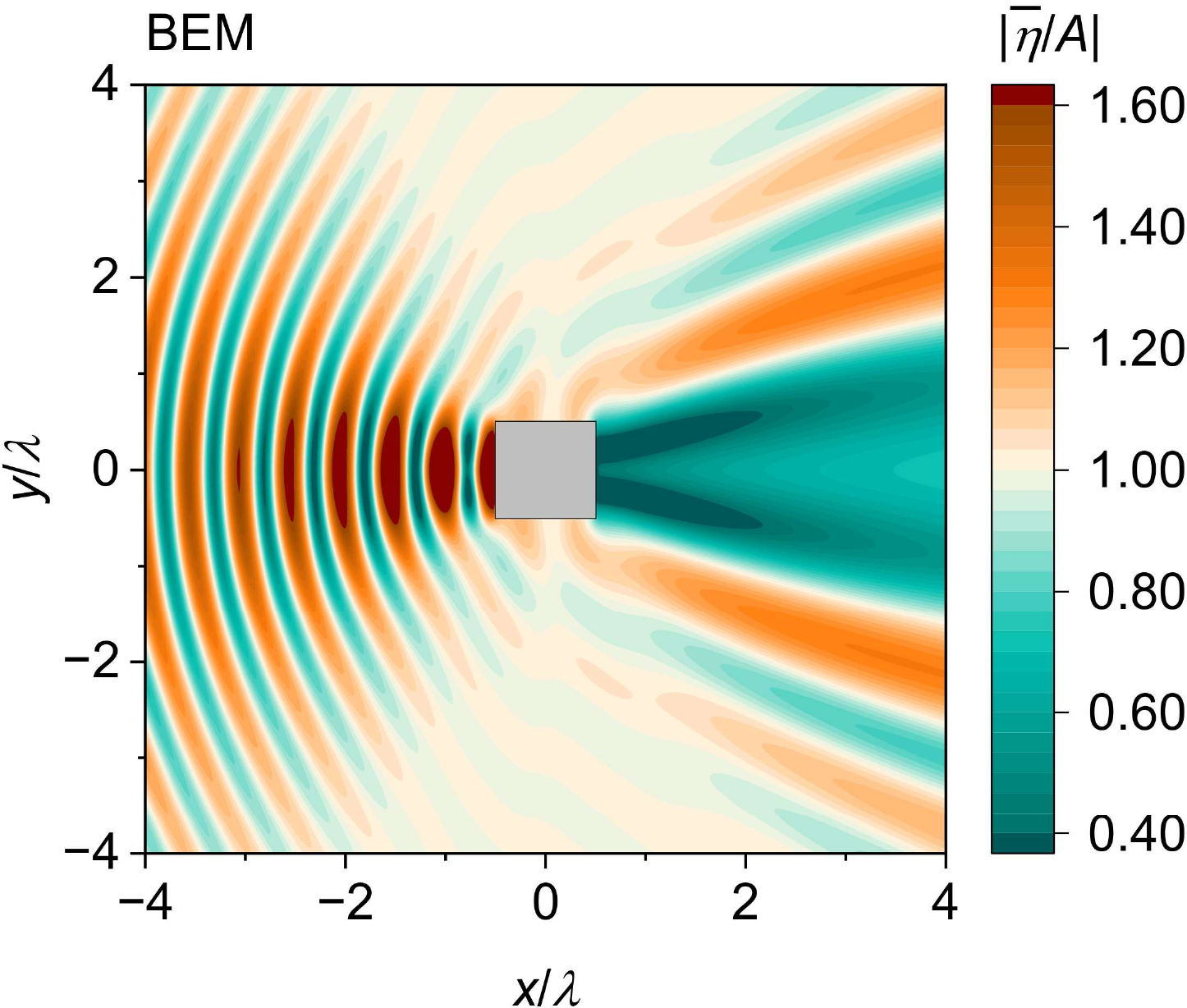}
  \subcaption{}
  \end{minipage}
}
{
  \begin{minipage}{0.32\linewidth}
  \centering
  \includegraphics[width=1.0\linewidth]{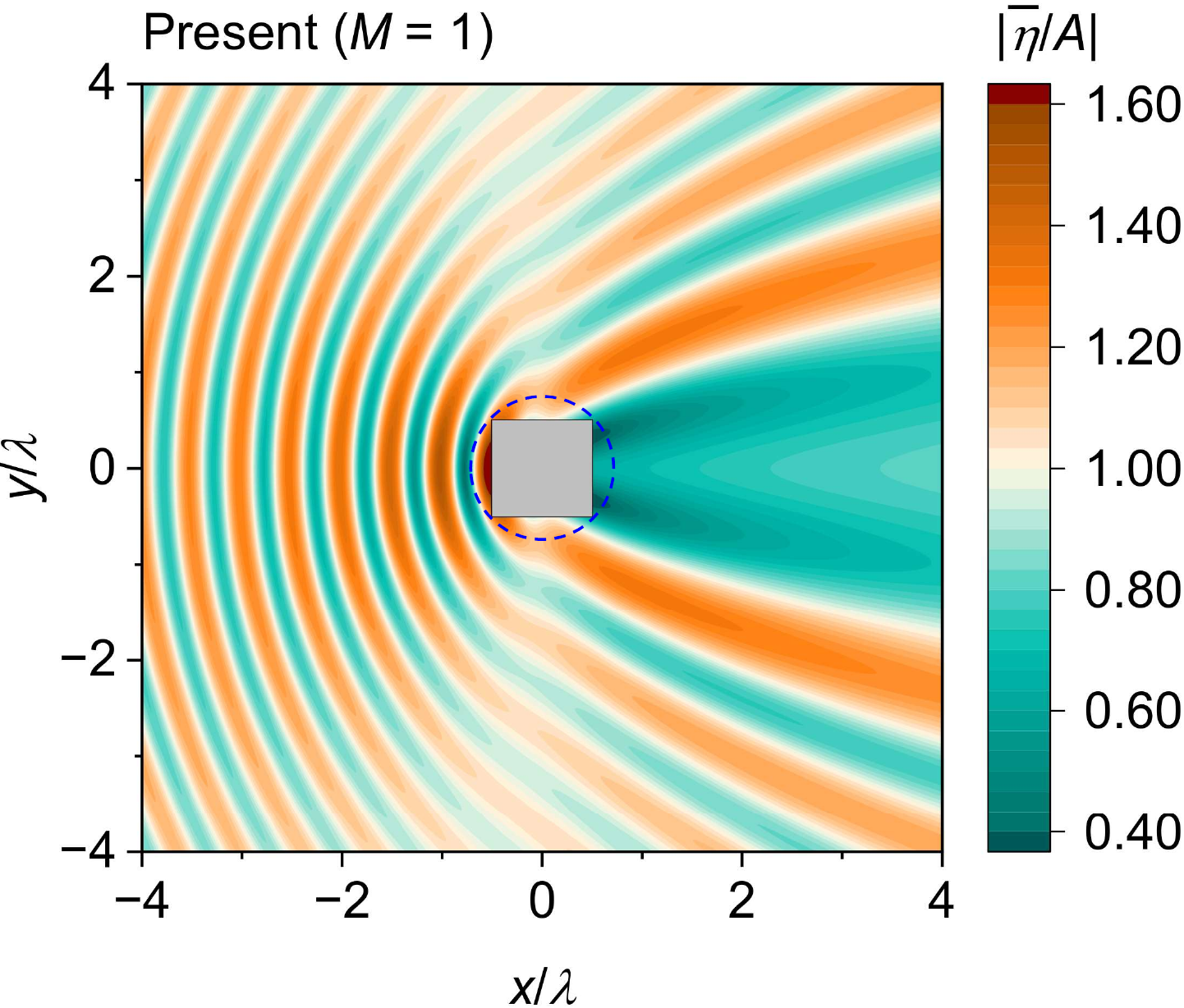}
  \subcaption{}
  \end{minipage}
}
{
  \begin{minipage}{0.32\linewidth}
  \centering
  \includegraphics[width=1.0\linewidth]{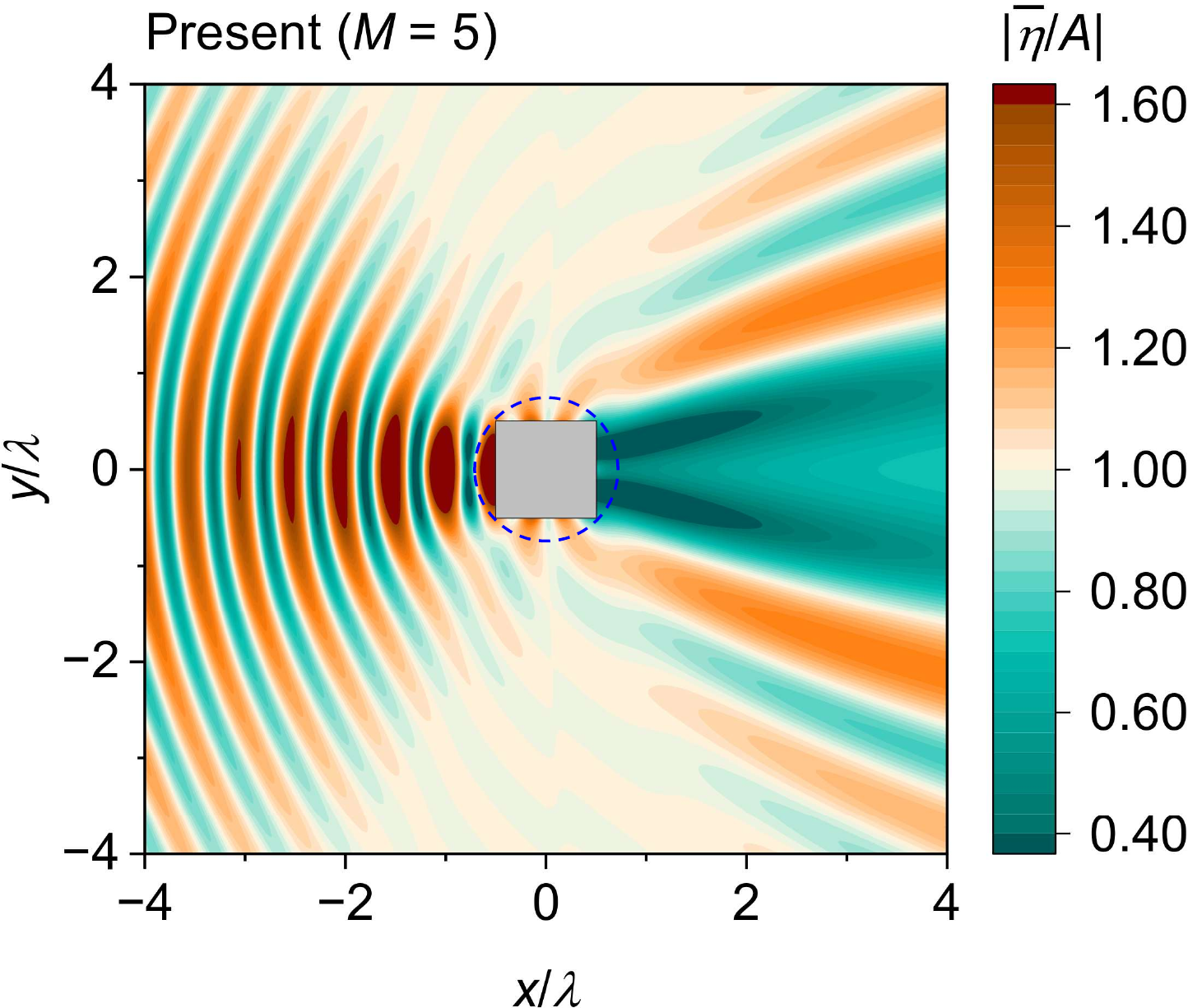}
  \subcaption{}
  \end{minipage}
}
\\
{
  \begin{minipage}{0.32\linewidth}
  \centering
  \includegraphics[width=1.0\linewidth]{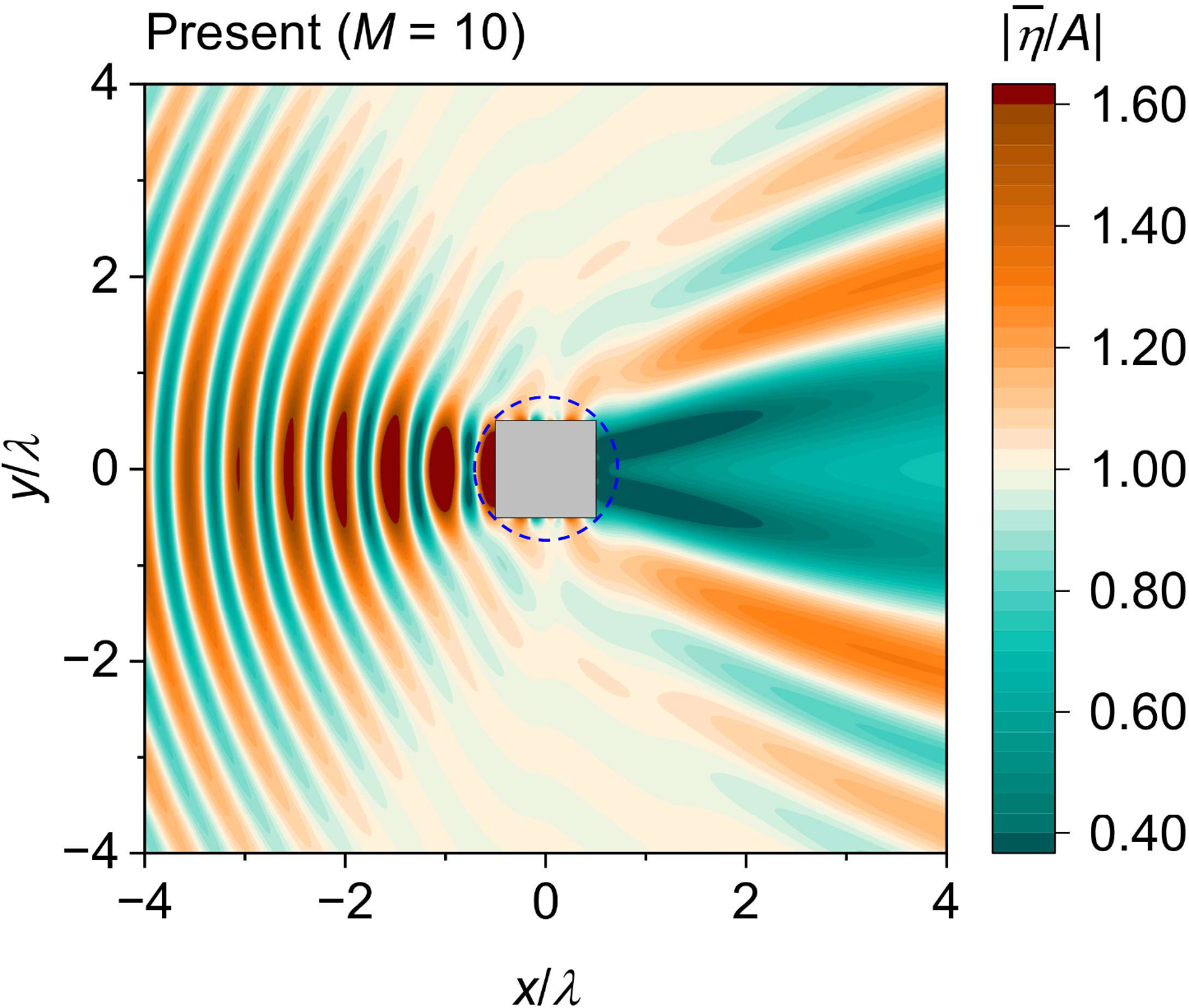}
  \subcaption{}
  \end{minipage}
}
{
  \begin{minipage}{0.32\linewidth}
  \centering
  \includegraphics[width=1.0\linewidth]{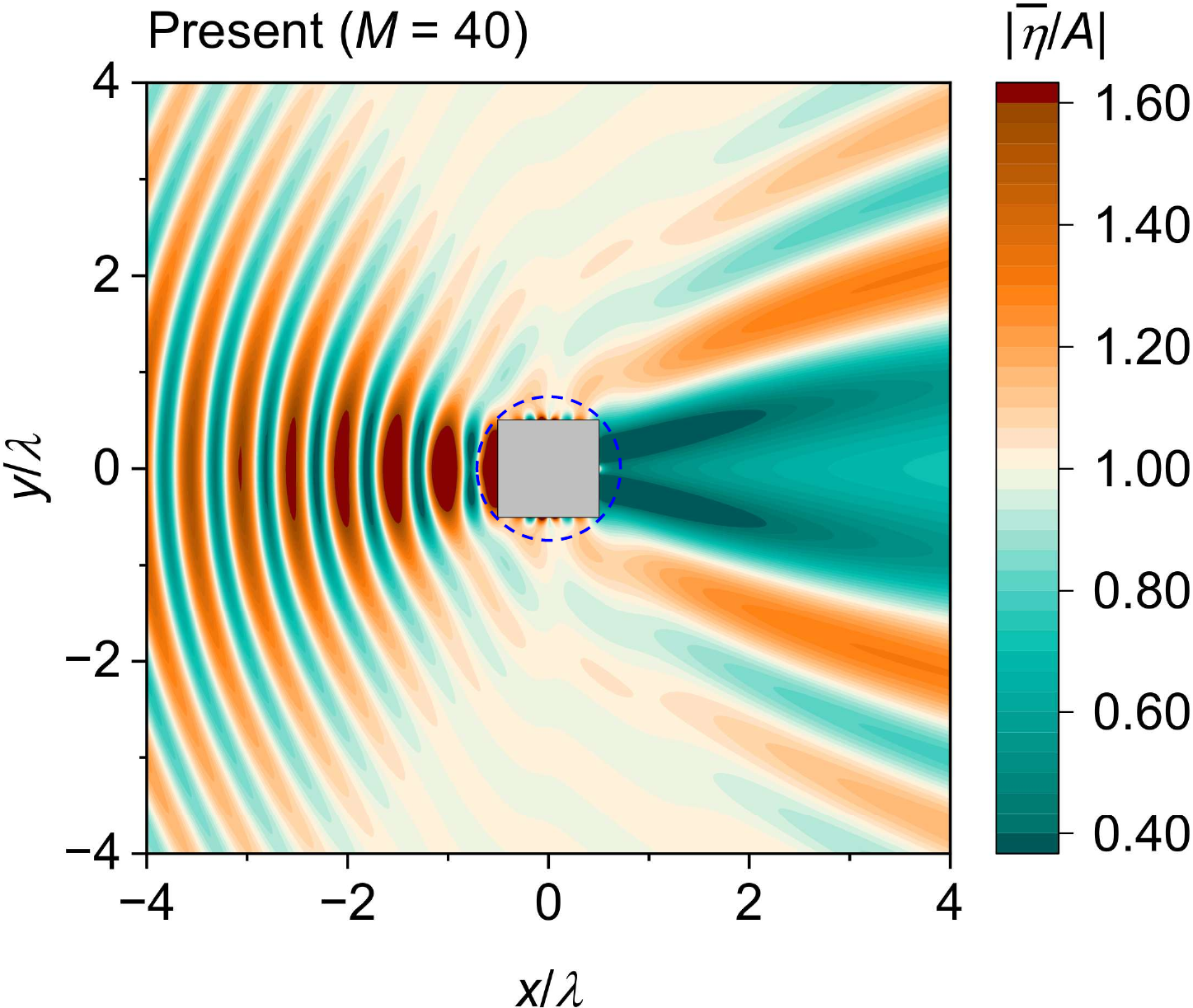}
  \subcaption{}
  \end{minipage}
}
{
  \begin{minipage}{0.32\linewidth}
  \centering
  \includegraphics[width=1.0\linewidth]{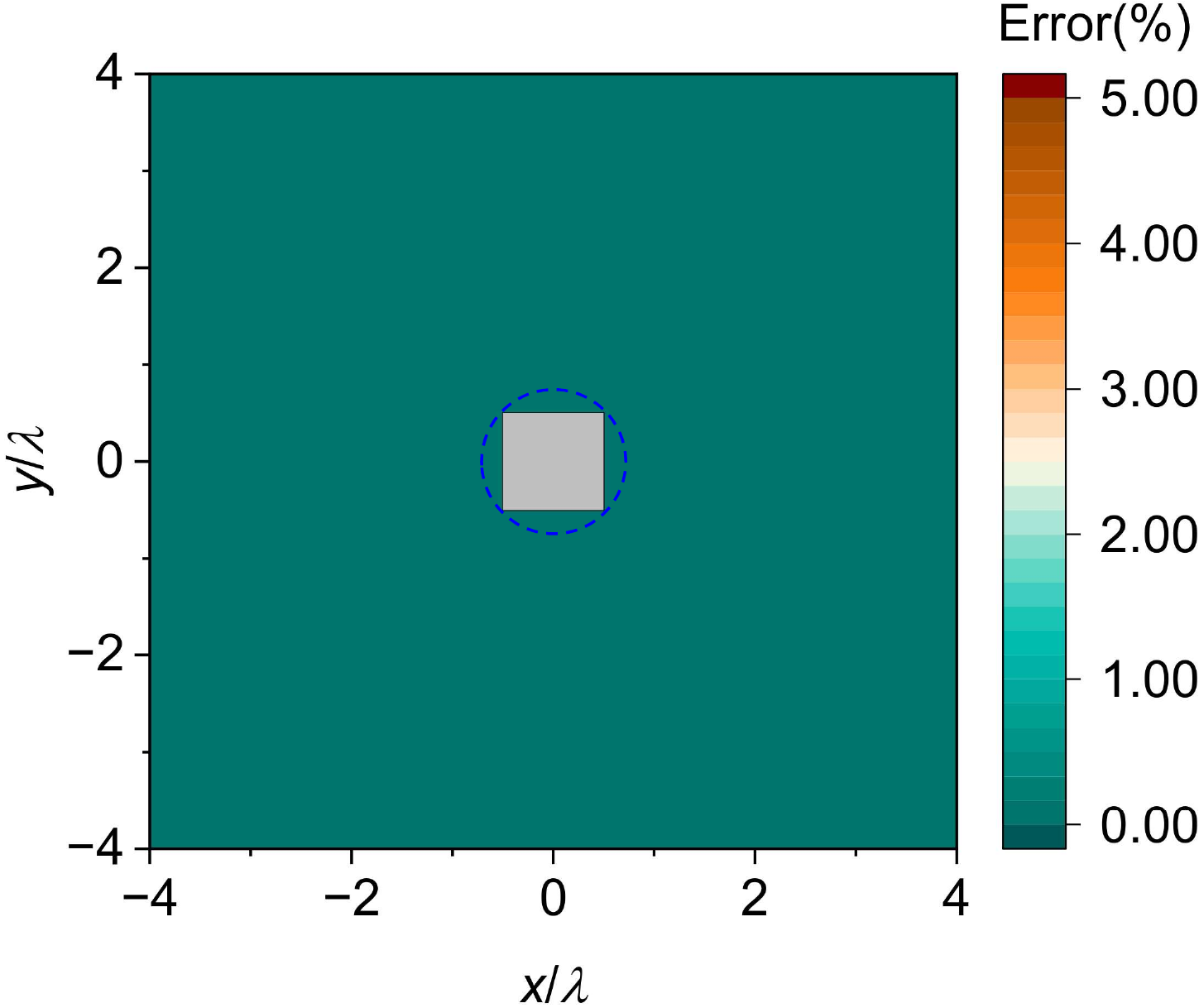}
  \subcaption{}
  \end{minipage}
}
\caption{
Amplitude distribution of wave field around isolate ice floe of Model-II in short waves: (a) BEM solution; (b) present solution with $M=1$; (c) present solution with $M=5$; (d) present solution with $M=10$; (e) present solution with $M=40$; and (f) relative error of present solution for $M=40$.}
\label{fig:diffracted_wave_3s}
\end{figure}

Fig. \ref{fig:diffracted_wave_3s} (a)-(e) shows the amplitude distribution of the wave field around an isolated ice floe of Model-II in short waves.
Fig. \ref{fig:diffracted_wave_3s} (a) shows the result calculated directly using BEM for comparison.
Figs. \ref{fig:diffracted_wave_3s} (b)-(e) show the amplitude distribution of the reconstructed wave fields with different values of $M$ in Eqs. \eqref{eq:plane_wave_neater} and \eqref{eq:scat_potent}.
The circumcircle of Model-II is plotted for reference.
Results with $M = 1$, 5, 10, and 40 are given for comparison.
As $M$ increases to $M = 5$, the reconstructed wave field outside the circumcircle is graphically indistinguishable from the BEM result.
Fig. \ref{fig:diffracted_wave_3s} (f) shows the relative error distribution of Fig. \ref{fig:diffracted_wave_3s} (e), compared to the BEM result in Fig. \ref{fig:diffracted_wave_3s} (a).
The relative error at any position in the wave field is calculated by dividing the absolute difference between the present solution and the BEM result by the amplitude of the ambient incident waves.
In a large field outside the circumcircle, the relative error of the current solution with $M = 40$ is visually negligible.

\begin{figure}[!htp]
\centering
{
  \begin{minipage}{0.32\linewidth}
  \centering
  \includegraphics[width=1.0\linewidth]{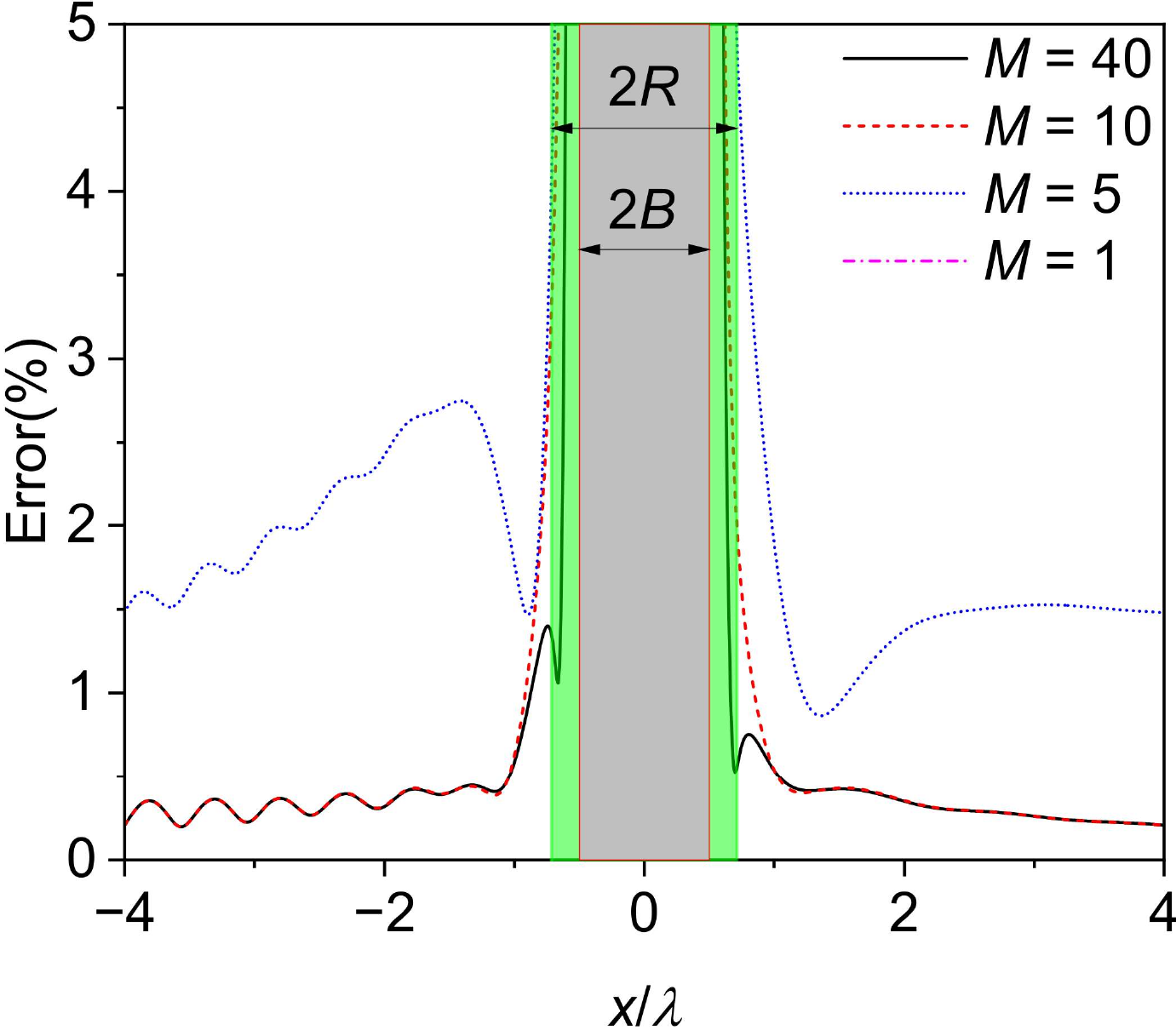}
  \subcaption{}
  \end{minipage}
  \begin{minipage}{0.32\linewidth}
  \centering
  \includegraphics[width=1.0\linewidth]{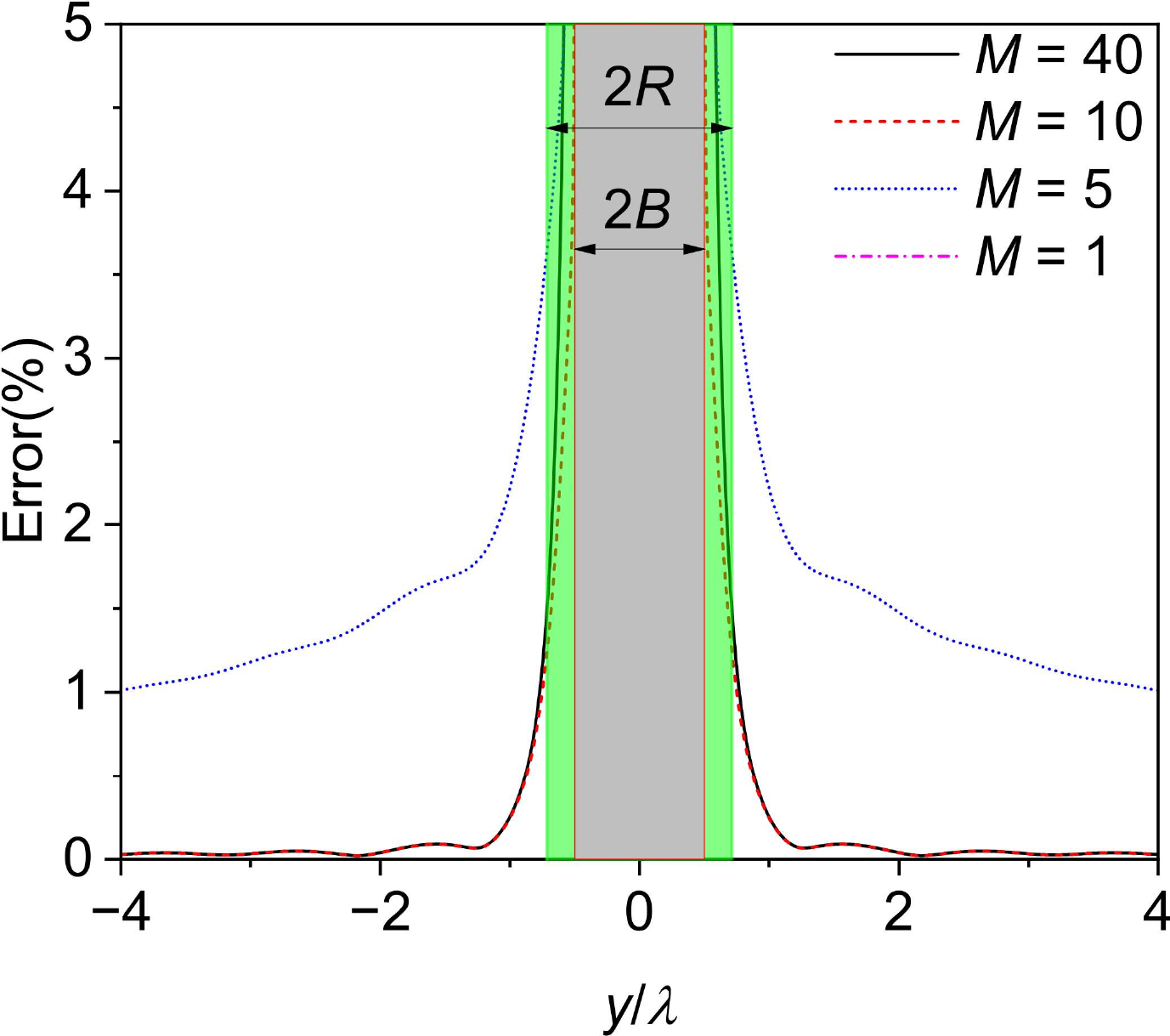}
  \subcaption{}
  \end{minipage}  
}
\caption{
Distribution of relative error along (a) $y=0$ and (b) $x=0$, for isolate ice floe of Model-II in short waves.
}
\label{fig:error_M_3s}
\end{figure}

Fig. \ref{fig:error_M_3s} further shows the distribution of the relative error of the reconstructed wave field along $y = 0$ and $x = 0$.
Here, the half side-length of Model-II is denoted by $B$.
When $M = 1$, the relative error at any position in the wave field is always greater than $8\%$, which is not acceptable.
With $M\geqslant 5$, at any location outside the circumcircle of the ice floe, the relative error of the reconstructed wave field is always less than $5\%$.
In other words, at a location inside the circumcircle of the ice floe, the accuracy of the reconstructed wave field cannot be guaranteed. 
Outside the circumcircle, as $M$ increases from 5 to 40, the maximum relative error drops significantly.
With $M = 40$, at any location more than $0.5\lambda$ away from the circumcircle of the ice floe, the relative error of the reconstructed wave field is always less than $0.5\%$.

\begin{figure}[htbp]
\centering
{
  \begin{minipage}{0.32\linewidth}
  \centering
  \includegraphics[width=1.0\linewidth]{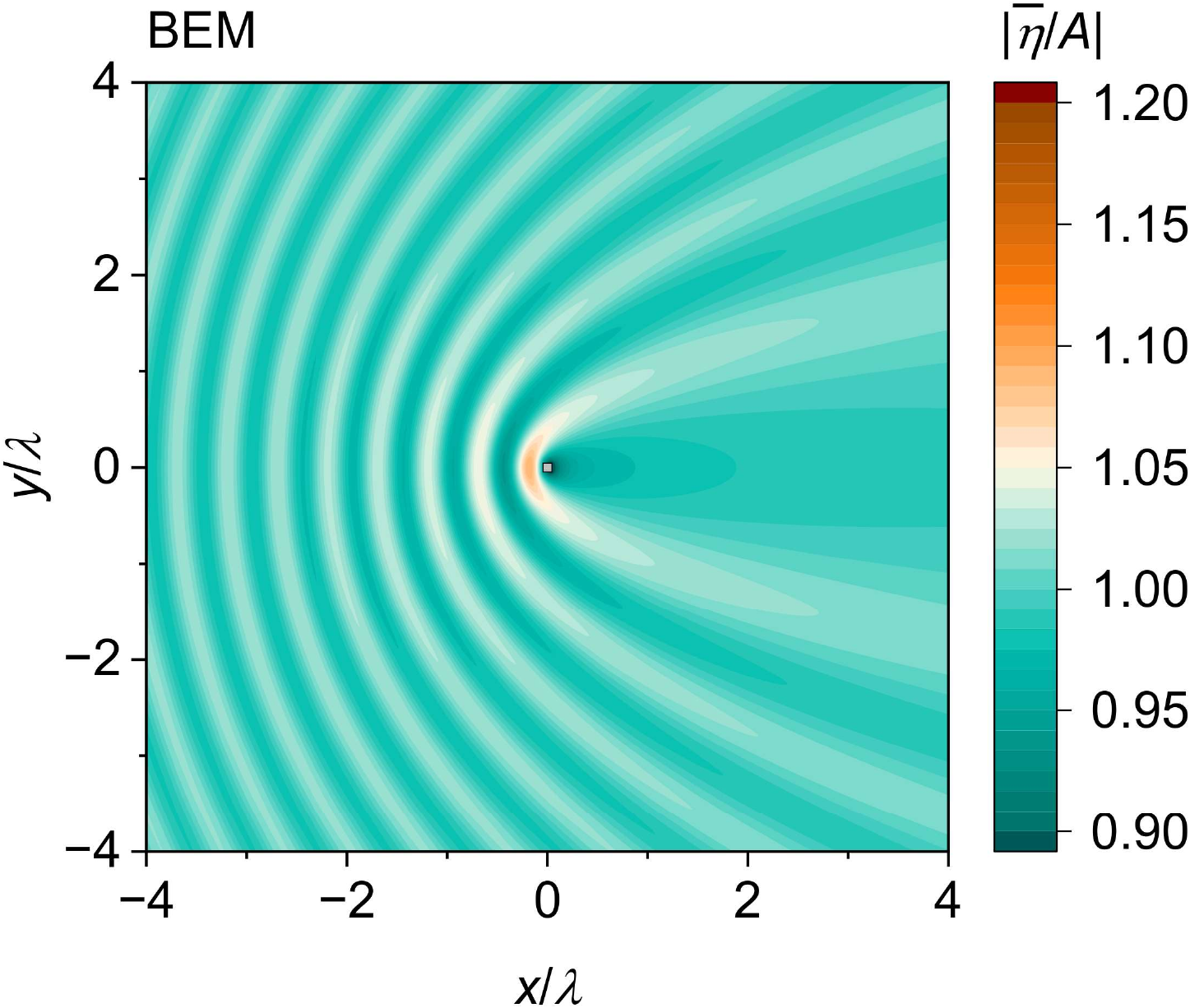}
  \subcaption{}
  \end{minipage}
}
{
  \begin{minipage}{0.32\linewidth}
  \centering
  \includegraphics[width=1.0\linewidth]{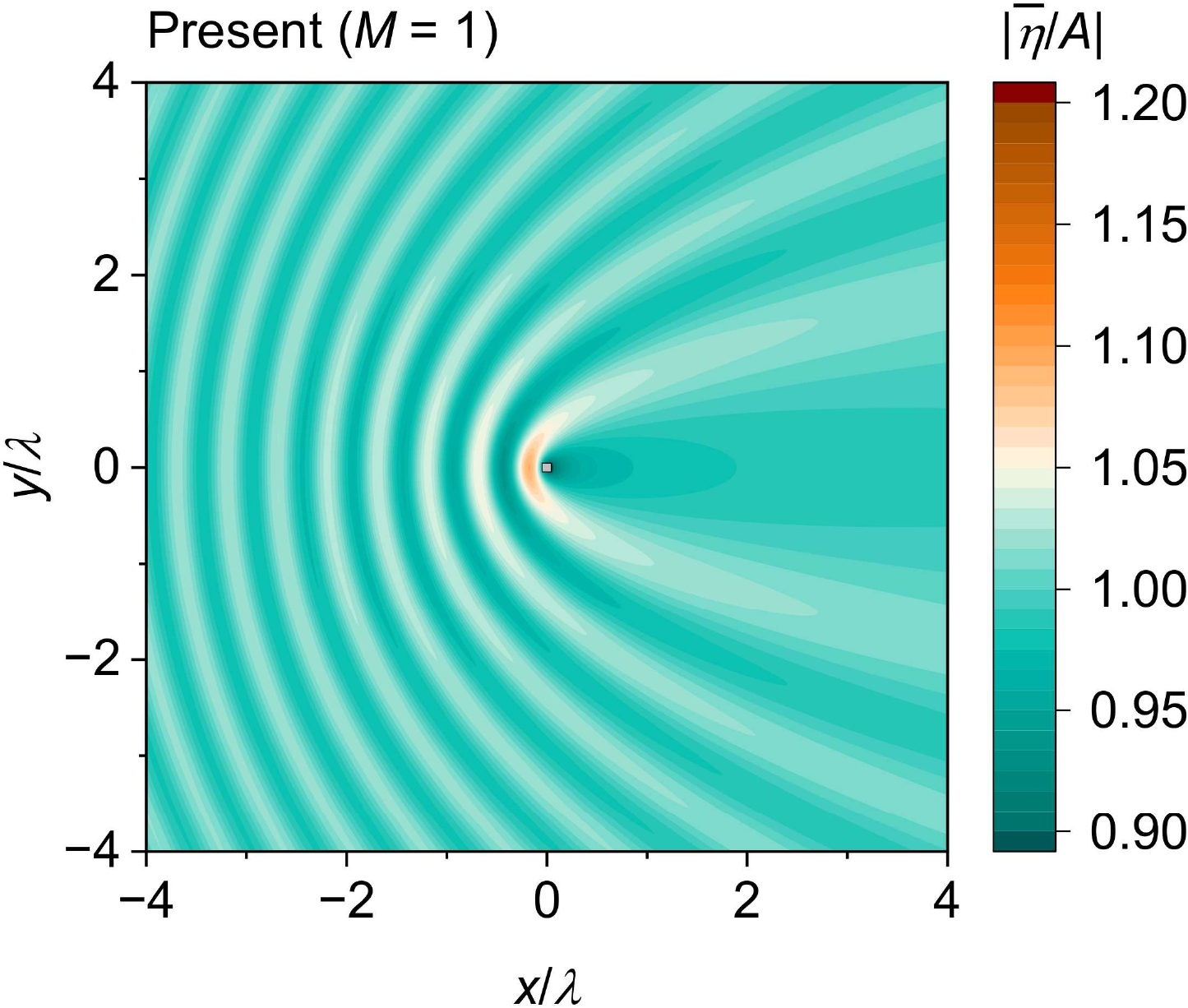}
  \subcaption{}
  \end{minipage}
}
{
  \begin{minipage}{0.32\linewidth}
  \centering
  \includegraphics[width=1.0\linewidth]{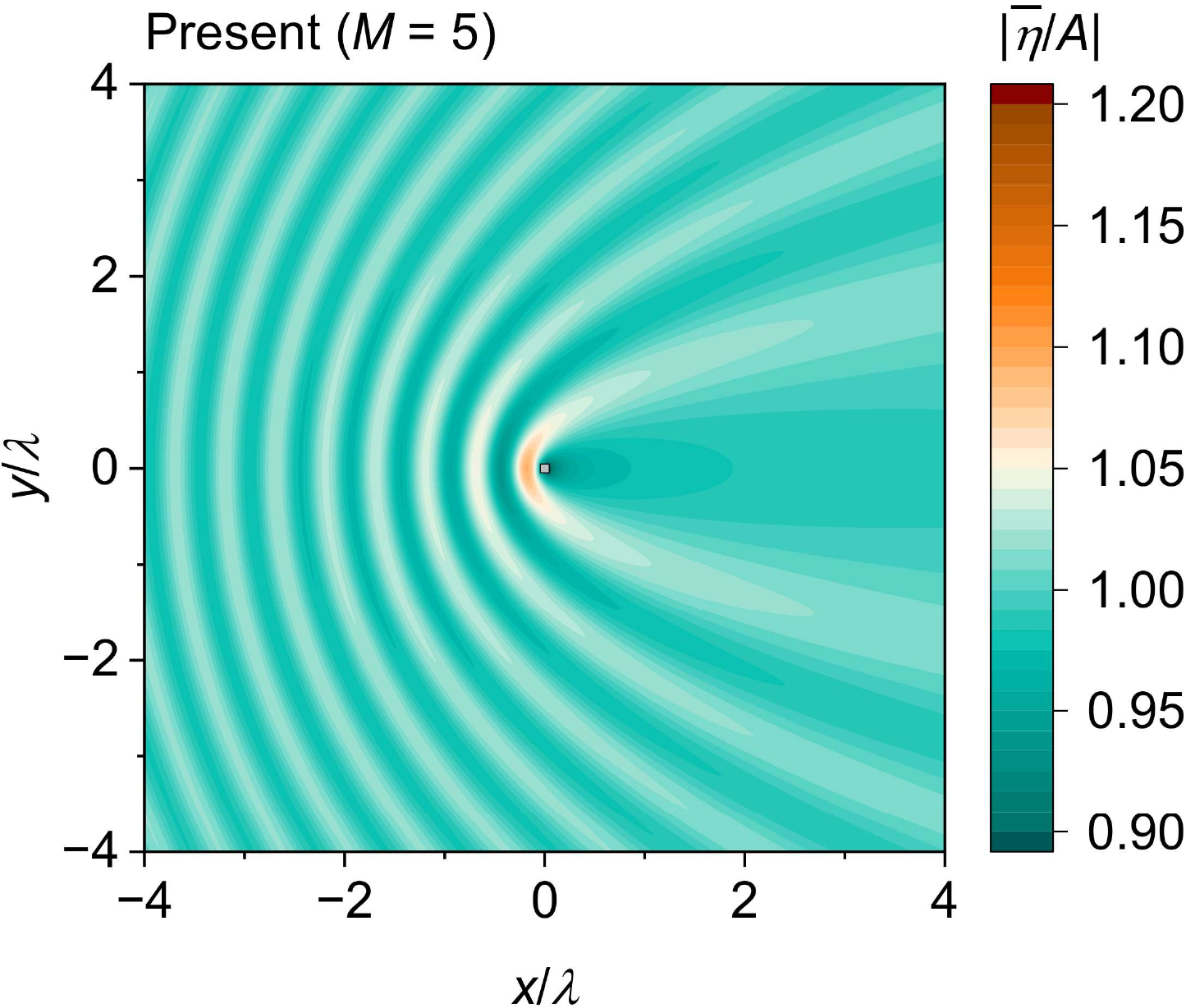}
  \subcaption{}
  \end{minipage}
}
\\
{
  \begin{minipage}{0.32\linewidth}
  \centering
  \includegraphics[width=1.0\linewidth]{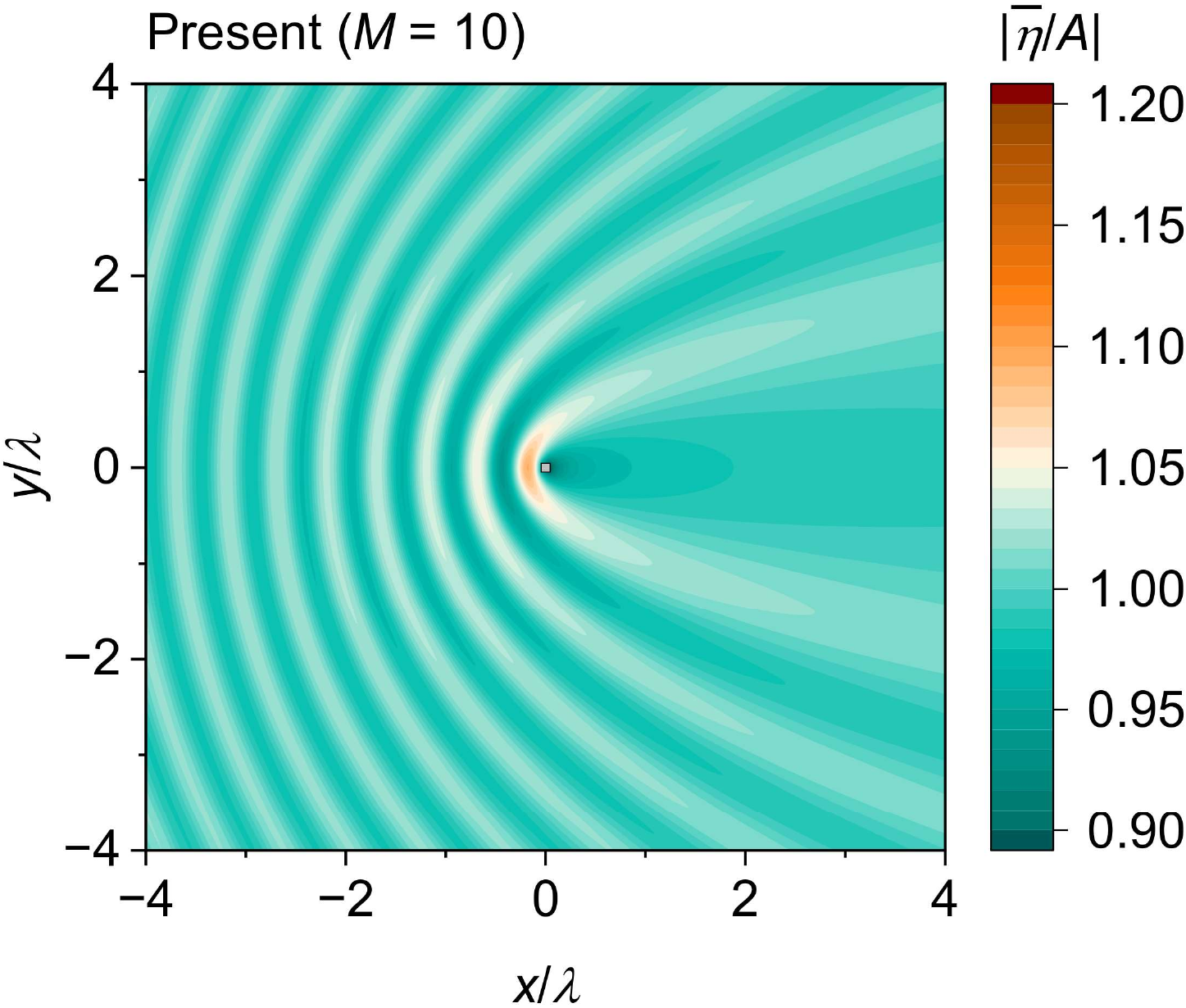}
  \subcaption{}
  \end{minipage}
}
{
  \begin{minipage}{0.32\linewidth}
  \centering
  \includegraphics[width=1.0\linewidth]{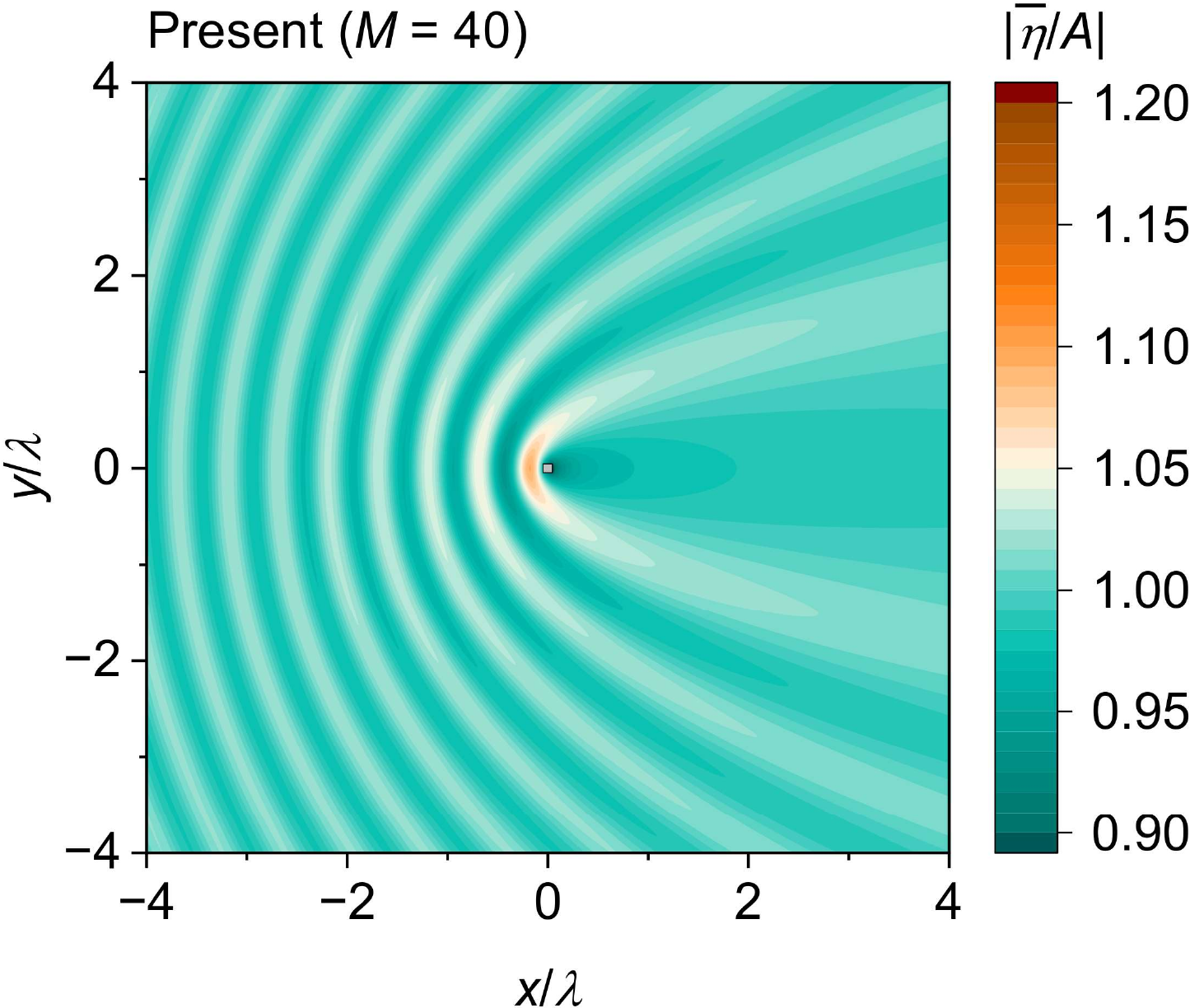}
  \subcaption{}
  \end{minipage}
}
{
  \begin{minipage}{0.32\linewidth}
  \centering
  \includegraphics[width=1.0\linewidth]{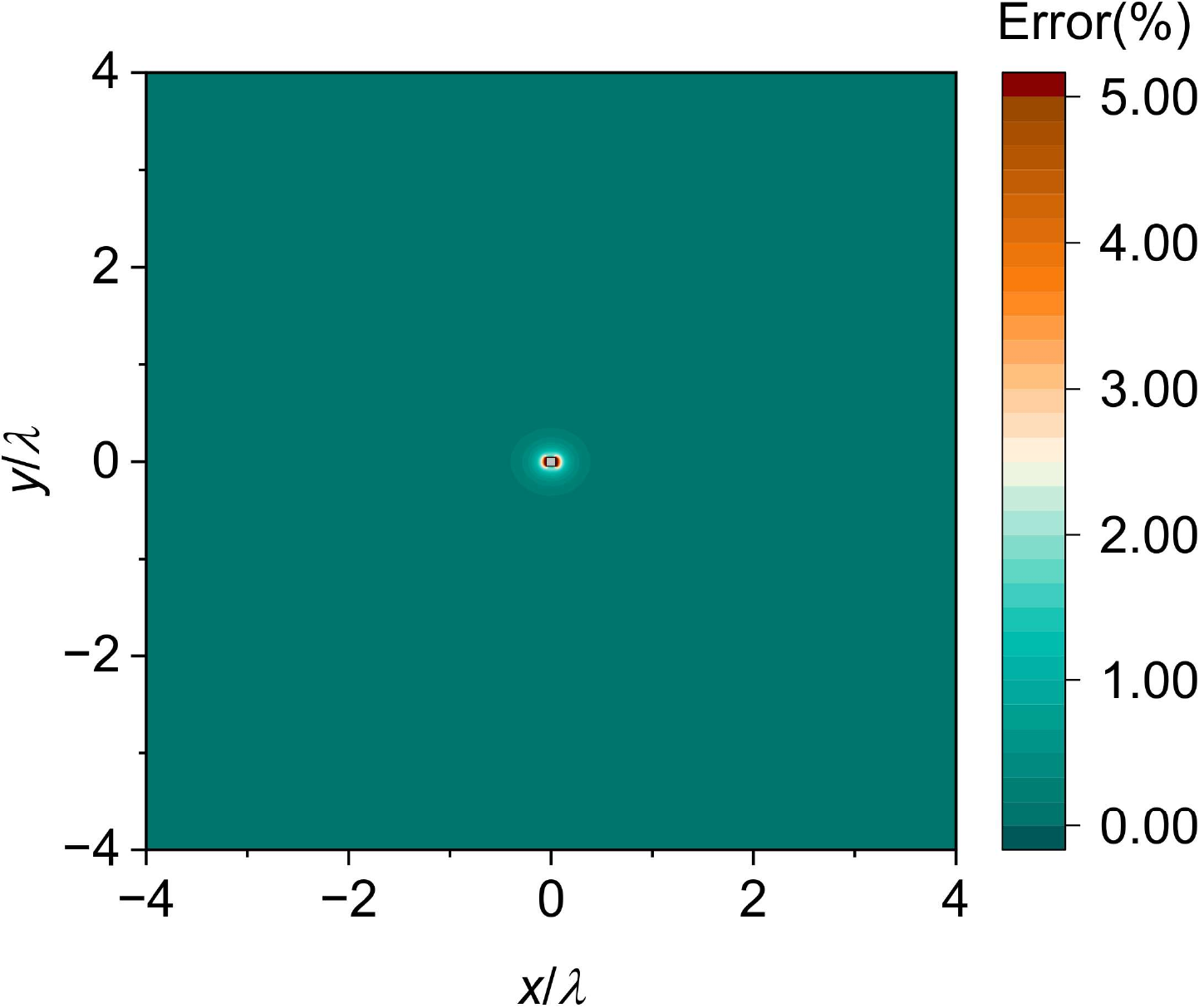}
  \subcaption{}
  \end{minipage}
}
\caption{
Amplitude distribution of wave field around isolate ice floe of Model-II in long waves: (a) BEM solution; (b) present solution with $M=1$; (c) present solution with $M=5$; (d) present solution with $M=10$; (e) present solution with $M=40$; and (f) relative error of present solution for $M=40$.}
\label{fig:diffracted_wave}
\end{figure}

\begin{figure}[!htp]
\centering
{
  \begin{minipage}{0.32\linewidth}
  \centering
  \includegraphics[width=1.0\linewidth]{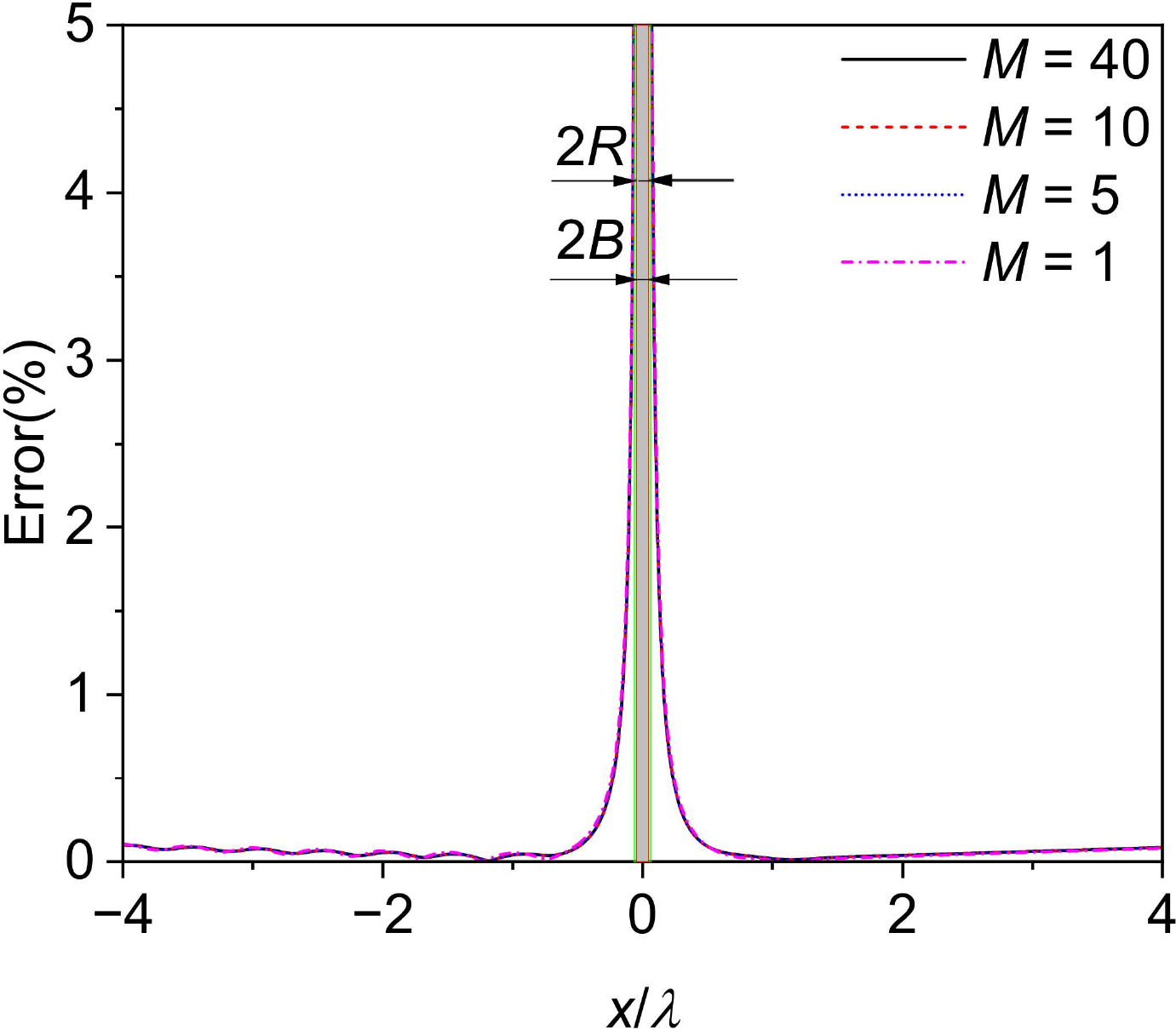}
  \subcaption{}
  \end{minipage}
  \begin{minipage}{0.32\linewidth}
  \centering
  \includegraphics[width=1.0\linewidth]{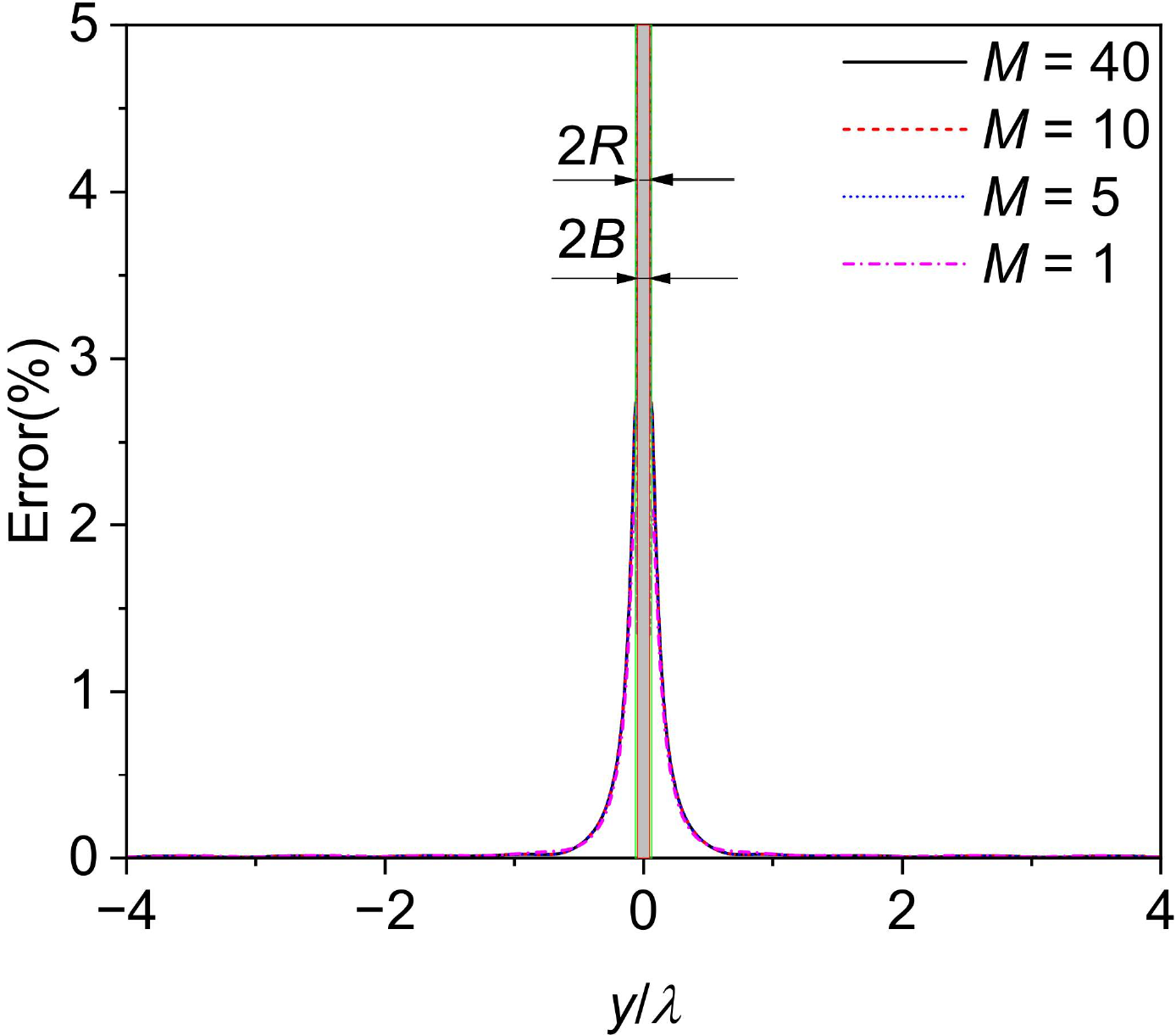}
  \subcaption{}
  \end{minipage}  
}
\caption{
Distribution of relative error along (a) $y=0$ and (b) $x=0$, for an isolate ice floe of Model-II in long waves.}
\label{fig:error_M}
\end{figure}

Similar to Fig. \ref{fig:diffracted_wave_3s}, Figs. \ref{fig:diffracted_wave} (a)-(e) show the amplitude distribution of the wave field around an isolated ice floe of Model-II in long waves.
Fig. \ref{fig:diffracted_wave} (a) shows the BEM result for comparison.
Figs. \ref{fig:diffracted_wave} (b)-(e) depict the present EI solutions with $M = 1$, 5, 10, and 40, respectively.
Fig. \ref{fig:diffracted_wave} (f) shows the relative error distribution of Fig. \ref{fig:diffracted_wave} (e), compared to the BEM result in Fig. \ref{fig:diffracted_wave} (a).
In long waves, even with $M=1$, the reconstructed wave field is graphically indistinguishable from the BEM result.
Specifically, the distributions of the relative error of the reconstructed wave field along $y = 0$ and $x = 0$ are shown in Figs. \ref{fig:error_M} (a) and (b), respectively.
The relative error of the reconstructed wave field is almost irrelevant to the value of $M$.
At any location more than $0.5 \lambda$ away from the circumcircle of the ice floe, the relative error of the reconstructed wave field is always less than $0.5\%$.
At a location close to but outside the circumcircle of the ice floe, the relative error in the reconstructed wave amplitude is always less than $5\%$.
In general, the present WCD approach is more effective for reconstructing the wave field of long waves.

\begin{figure}[htbp] 
	\centering
	\includegraphics[width=0.4\linewidth]{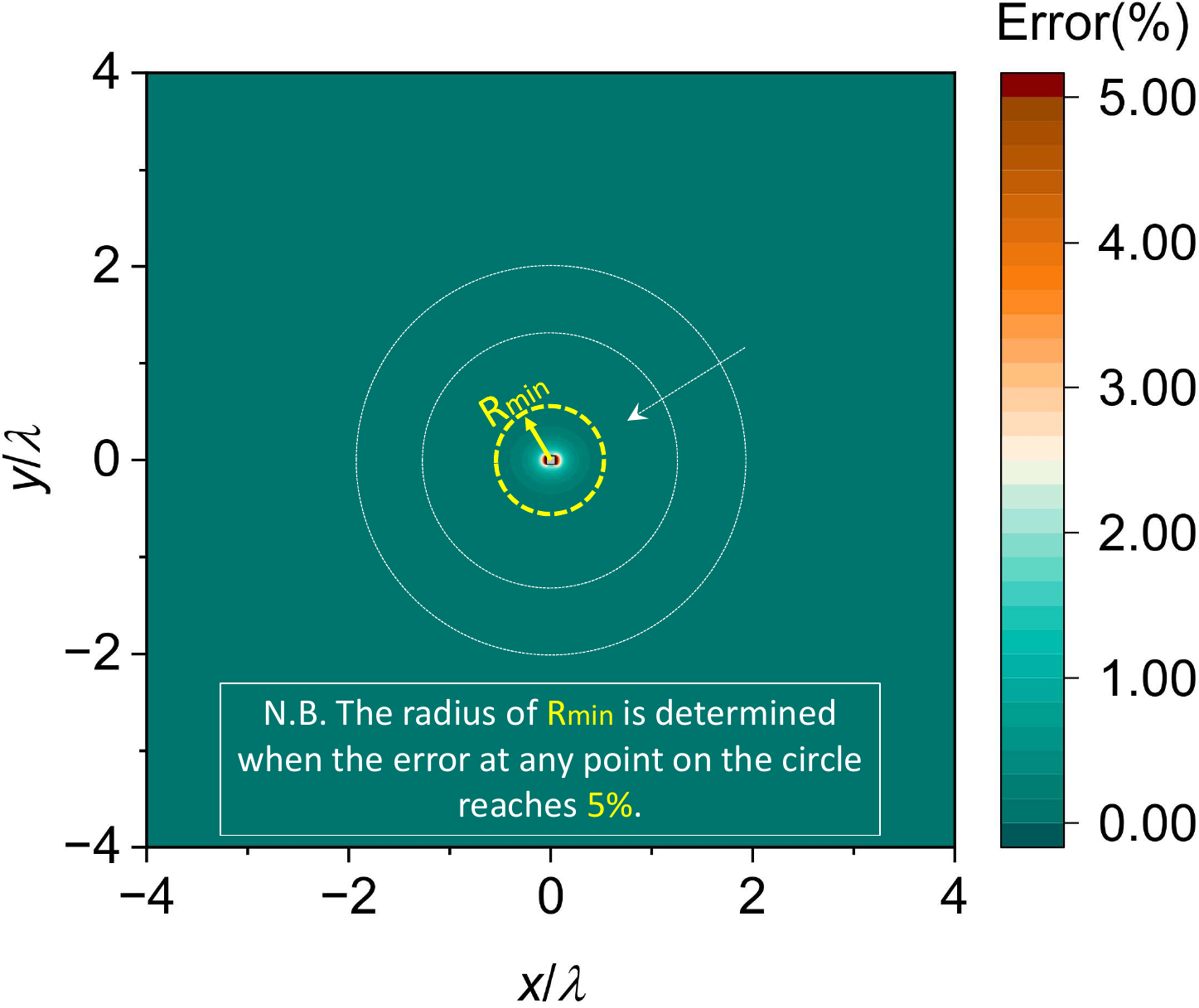}
	\caption{Definition of error index $R_{\rm{min}}$ to evaluate error level of reconstructed wave field, where $R_{\rm{min}}$ is radius of a concentric circle of ice floe's circumcircle such that the maximum error at any point on this circle attains $5\%$.	
	}
	\label{fig:R_min}
\end{figure}

According to Figs. \ref{fig:diffracted_wave_3s} (f) and \ref{fig:diffracted_wave} (f), it can intuitively seen that the relative error of the WCD approach distributes radially from the circumcircle of the ice floe.
In general, the relative error decreases rapidly in the radial direction away from the ice floe.
This study defines an error index $R_{\rm{min}}$ to evaluate the error level of the reconstructed wave field.
As demonstrated in Fig. \ref{fig:R_min}, the value of $R_{\rm{min}}$ is the radius of a concentric circle of the ice floe's circumcircle such that the maximum error at any point on this circle attains $5\%$. 
The relative error over $5\%$ is restricted within the circle with a radius of $R_{\rm{min}}$.
Therefore, a smaller $R_{\rm{min}}$ indicates less error in wave-field reconstruction.

\begin{figure}[!htp]
\centering
{
  \begin{minipage}{0.32\linewidth}
  \centering
  \includegraphics[width=1.0\linewidth]{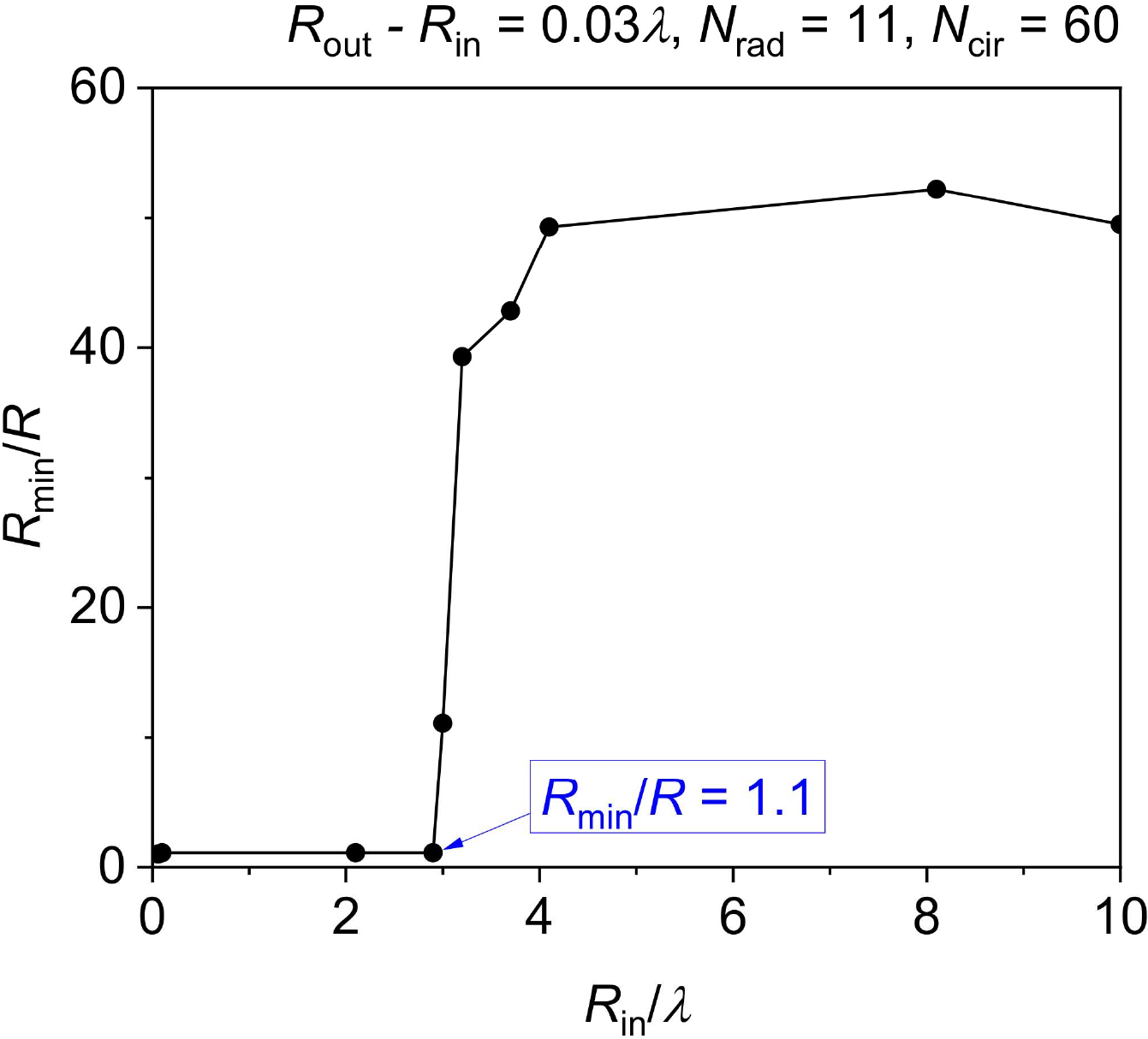}
  \subcaption{}
  \end{minipage}
}
{
  \begin{minipage}{0.32\linewidth}
  \centering
  \includegraphics[width=1.0\linewidth]{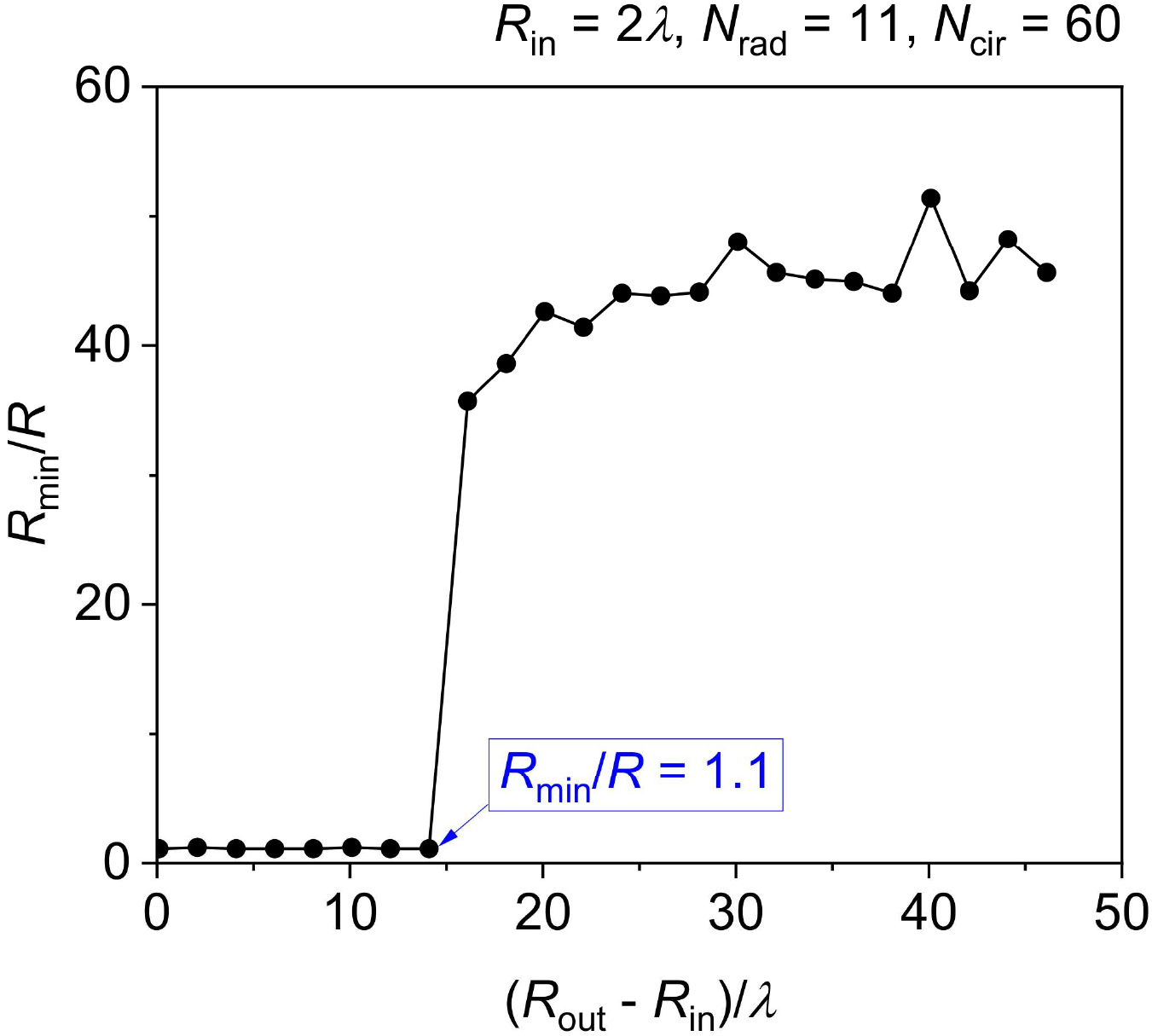}
  \subcaption{}
  \end{minipage}
}
{
  \begin{minipage}{0.32\linewidth}
  \centering
  \includegraphics[width=1.0\linewidth]{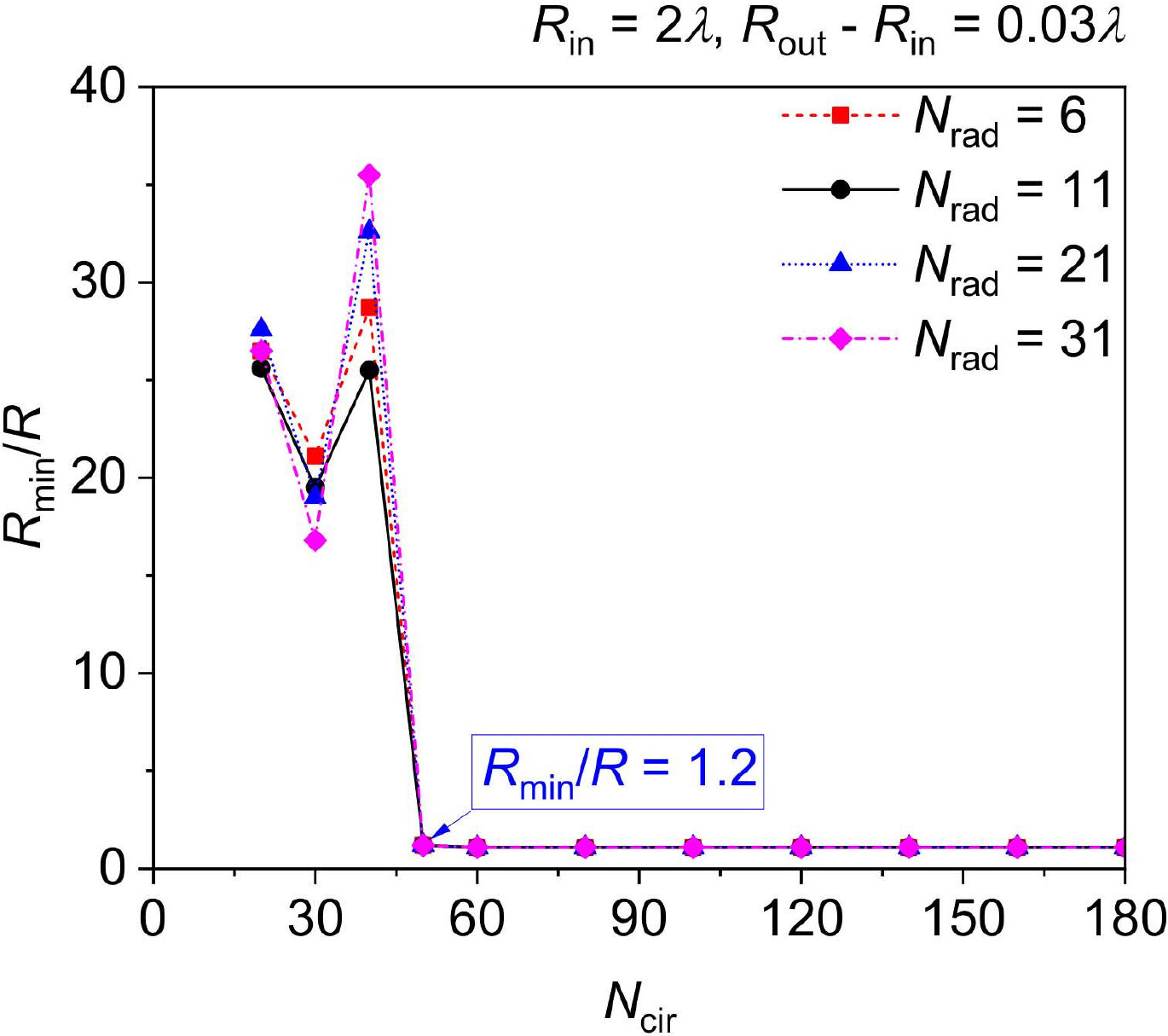}
  \subcaption{}
  \end{minipage}
}
\caption{
Effects of (a) $R_{\rm{in}}$, (b) $R_{\rm{out}}-R_{\rm{in}}$, and (c) $N_{\rm{cir}}$ and $N_{\rm{rad}}$ on $R_{\rm{min}}$, for wave field around isolate ice floe of Model-II in long waves.
}
\label{fig:parameter}
\end{figure}

In the present WCD approach, wave gauges are placed in an annular belt region around the ice floe.
How to deploy the array of wave gauges needs to be determined for practical purposes.
The scale of the wave-gauge belt is determined by the inner radius $R_{\rm{in}}$ and the belt width $R_{\rm{out}}-R_{\rm{in}}$.
There are $N_{\rm{cir}}$ wave gauges regularly placed along the circumferential direction of the belt, and $N_{\rm{rad}}$ wave gauges along the radial direction.
Here, the reconstructed wave fields around an isolated ice floe of Model-II in long waves are taken as an example.
Fig. \ref{fig:parameter} tests the effects of $R_{\rm{in}}$, $R_{\rm{out}}-R_{\rm{in}}$, $N_{\rm{cir}}$, and $N_{\rm{rad}}$ on the error index $R_{\rm{min}}$.
In Fig. \ref{fig:parameter} (a), $60 \times 11$ wave gauges are placed in the annular belt, and the belt width is $R_{\rm{out}} - R_{\rm{in}} = 0.03\lambda=0.5R$.
The minimum inner radius coincides with the circumcircle of the ice floe, i.e., $R_{\rm{in}} = R$.
When the inner radius $R_{\rm{in}}$ increases to be close to $3\lambda$, the value of $R_{\rm{min}}$ stays below $1.1R$.
This means that, beyond a distance of $0.1R$ from the circumcircle of the ice floe, the wave amplitude of the reconstructed wave field has an error of less than $5\%$.
As $R_{\rm{in}}$ exceeds three times the wavelength, the value of $R_{\rm{min}}$ increases rapidly, reaching about $50R$.
Fig. \ref{fig:parameter} (b) shows the effects of the belt width $R_{\rm{out}} - R_{\rm{in}}$ on the error index $R_{\rm{min}}$. 
Here, $R_{\rm{in}} = 2\lambda$ and $N_{\rm{cir}} \times N_{\rm{rad}} = 60 \times 11$.
As the belt width increases from $0.03\lambda$ (i.e., $0.5R$) to about $14\lambda$, $R_{\rm{min}}$ remains below $1.1R$.
When the belt width is further increased, the error index can rise to over $40R$.
Fig. \ref{fig:parameter} (c) considers the effects of the number of wave gauges on the error level of the WCD approach.
The accuracy of the WCD approach is more sensitive to the circular number of wave gauges.
As long as $N_{\rm{cir}}$ reaches 60, further increasing wave gauges in the radial direction does not improve the accuracy efficiently.
With $R_{\rm{in}} = 2\lambda$ and $R_{\rm{out}}-R_{\rm{in}}=0.5R$, a practical arrangement of $N_{\rm{cir}} \times N_{\rm{rad}} = 60 \times 6$ can guarantee $R_{\rm{min}}/R<1.1$.

The operation rules of the WCD approach are recommended as follows: (i) the inner radius of the wave-gauge belt should satisfy $R<R_{\rm{in}}\leqslant 2\lambda$; (ii) the belt width should meet the condition $0.5R<(R_{\rm{out}}-R_{\rm{in}})\leqslant 10\lambda$; and (iii) the number of wave gauges should follow $N_{\rm{cir}} \times N_{\rm{rad}} = 60 \times 6$.
Following the operation rules, the following error - range statements can be made conservatively:
(i) At a location more than $0.2R$ away from the circumcircle of an ice floe, the relative error in the reconstructed wave amplitude is always less than $5\%$; and (ii) at any location more than $0.5\lambda$ away from the circumcircle, the relative error is always less than $0.5\%$.

\begin{figure}[!htp]
\centering
{
  \begin{minipage}{0.32\linewidth}
  \centering
  \includegraphics[width=1.0\linewidth]{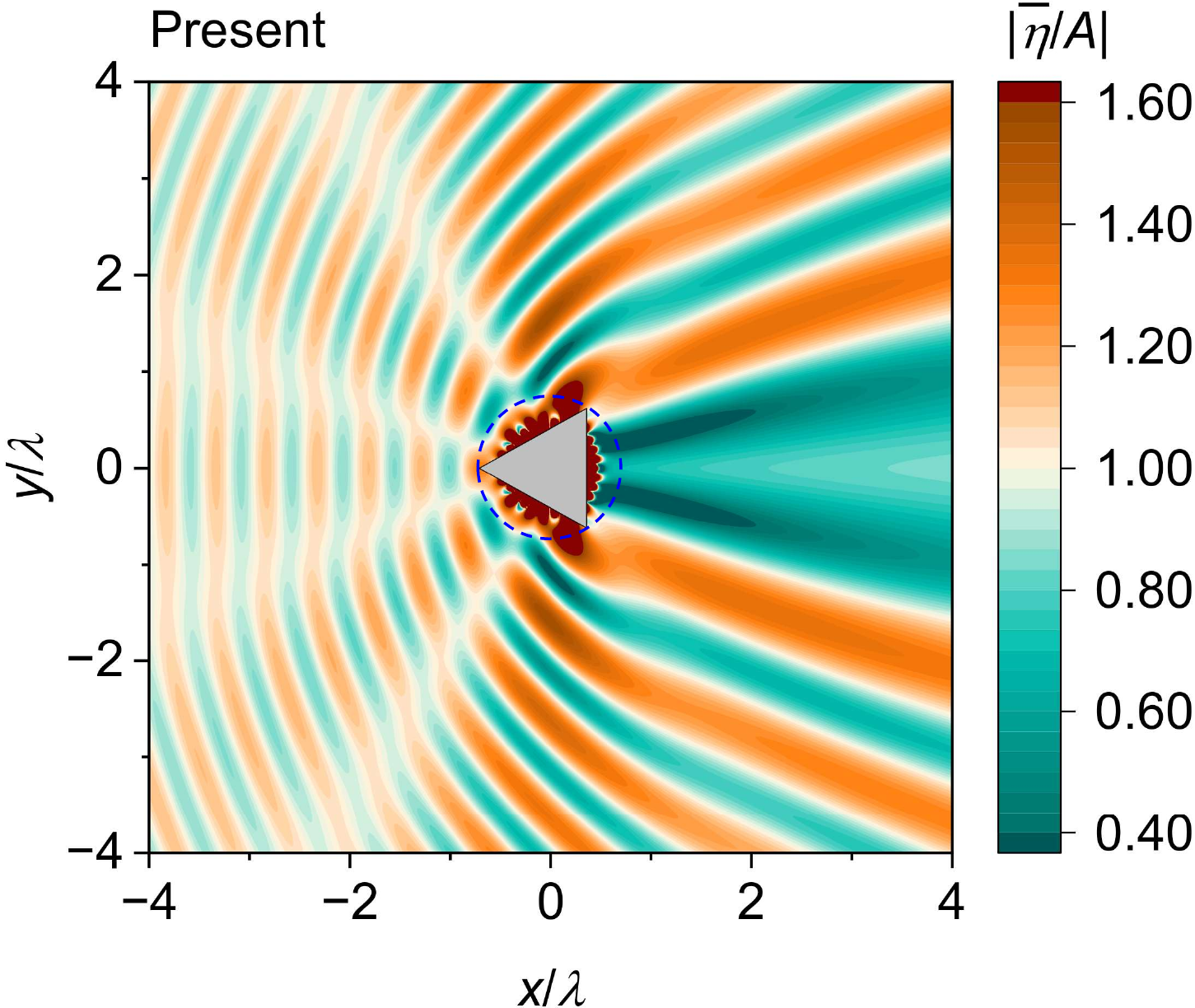}
  \subcaption{}
  \end{minipage}
}
{
  \begin{minipage}{0.32\linewidth}
  \centering
  \includegraphics[width=1.0\linewidth]{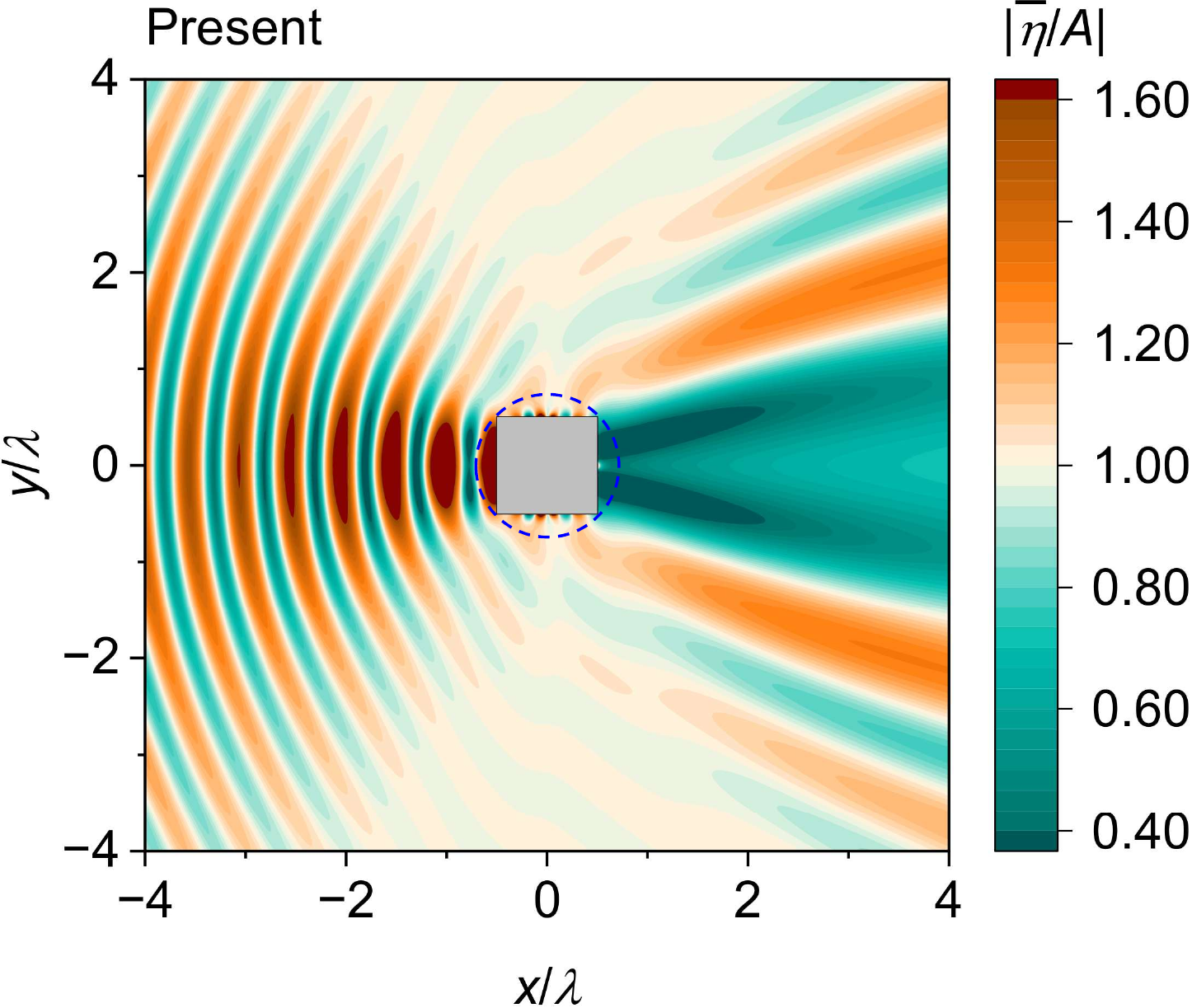}
  \subcaption{}
  \end{minipage}
}\\
{
  \begin{minipage}{0.32\linewidth}
  \centering
  \includegraphics[width=1.0\linewidth]{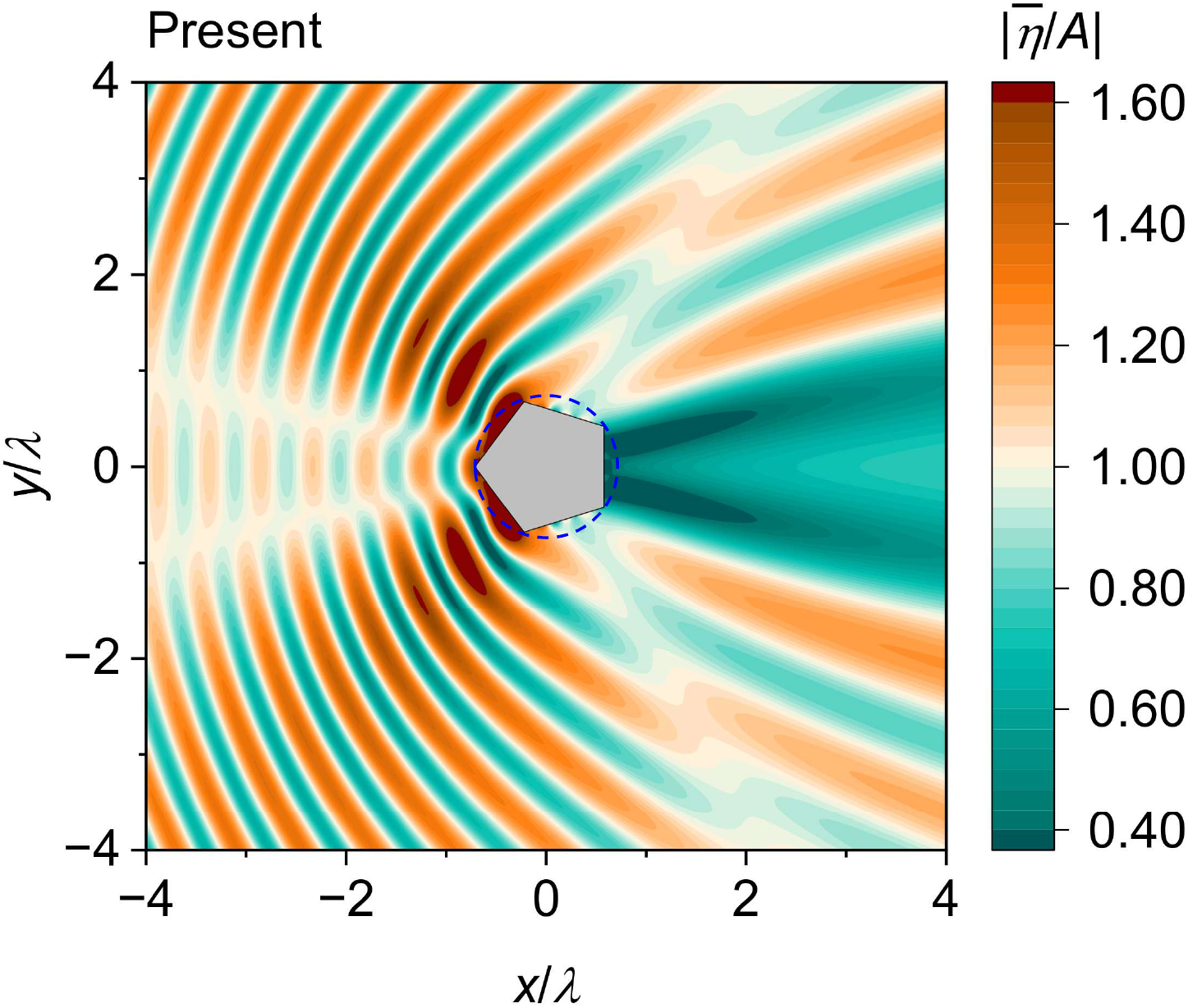}
  \subcaption{}
  \end{minipage}
}
{
  \begin{minipage}{0.32\linewidth}
  \centering
  \includegraphics[width=1.0\linewidth]{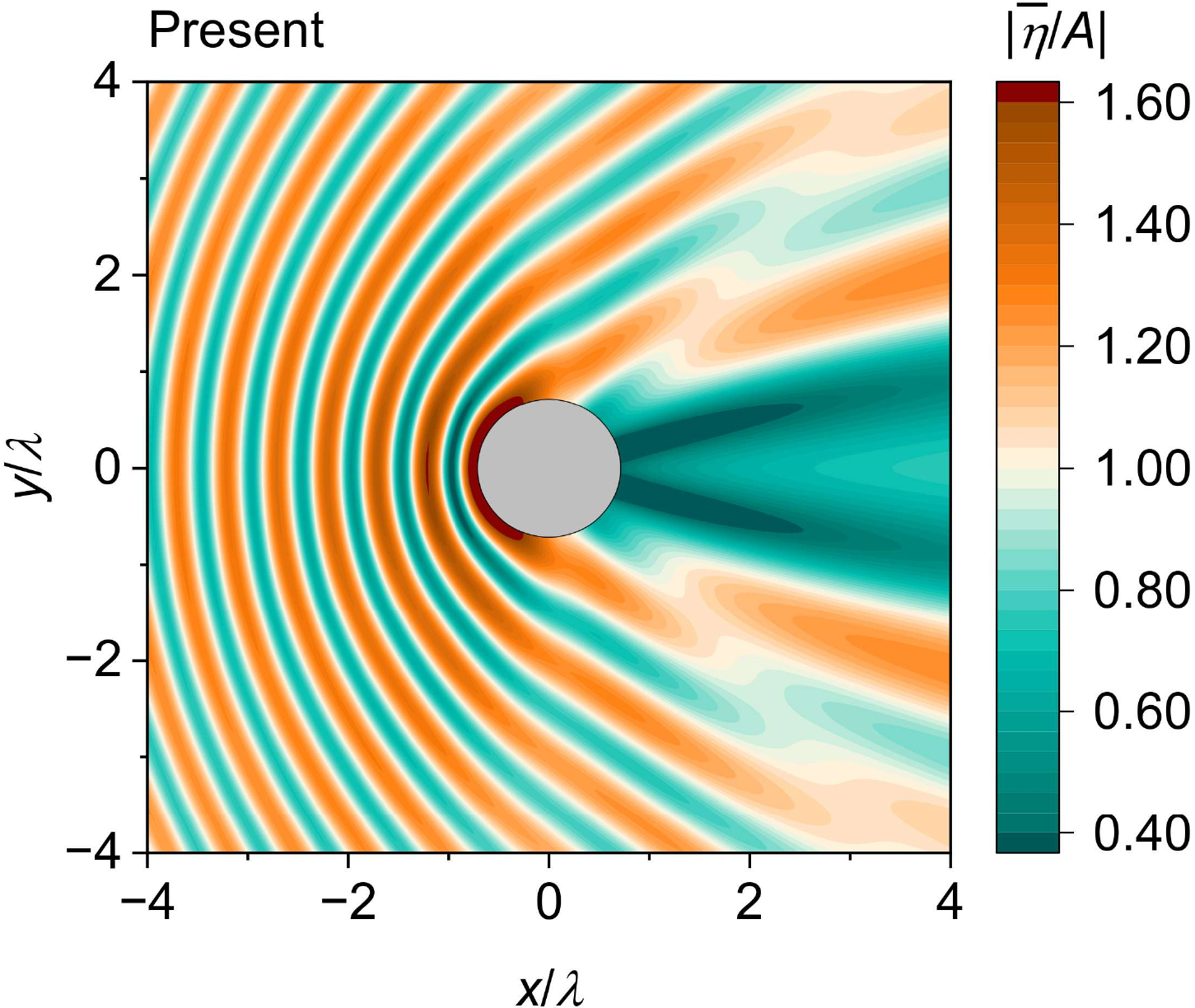}
  \subcaption{}
  \end{minipage}
}
\caption{
Amplitude distribution of wave field around isolate ice floe of (a) Model-I, (b) Model-II, (c) Model-III, and (d) Model-IV in short waves.}
\label{fig:wave_field_shape}
\end{figure}

\begin{figure}[!htp]
\centering
{
  \begin{minipage}{0.35\linewidth}
  \centering
  \includegraphics[width=1.0\linewidth]{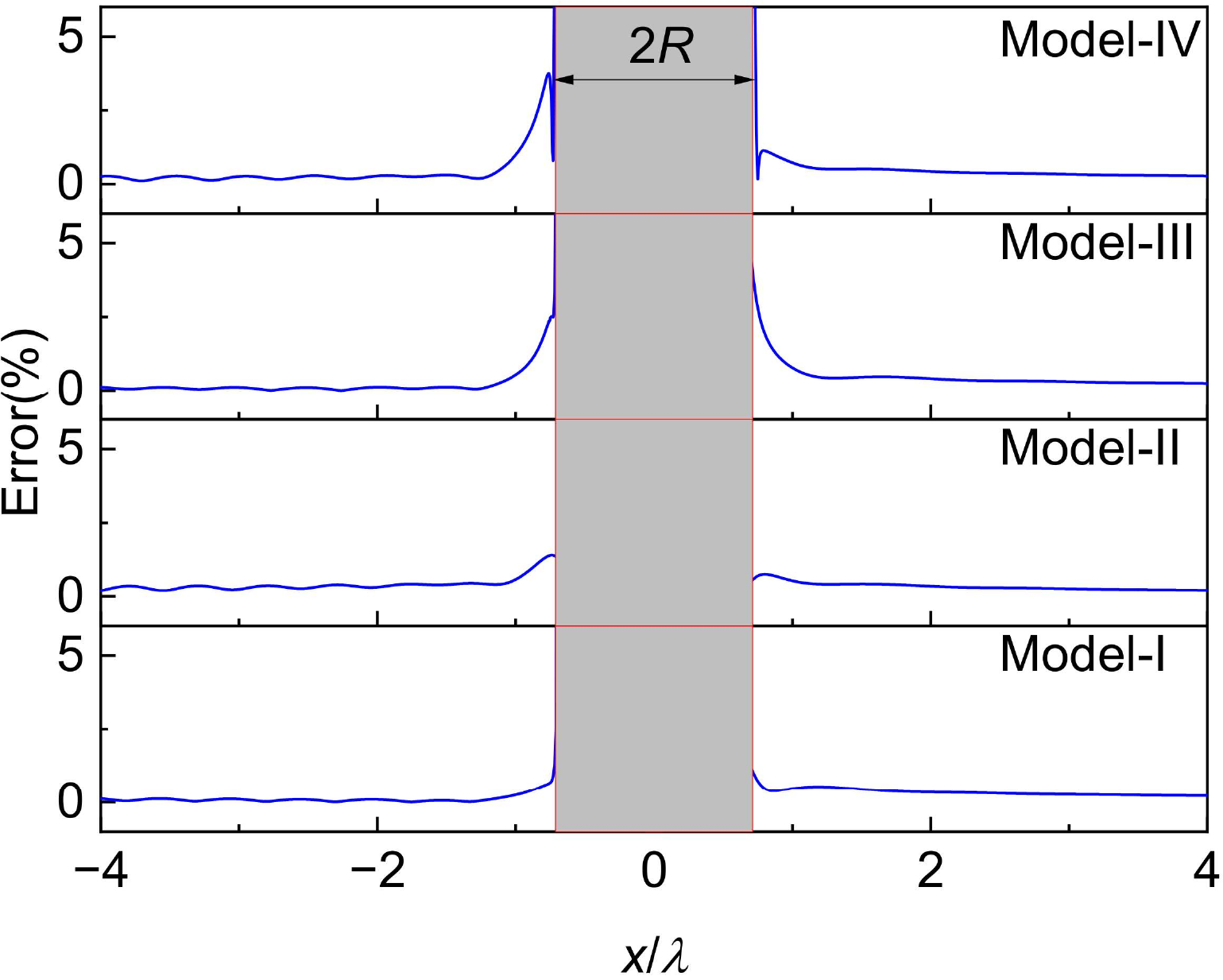}
  \subcaption{}
  \end{minipage}
}
{
  \begin{minipage}{0.35\linewidth}
  \centering
  \includegraphics[width=1.0\linewidth]{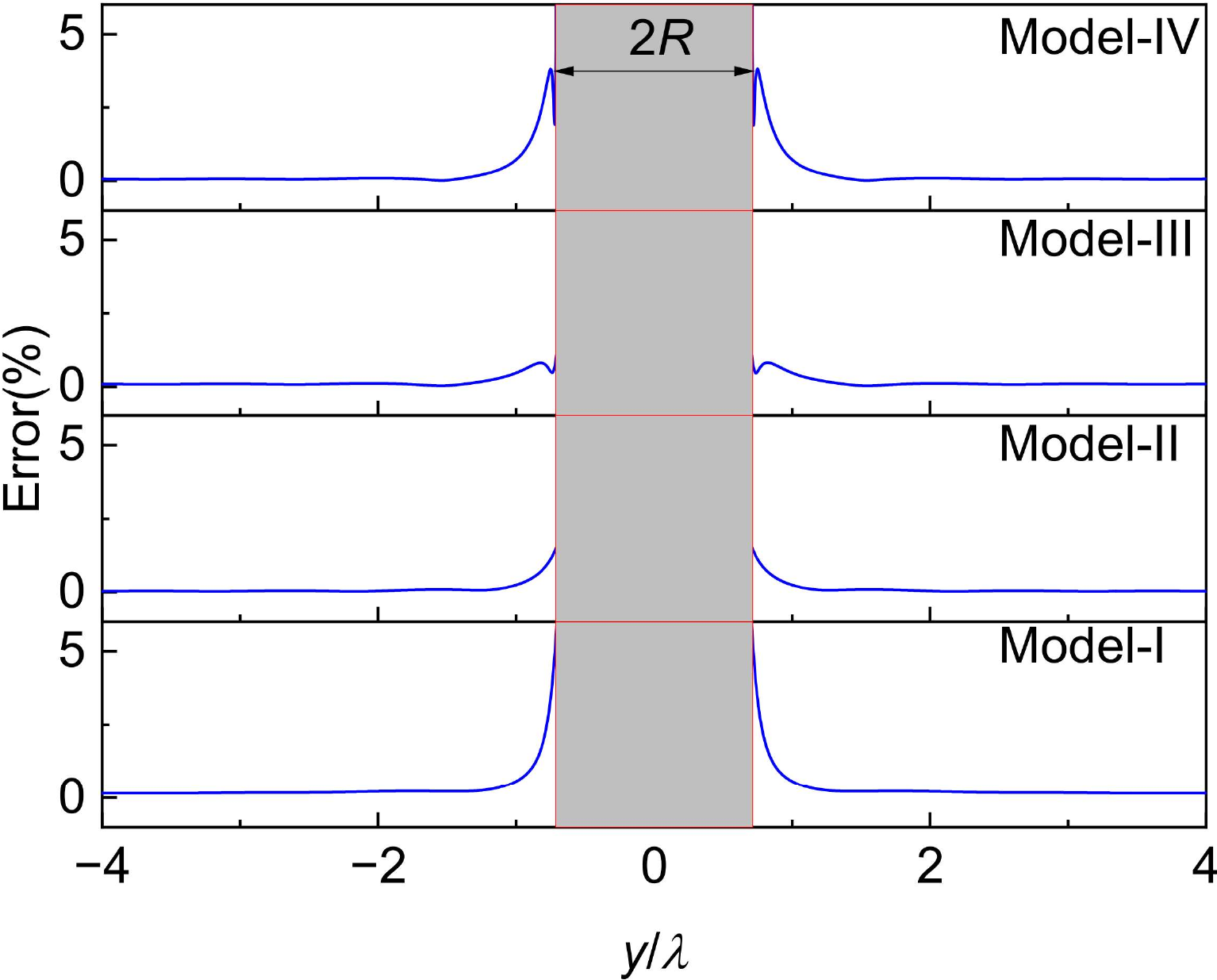}
  \subcaption{}
  \end{minipage}
}
\caption{
Distribution of relative error along (a) $y=0$ and (b) $x=0$, for isolate ice floe of Model-I, Model-II, Model-III, and Model-IV in short waves.}
\label{fig:error_shape_compare}
\end{figure}

Based on the operation rules derived above, the WCD approach is further applied to wave field reconstruction around Model-I, Model-III, and Model-IV.
Fig. \ref{fig:wave_field_shape} shows the amplitude distribution of the wave field around the isolated ice floe of Model-I, Model-II, Model-III, and Model-IV in long waves.
The circumcircle of each ice floe is plotted for reference.
Outside the circumcircle, the amplitude distribution is smooth and regular.
For each ice floe model, the reconstructed wave field is compared with the one calculated directly by the BEM code HAMS.
The distribution of relative error along $y = 0$ and $x = 0$ for each isolated ice floe is shown in Fig. \ref{fig:error_shape_compare}.
The error-range statement still holds for all these ice floe shapes that: the error of the reconstructed wave field is always less than $0.5\%$ at any location more than $0.5\lambda$ away from the circumcircle of the ice floe; and the error is less than $5\%$ at a location more than $0.2R$ away from the circumcircle.

\subsection{Error analysis of enhanced interaction method for multiple ice floes}\label{subsec:exam_EI}

This subsection further considers the EI method for scenarios with multiple ice floes.
The error-range statement of the WCD approach, derived from the isolate-ice-floe condition, is examined in the multiple-ice-floe scenario.
The wave fields of two or three ice floes are given for demonstration.
The circumcircle distance between adjacent ice floes is defined as the shortest distance between the circumcircles of the ice floes.
Referring to Fig. \ref{fig:coord_sys}, the circumcircle distance between ice-$i$ and ice-$j$ is defined as $d = L_{ij}-R_i - R_j$. 
Here, $R_i$ and $R_j$ are the radii of the circumcircles of two ice floes.

\begin{figure}[!htp]
\centering
{
  \begin{minipage}{0.32\linewidth}
  \centering
  \includegraphics[width=1.0\linewidth]{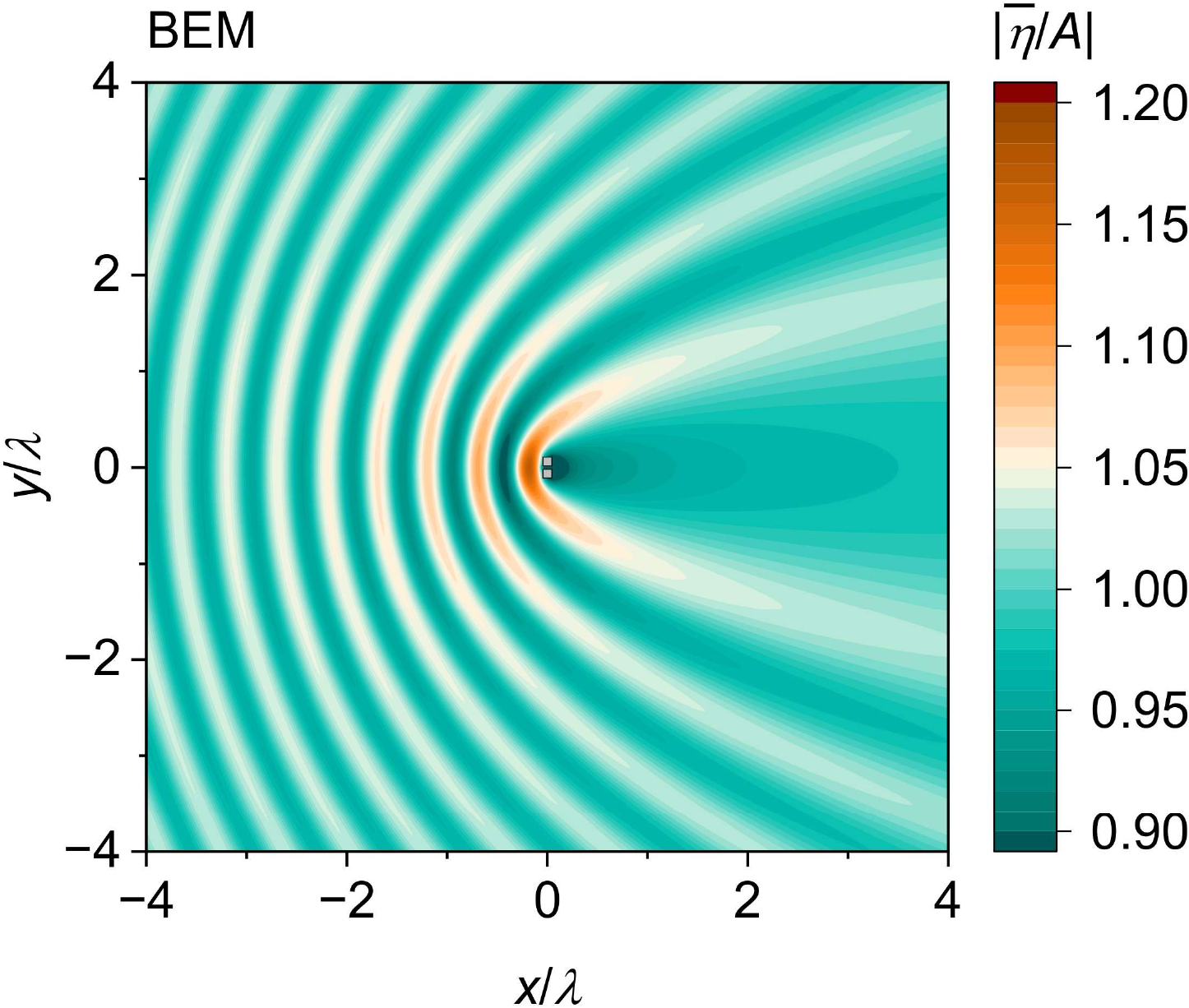}
  \subcaption{}
  \end{minipage}
}
{
  \begin{minipage}{0.32\linewidth}
  \centering
  \includegraphics[width=1.0\linewidth]{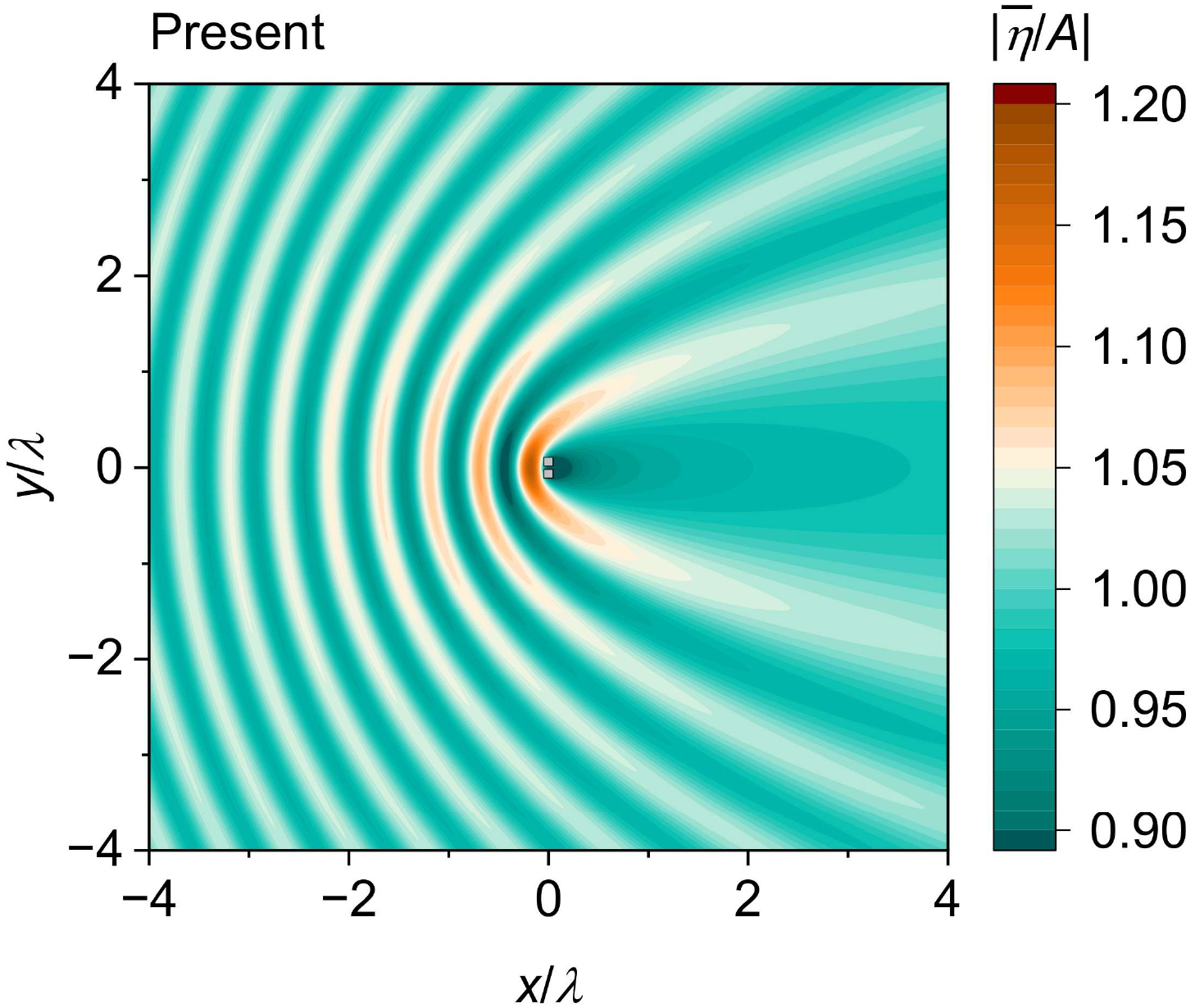}
  \subcaption{}
  \end{minipage}
}
{
  \begin{minipage}{0.32\linewidth}
  \centering
  \includegraphics[width=1.0\linewidth]{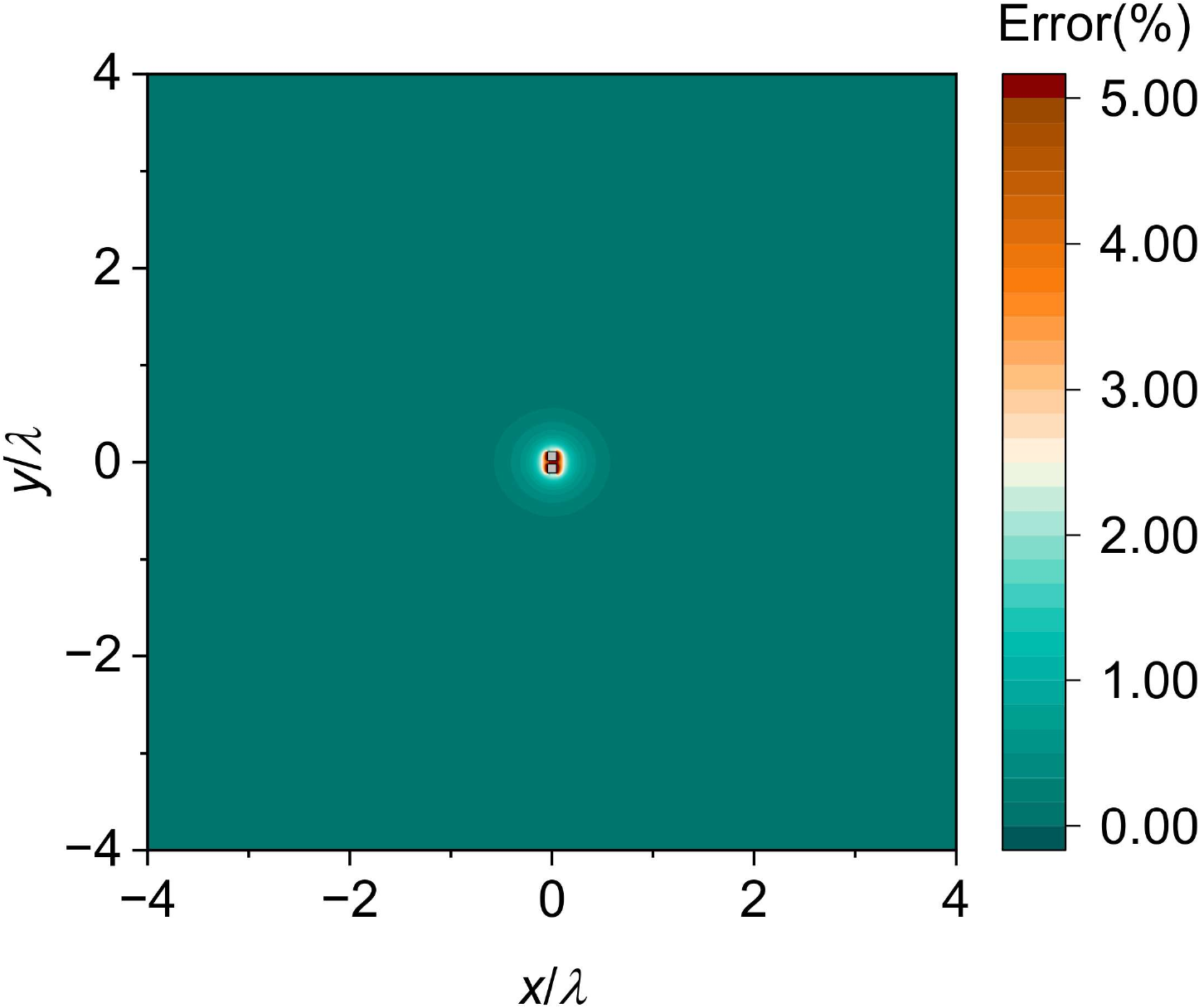}
  \subcaption{}
  \end{minipage}
}
\caption{
Amplitude distribution of wave field around two identical ice floes of Model-II with distance $d=0$ in short waves: (a) BEM solution; (b) present EI solution; and (c) relative error of present solution in long waves.
}
\label{fig:distance_effect_L0}
\end{figure}

\begin{figure}[!htp]
\centering
{
  \begin{minipage}{0.32\linewidth}
  \centering
  \includegraphics[width=1.0\linewidth]{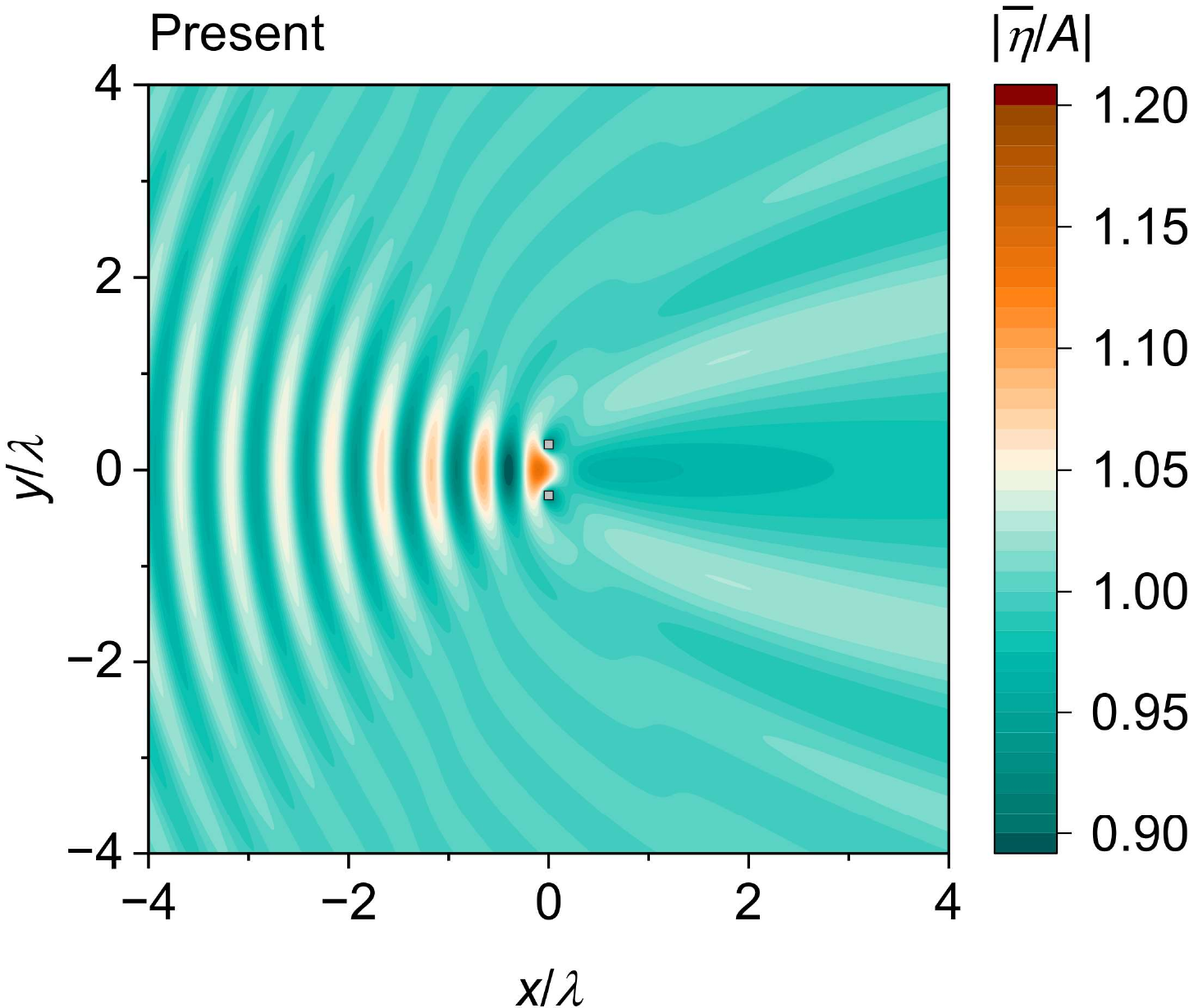}
  \subcaption{}
  \end{minipage}
}
{
  \begin{minipage}{0.32\linewidth}
  \centering
  \includegraphics[width=1.0\linewidth]{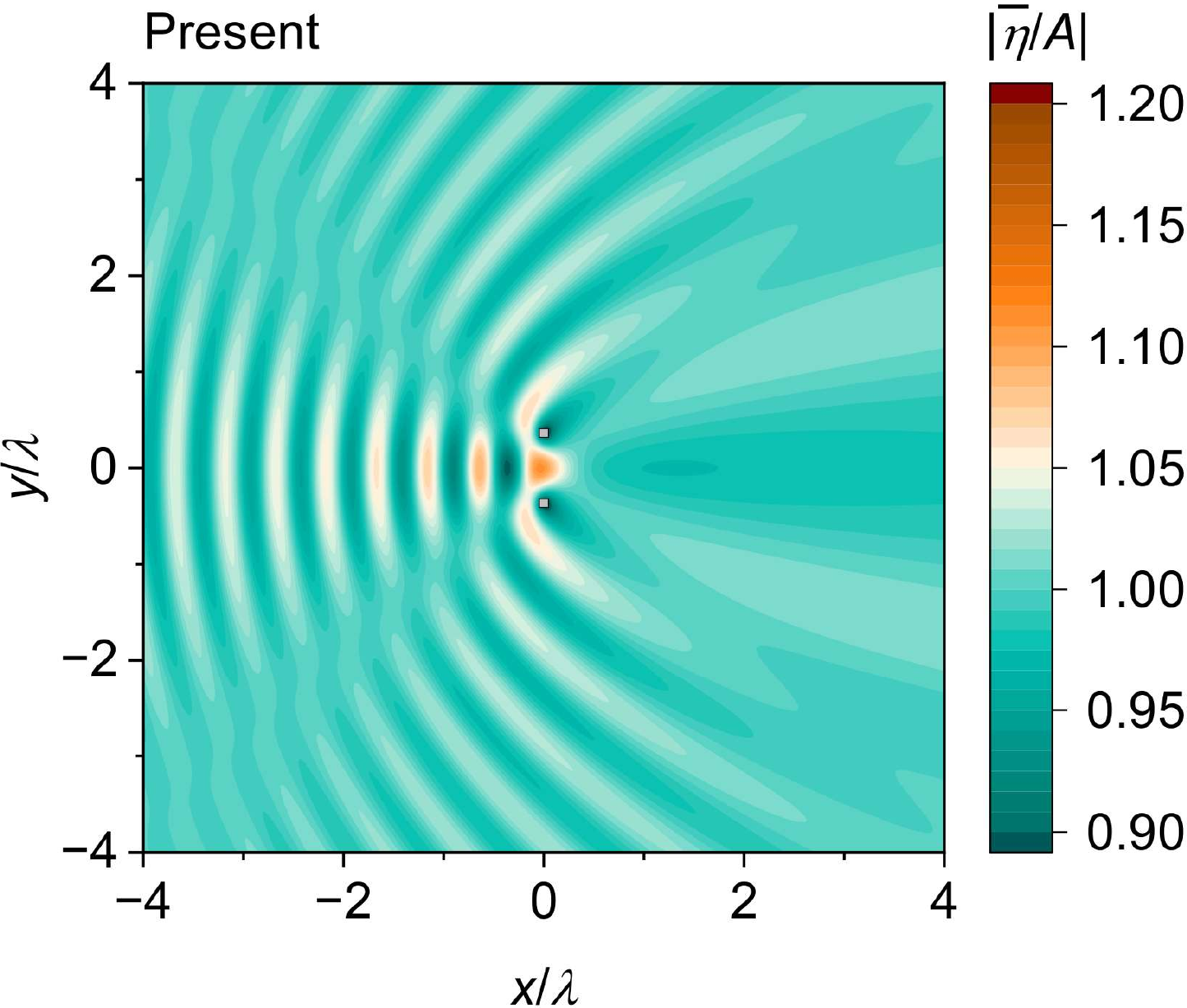}
  \subcaption{}
  \end{minipage}
}
{
  \begin{minipage}{0.32\linewidth}
  \centering
  \includegraphics[width=1.0\linewidth]{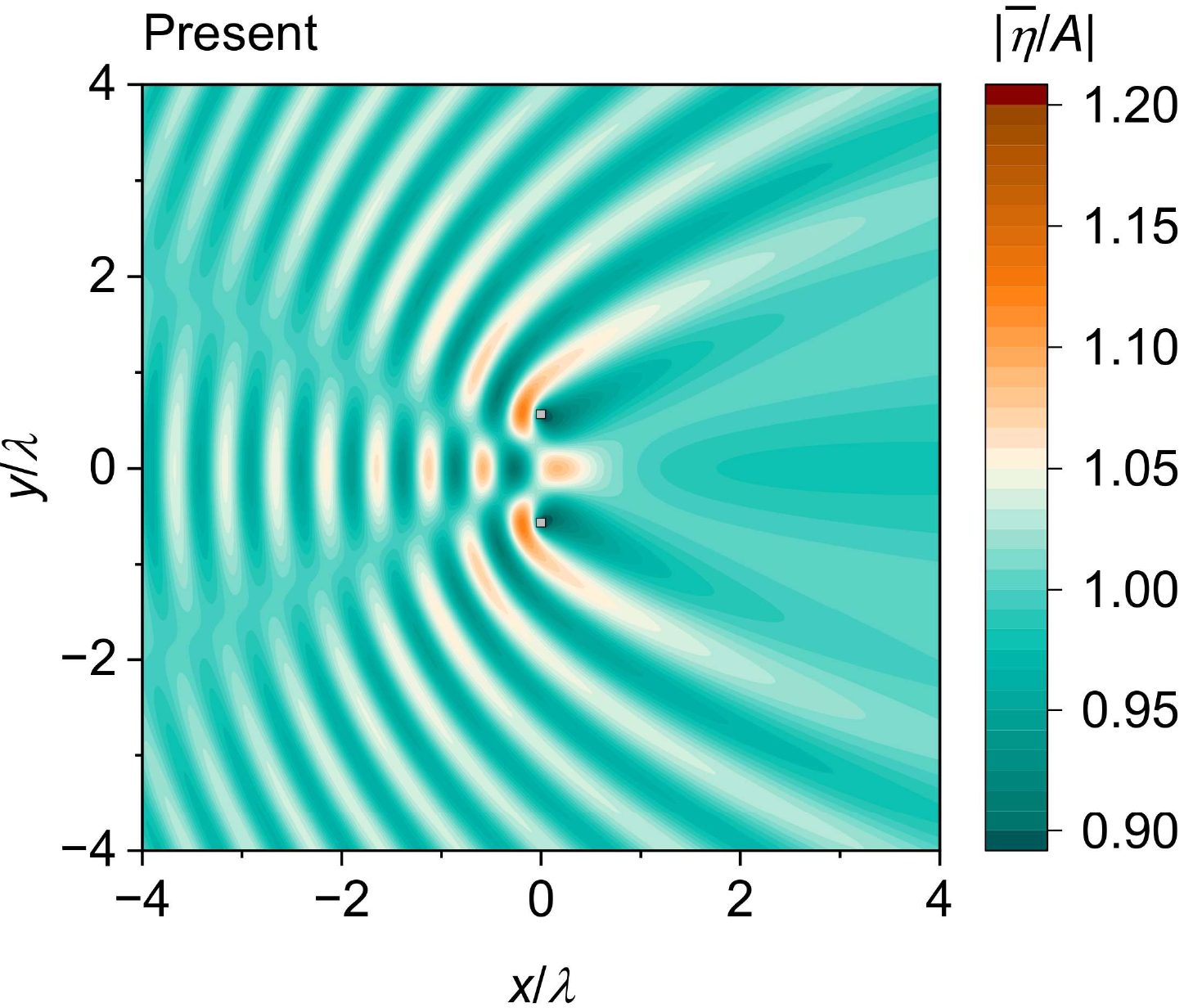}
  \subcaption{}
  \end{minipage}
}
\caption{
Amplitude distribution of wave field around two identical ice floes of Model-II with circumcircle distance of (a) $d=0.4\lambda$, (b) $d=0.6\lambda$, and (c) $d=1.0\lambda$ in long waves.}
\label{fig:distance_effect}
\end{figure}

Fig. \ref{fig:distance_effect_L0} shows the amplitude distribution of the wave field around two identical ice floes of Model-II with distance $d = 0$ in long waves. 
Fig. \ref{fig:distance_effect_L0} (a) and (b) show the results calculated by the BEM and the present EI method, respectively.
Fig. \ref{fig:distance_effect_L0} (c) shows the relative error of the wave field predicted by the EI method compared to the BEM solution.
Since the two ice floes are close enough, the generated wave field is very similar to that caused by an isolated ice floe in Fig. \ref{fig:diffracted_wave}.
In Fig. \ref{fig:distance_effect}, the circumcircle distance is further increased to $d = 0.4\lambda$, $0.6\lambda$, and $1.0\lambda$, respectively.
The distributions of relative error along $y = 0$ and $x = 0$ of the wave field are depicted in Fig. \ref{fig:distance_effect}.
As the circumcircle distance increases from 0 to $1.0\lambda$, the relative error along the mid-line between two ice floes decreases significantly from about $5\%$ to $0.1\%$.
Along the line of the centres of the two circumcircles, the relative error of the predicted wave amplitude is always less than $5\%$.
The error-range statement of the WCD approach for the isolate-ice-floe case still applies to the EI method in the current multiple-ice-floe scenario.
\begin{figure}[!htp]
\centering
{
  \begin{minipage}{0.35\linewidth}
  \centering
  \includegraphics[width=1.0\linewidth]{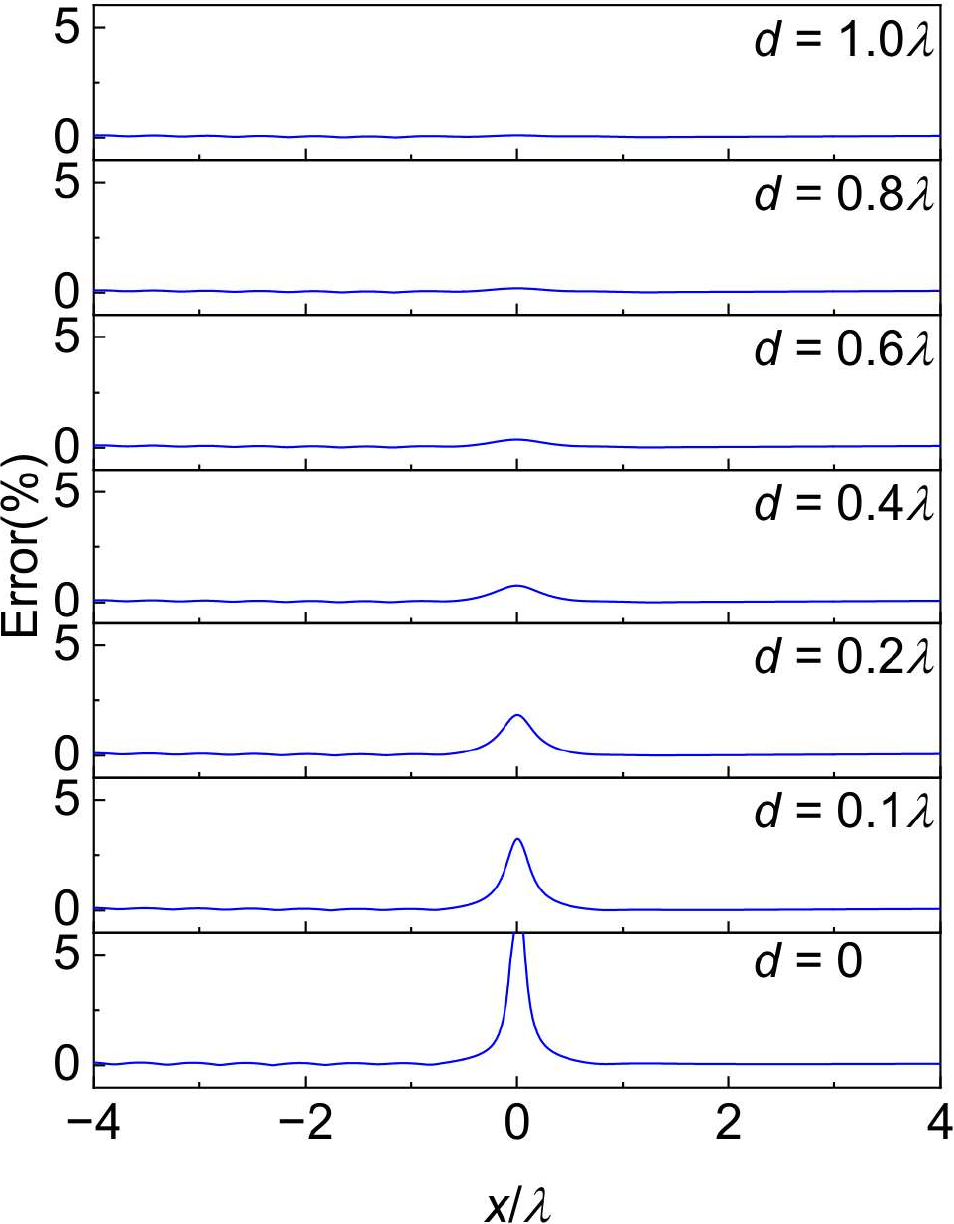}
  \subcaption{}
  \end{minipage}
}
{
  \begin{minipage}{0.35\linewidth}
  \centering
  \includegraphics[width=1.0\linewidth]{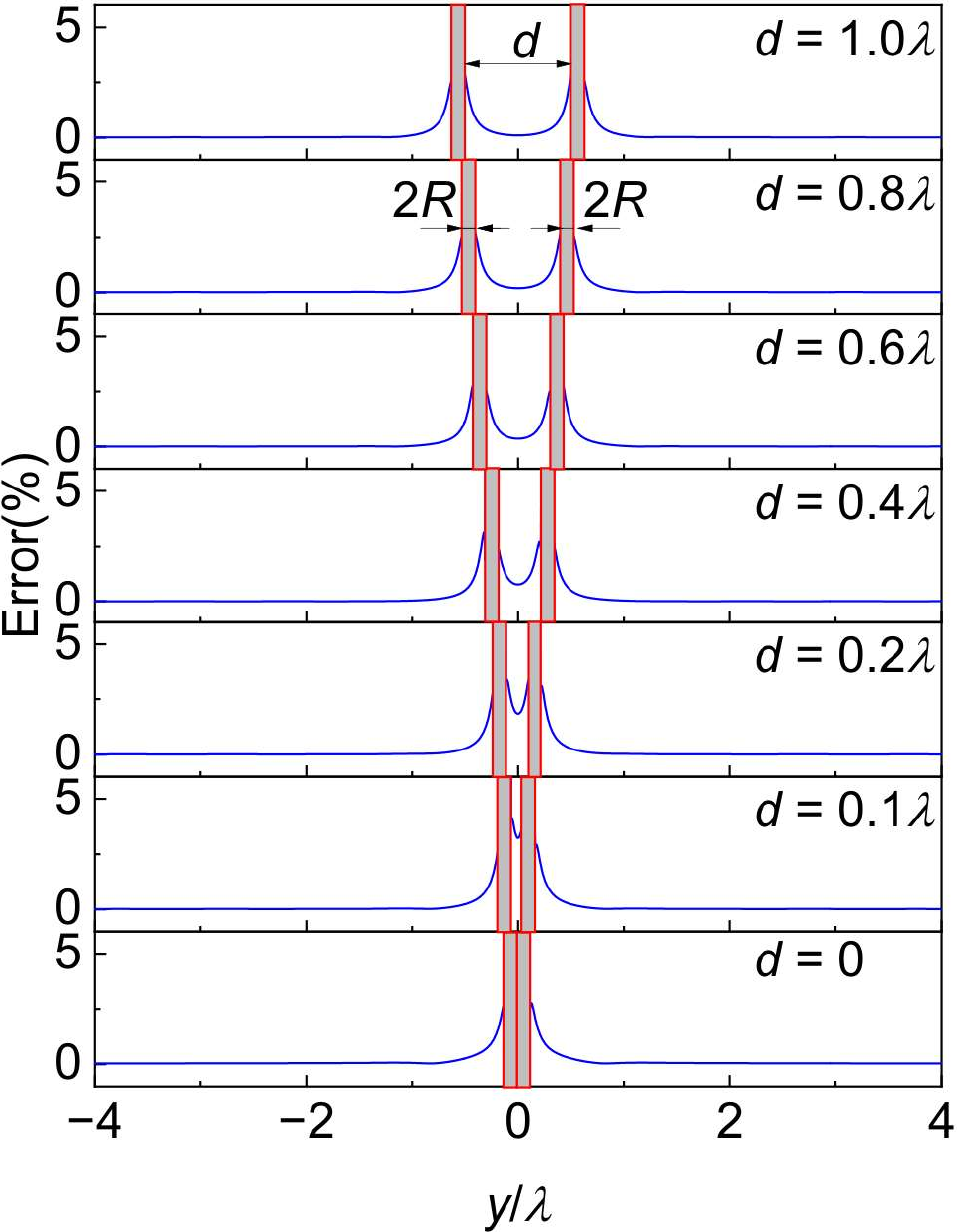}
  \subcaption{}
  \end{minipage}
}
\caption{
Distribution of relative error along (a) $y=0$ and (b) $x=0$, for two identical ice floes of Model-II with circumcircle distance $d$ ranging from 0 to $1.0\lambda$ in long waves.}
\label{fig:distance_error}
\end{figure}

\begin{figure}[!htp]
\centering
{
  \begin{minipage}{0.32\linewidth}
  \centering
  \includegraphics[width=1.0\linewidth]{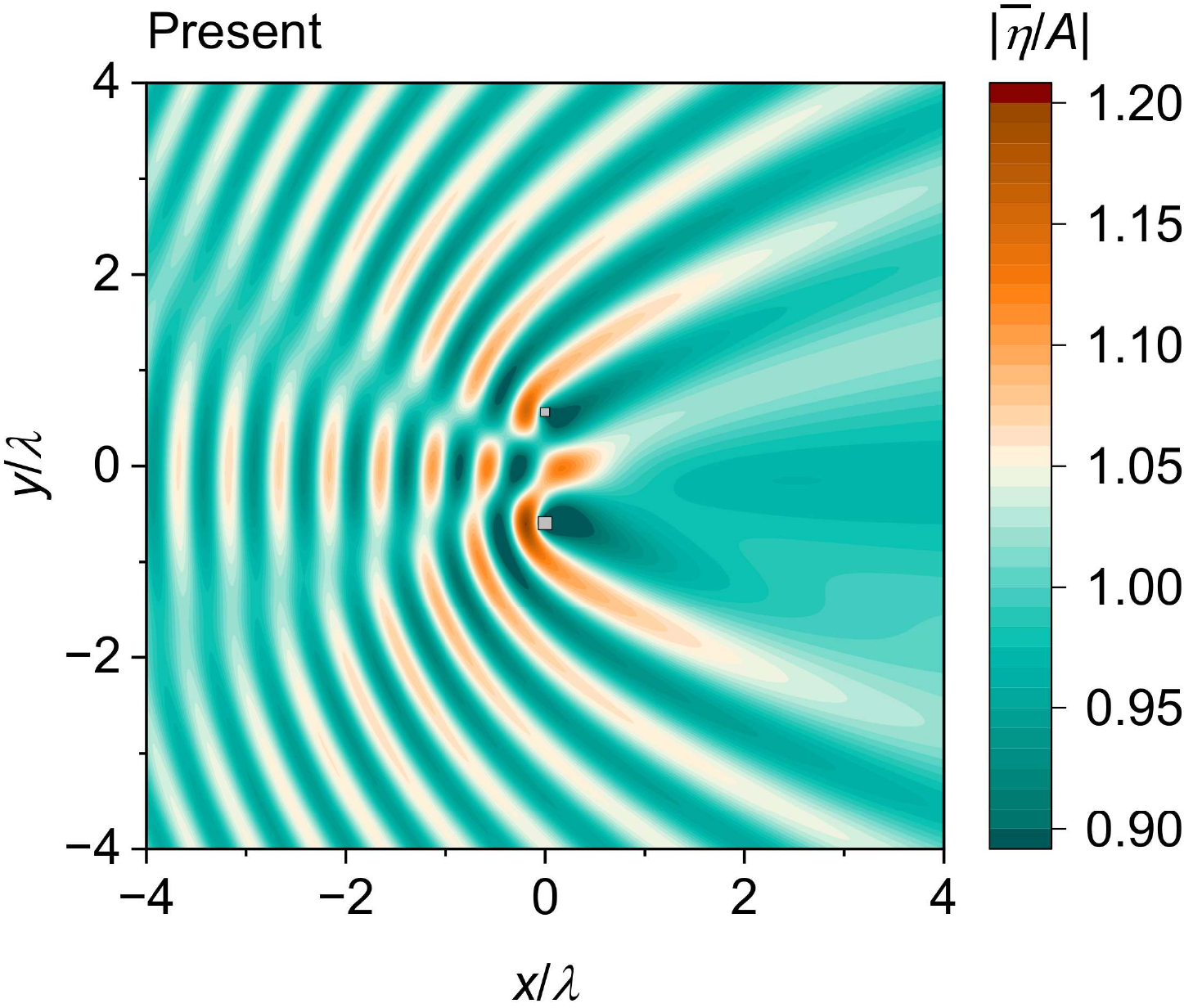}
  \subcaption{}
  \end{minipage}
}
{
  \begin{minipage}{0.32\linewidth}
  \centering
  \includegraphics[width=1.0\linewidth]{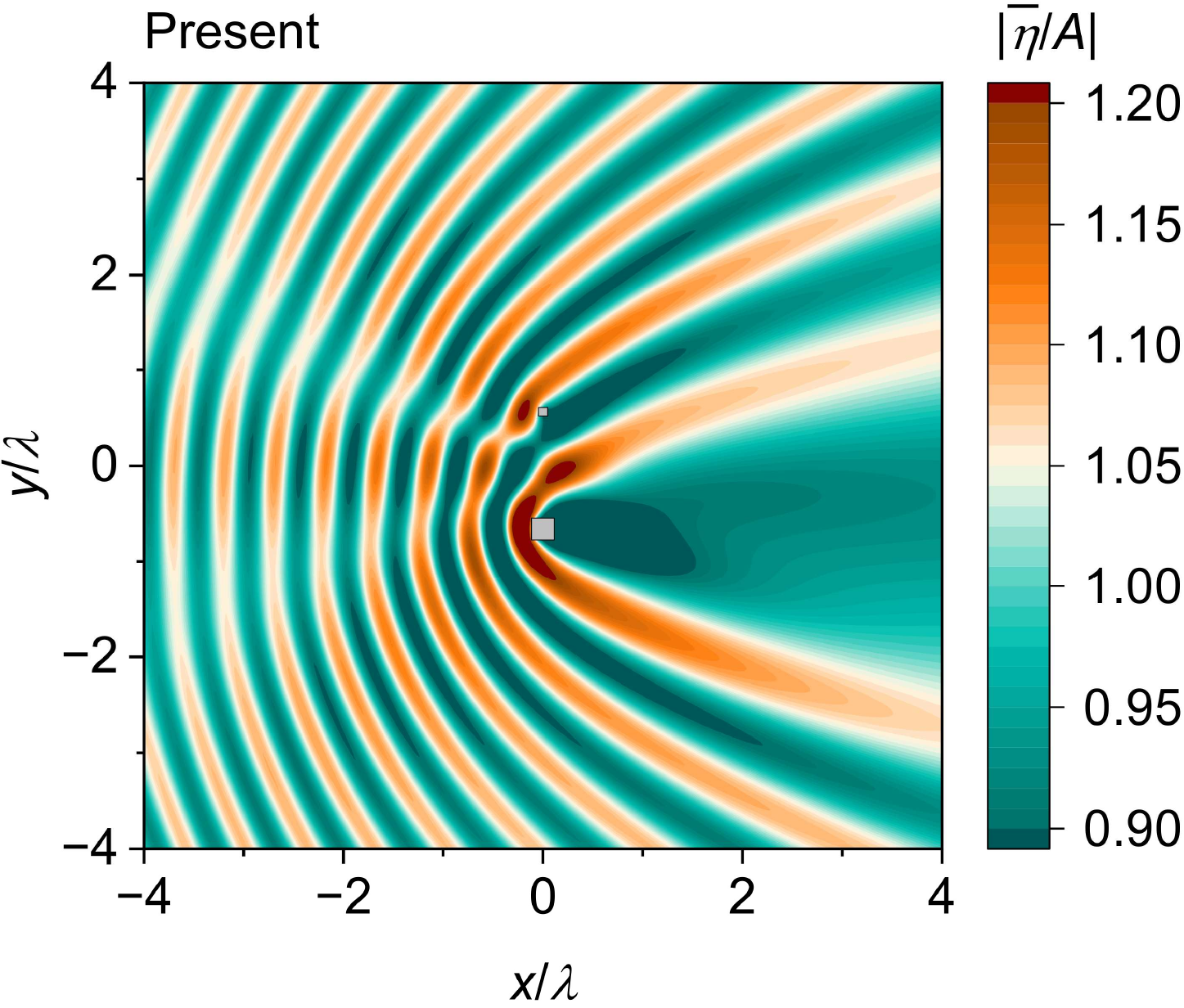}
  \subcaption{}
  \end{minipage}
}
{
  \begin{minipage}{0.32\linewidth}
  \centering
  \includegraphics[width=1.0\linewidth]{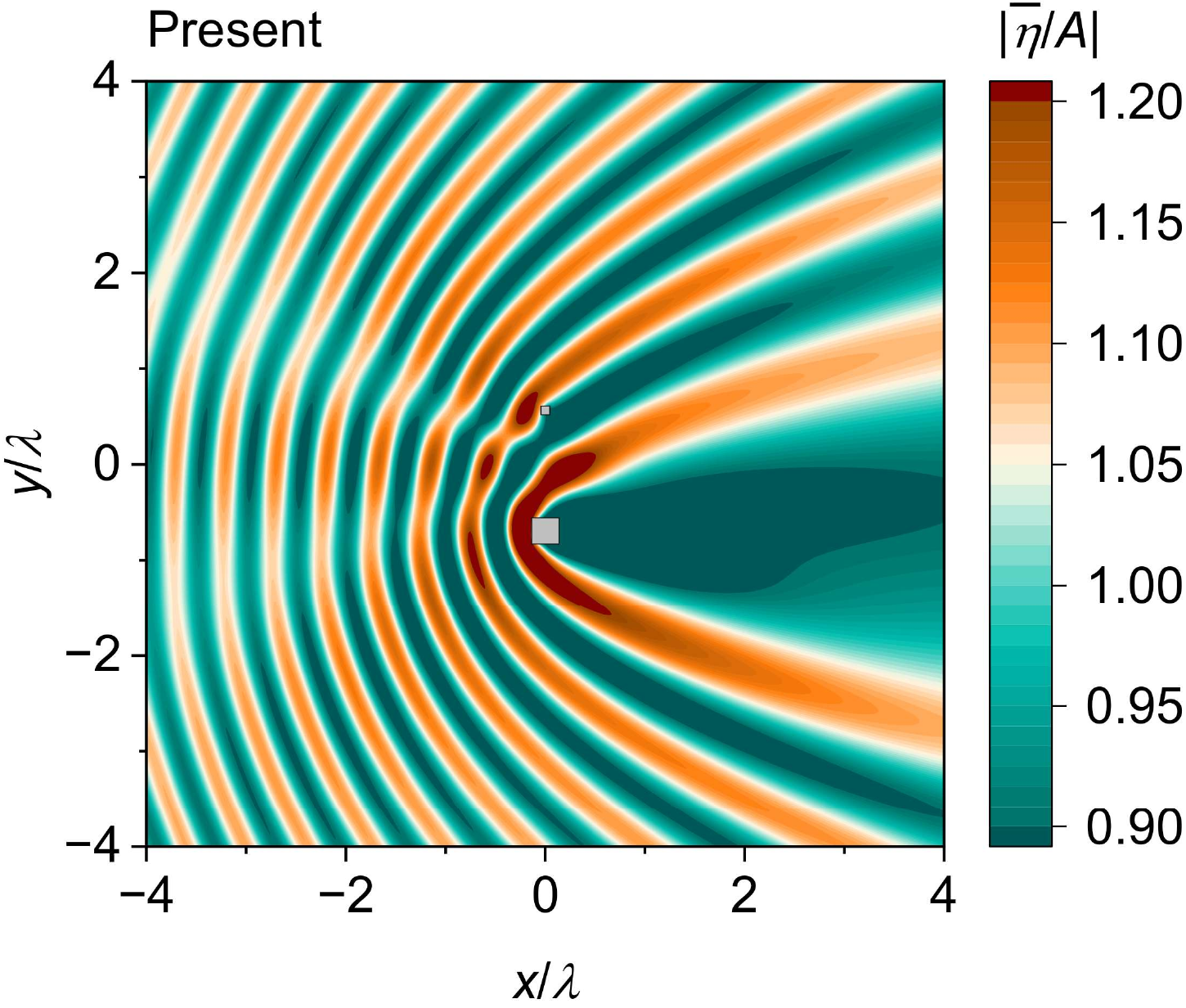}
  \subcaption{}
  \end{minipage}
}
\caption{
Amplitude distribution of wave field around two ice floes of Model-II with size difference of (a) $R_2=1.5R_1$, (b) $R_2=2.0R_1$, and (c) $R_2=3.0R_1$ in long waves.
}
\label{fig:size_effect}
\end{figure}

Fig. \ref{fig:size_effect} further considers the scenario of two ice floes of Model-II with different sizes in long waves.
The circumradii of two ice floes are $R_1 = R$ and $R_2$, respectively.
The circumcircle distance between two ice floes is fixed at $d = 1.0\lambda$.
As $R_2$ increases from $1.0R_1$ to $3.0R_1$, the disturbance of the second ice floe to the water surface intensifies.
The difference between the maximum and minimum wave amplitudes is evidently enlarged.
The relative error of the present EI solution and the BEM result can be obtained in a similar way to those in Fig. \ref{fig:distance_effect_L0}.
Fig. \ref{fig:size_error} shows the distribution of the relative error along $y = 0$ and $x = 0$, where the size difference of two ice floes ranges from $R_2 = R_1$ to $R_2 = 3R_1$.
Along the mid-line between two ice floes, the relative error is generally less than $2\%$.
Along the line connecting the centers of the two circumcircles, outside the circumcircles of both ice floes, the relative error of the predicted wave amplitude is always less than $5\%$.
\begin{figure}[!htp]
\centering
{
  \begin{minipage}{0.35\linewidth}
  \centering
  \includegraphics[width=1.0\linewidth]{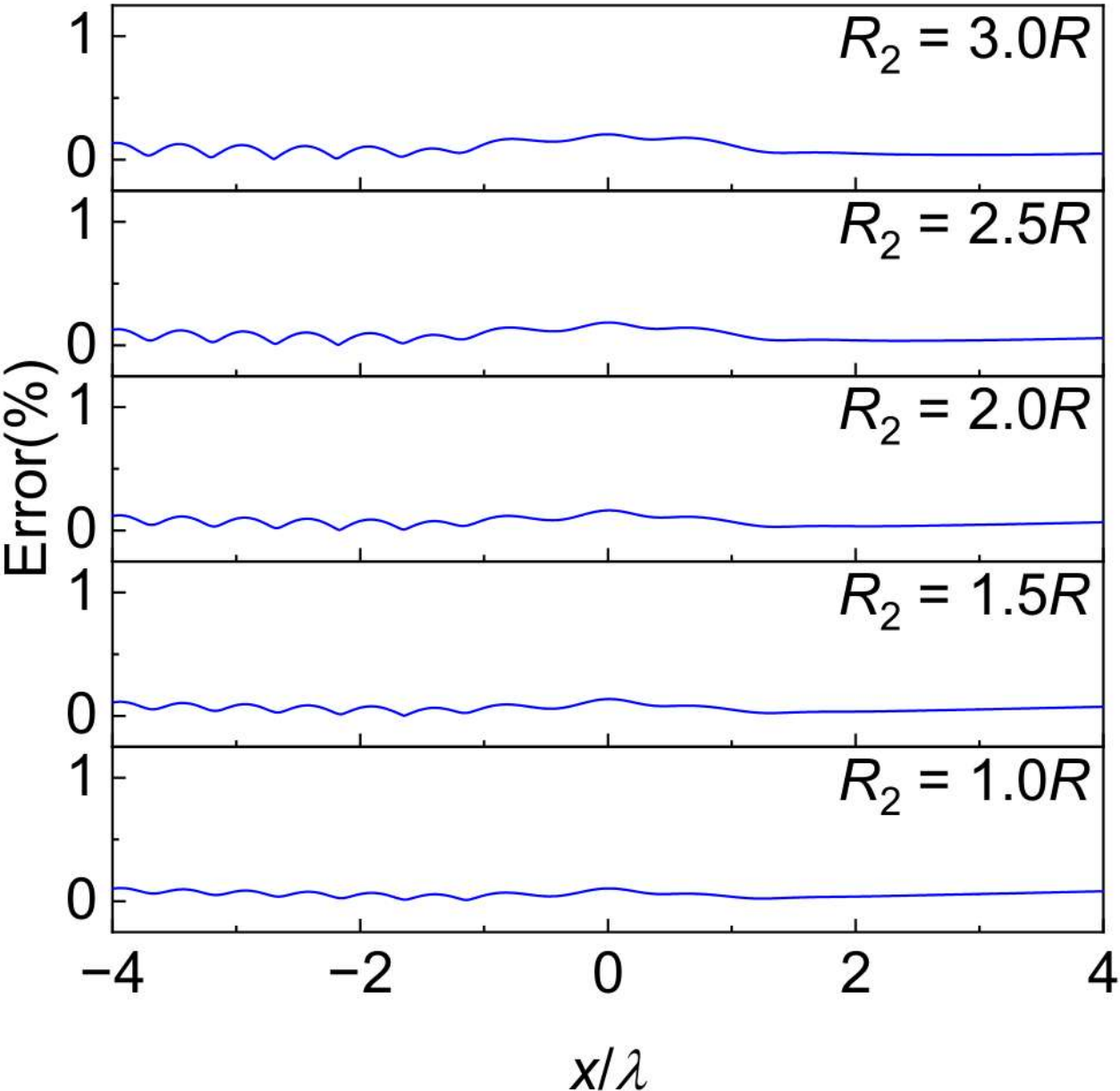}
  \subcaption{}
  \end{minipage}
}
{
  \begin{minipage}{0.35\linewidth}
  \centering
  \includegraphics[width=1.0\linewidth]{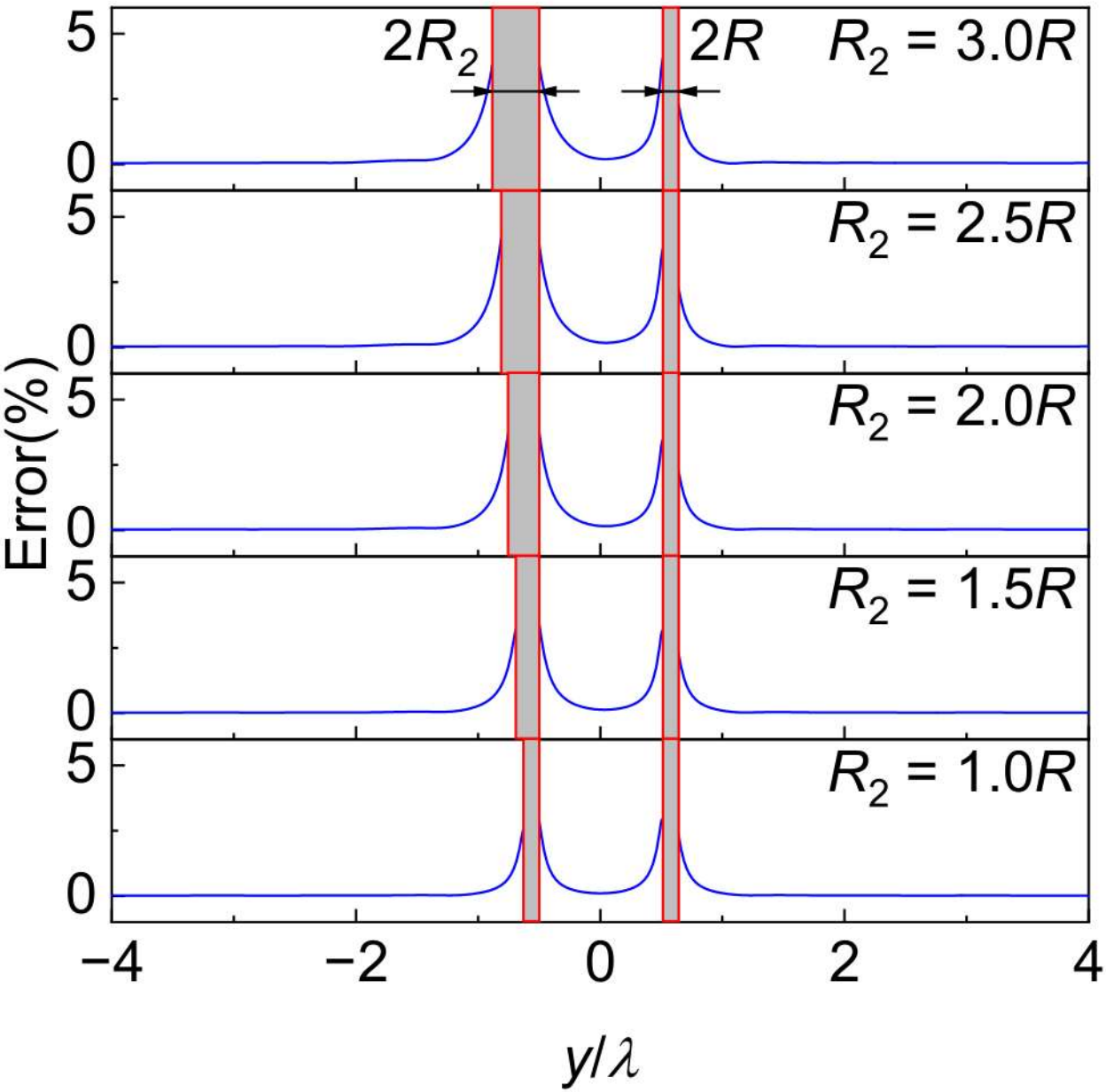}
  \subcaption{}
  \end{minipage}
}
\caption{
Distribution of relative error along (a) $y=0$ and (b) $x=0$, for two ice floes of Model-II with size difference ranging from $R_2=R_1$ to $R_2=3R_1$ in long waves.
}
\label{fig:size_error}
\end{figure}

\begin{figure}[!htp]
\centering
{
  \begin{minipage}{0.32\linewidth}
  \centering
  \includegraphics[width=1.0\linewidth]{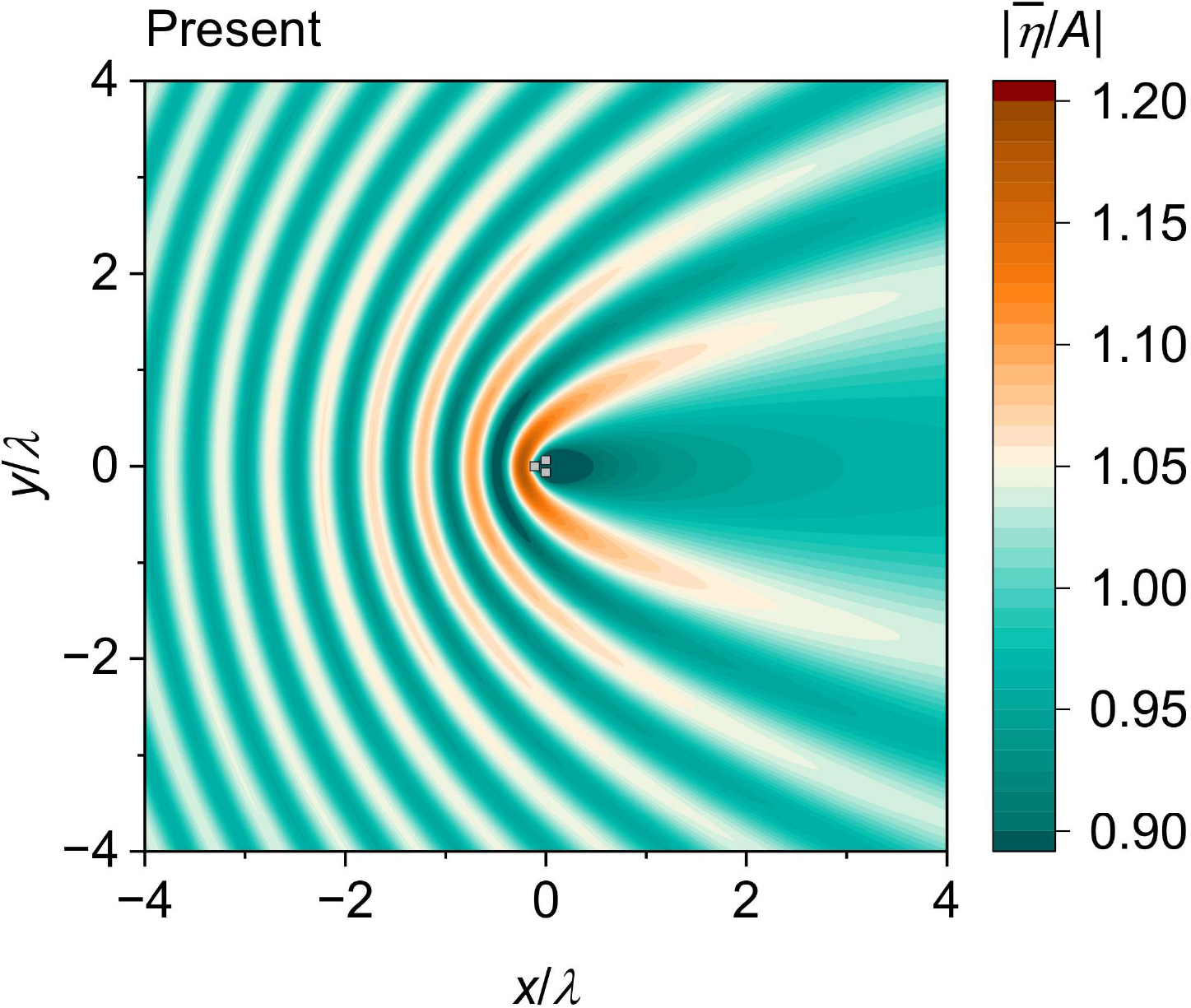}
  \subcaption{}
  \end{minipage}
}
{
  \begin{minipage}{0.32\linewidth}
  \centering
  \includegraphics[width=1.0\linewidth]{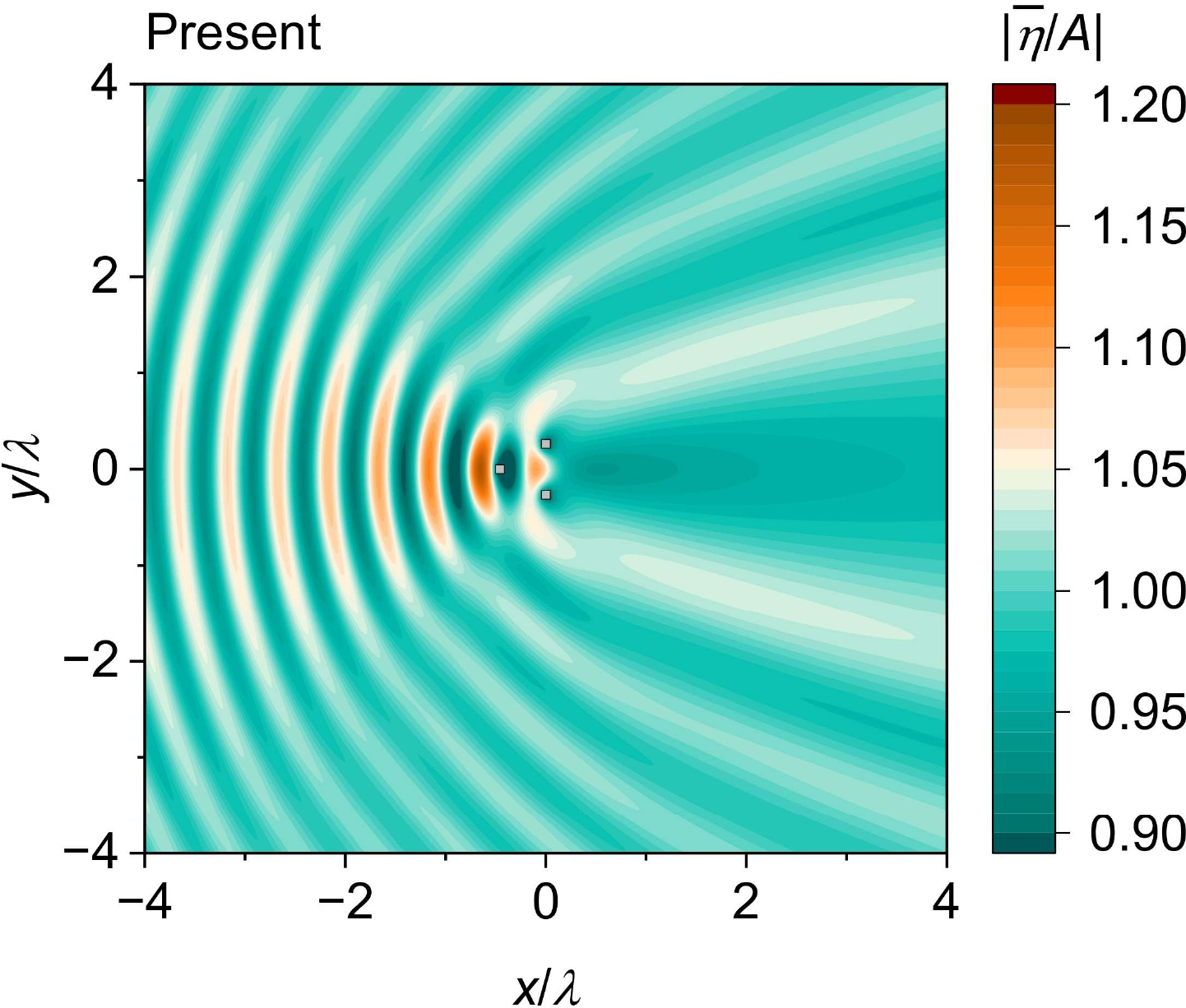}
  \subcaption{}
  \end{minipage}
}\\
{
  \begin{minipage}{0.32\linewidth}
  \centering
  \includegraphics[width=1.0\linewidth]{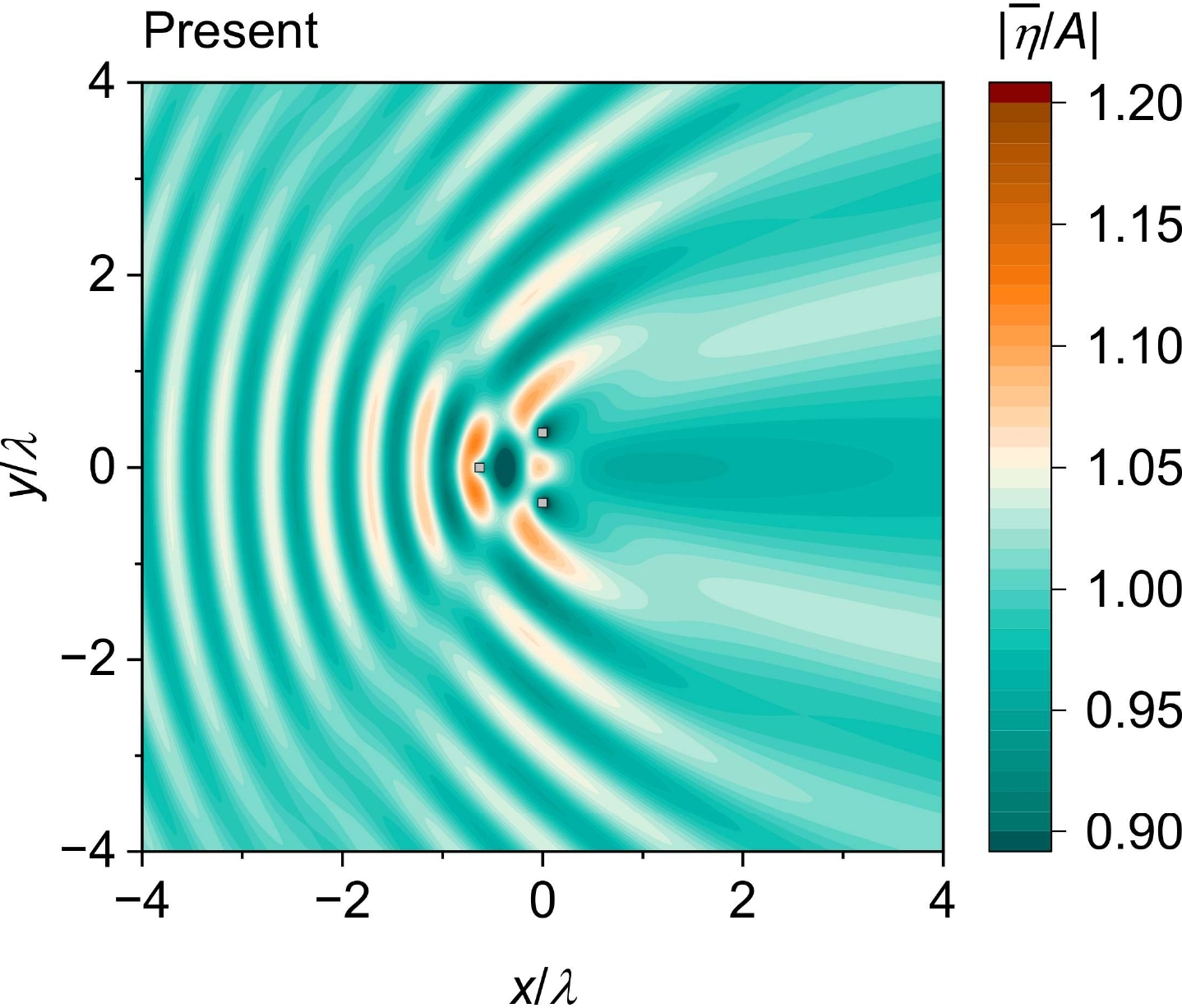}
  \subcaption{}
  \end{minipage}
}
{
  \begin{minipage}{0.32\linewidth}
  \centering
  \includegraphics[width=1.0\linewidth]{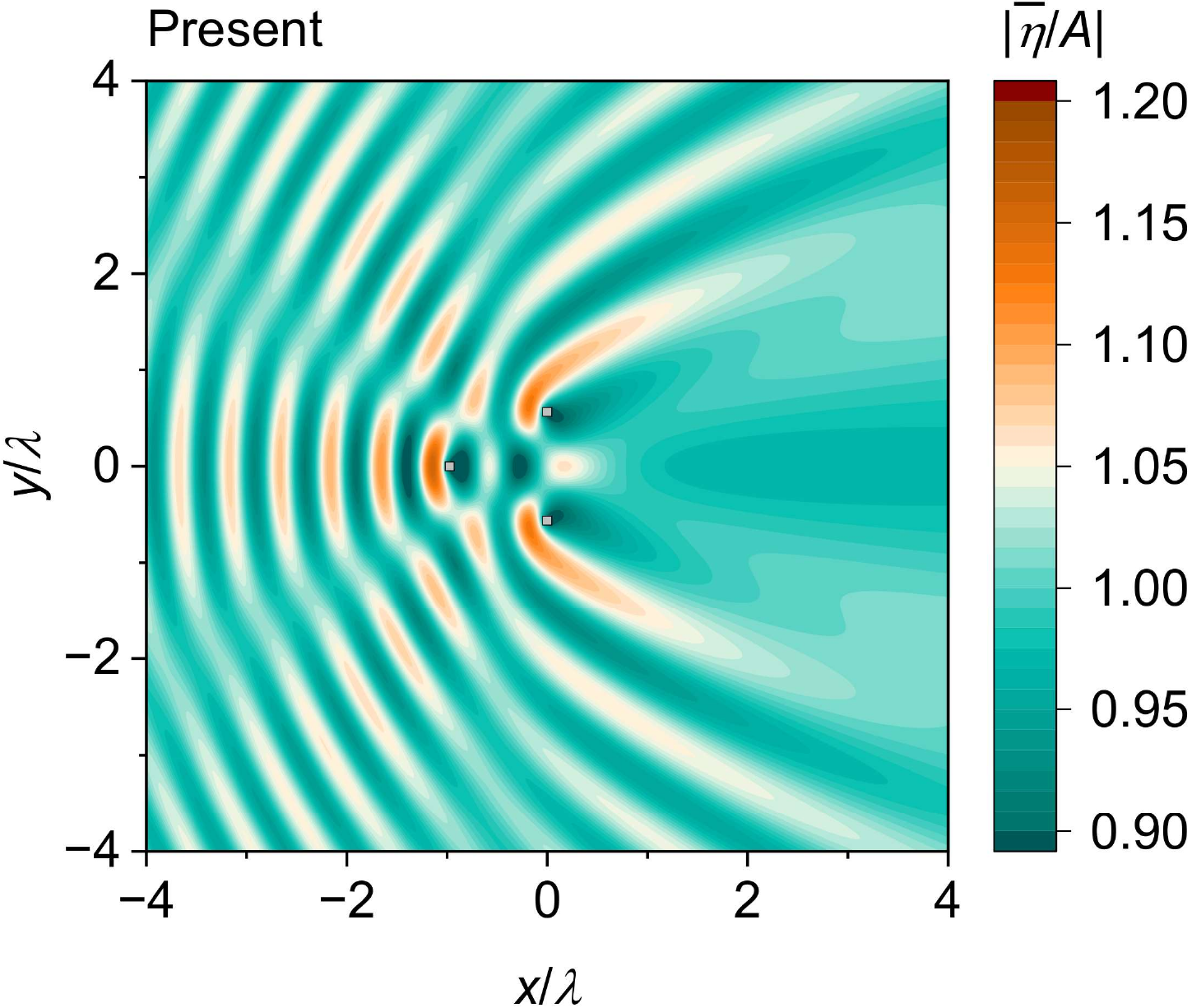}
  \subcaption{}
  \end{minipage}
}
\caption{
Amplitude distribution of wave field around three identical ice floes of Model-II with circumcircle distance of (a) $d=0$, (b) $d=0.4\lambda$, (c) $d=0.6\lambda$, and (d) $d=1.0\lambda$ in long waves.}
\label{fig:distance_effect_3D}
\end{figure}

\begin{figure}[!htp]
\centering
{
  \begin{minipage}{0.35\linewidth}
  \centering
  \includegraphics[width=1.0\linewidth]{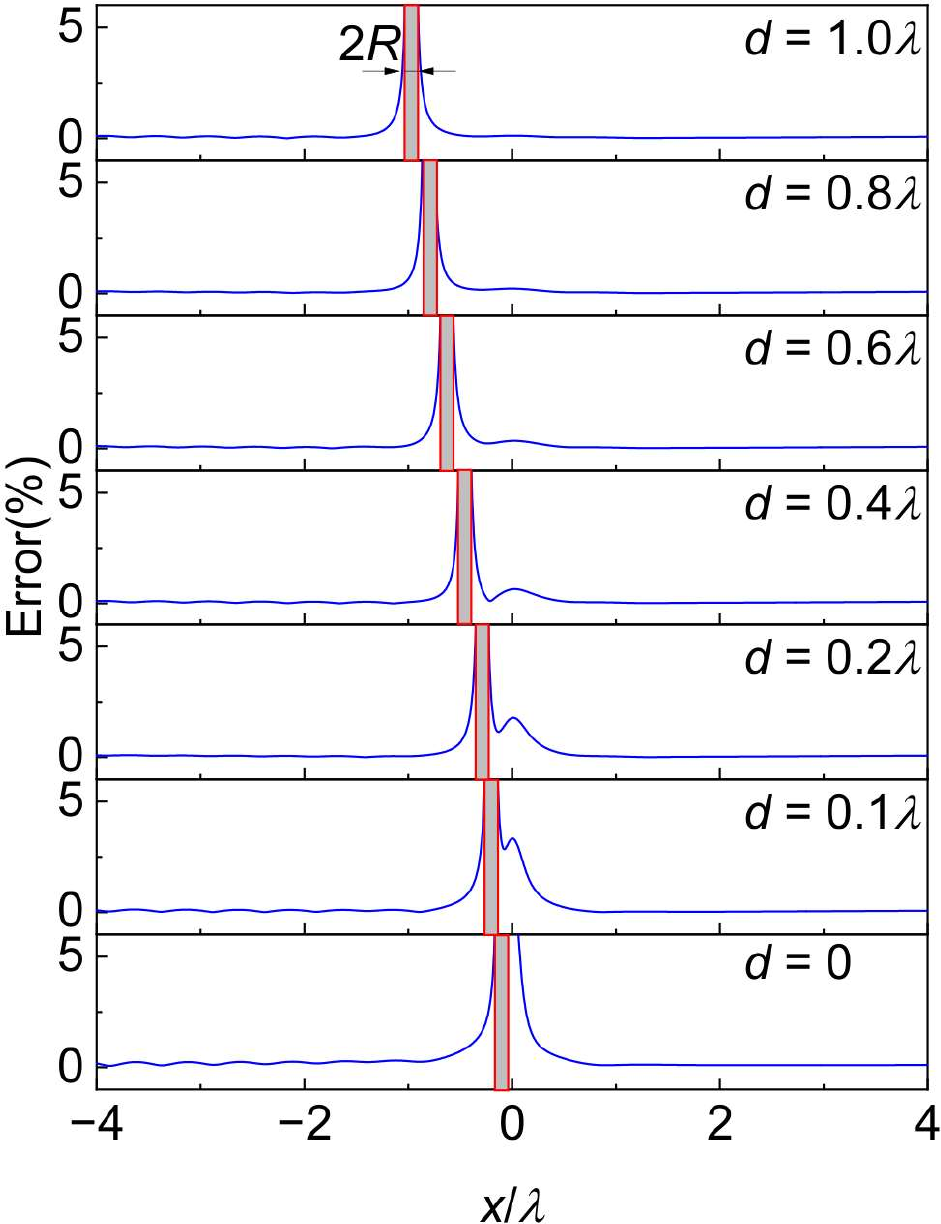}
  \subcaption{}
  \end{minipage}
}
{
  \begin{minipage}{0.35\linewidth}
  \centering
  \includegraphics[width=1.0\linewidth]{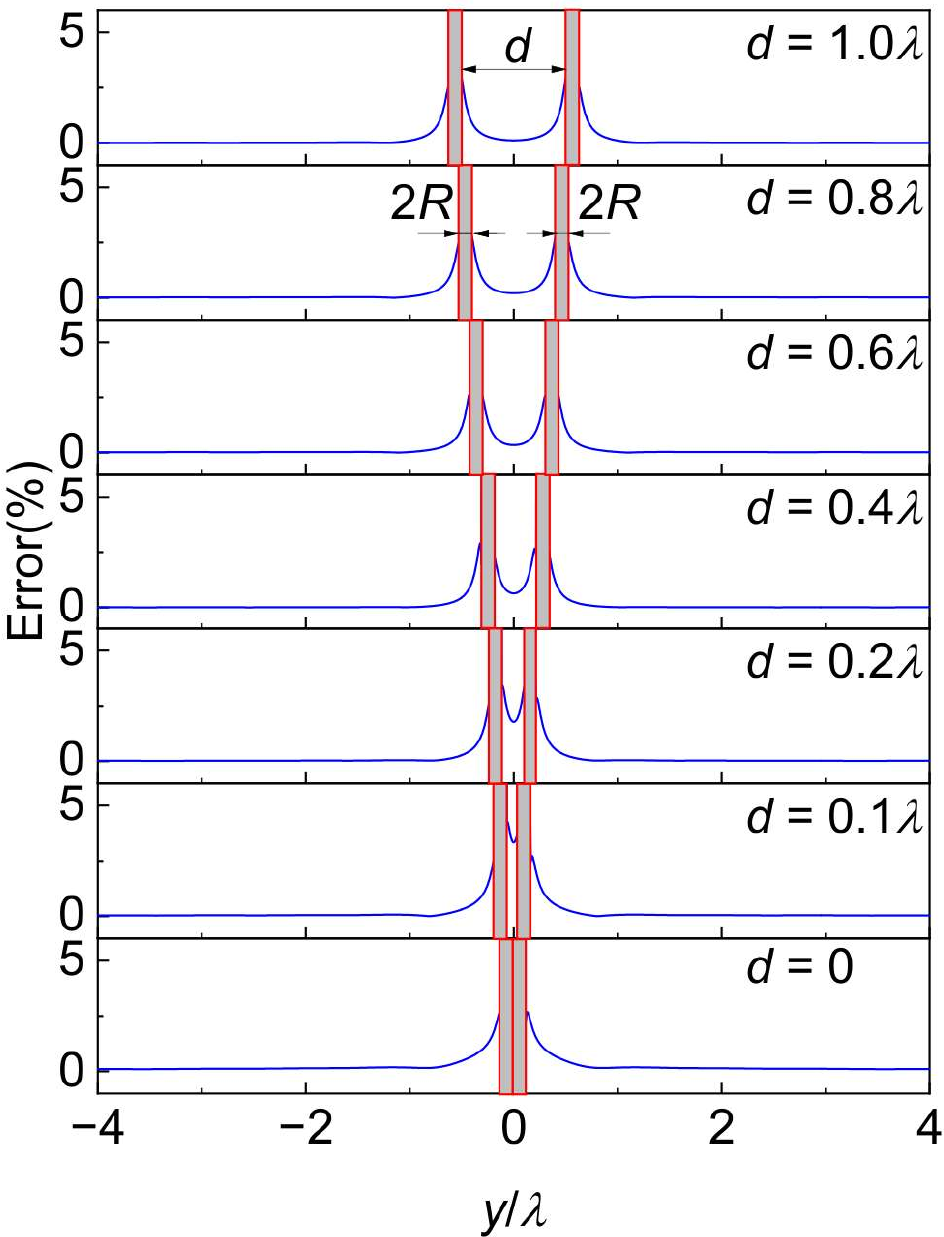}
  \subcaption{}
  \end{minipage}
}
\caption{
Distribution of relative error along (a) $y=0$ and (b) $x=0$, for three identical ice floes of Model-II with circumcircle distance $d$ ranging from 0 to $1.0\lambda$ in long waves.}
\label{fig:distance_error_3D}
\end{figure}

Fig. \ref{fig:distance_effect_3D} shows the amplitude distribution of the wave field around three identical ice floes of Model-II in long waves. 
The circumcircle distances are $d = 0$, $0.4\lambda$, $0.6\lambda$, and $1.0\lambda$, respectively.
The irregularity of the wave amplitude distribution increases as the circumcircle distance increases.
The relative error of the present EI solution is obtained by comparing it with the BEM result.
Fig. \ref{fig:distance_error_3D} specifically compares the distributions of the relative error along $y = 0$ and $x = 0$, with the circumcircle distance $d$ increasing from 0 to $1.0\lambda$.
Except for the case of $d = 0$, the relative error of the EI method is always less than $5\%$ at a location outside the circumcircle of each ice floe.

\begin{figure}[!htp]
\centering
{
  \begin{minipage}{0.32\linewidth}
  \centering
  \includegraphics[width=1.0\linewidth]{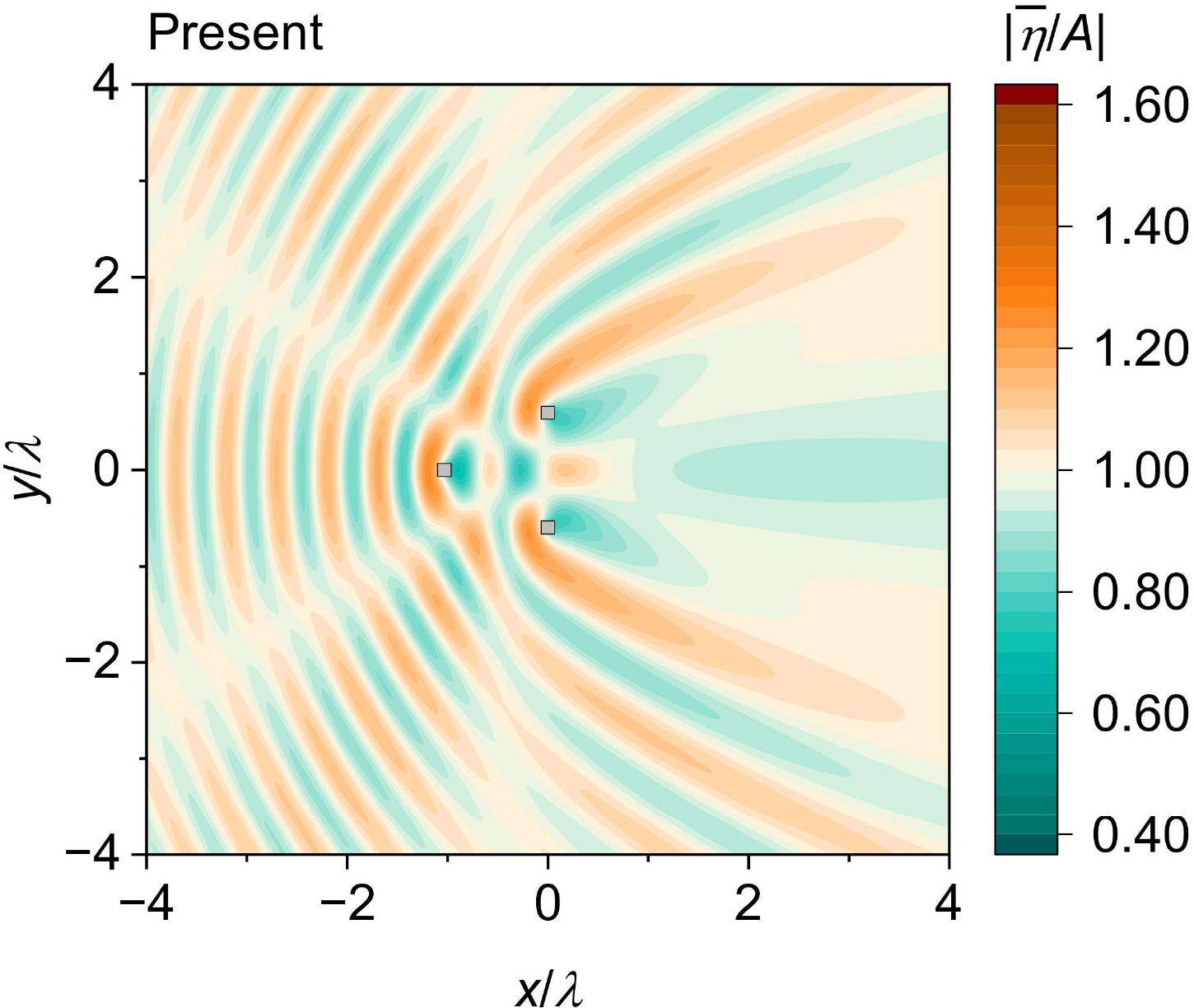}
  \subcaption{}
  \end{minipage}
}
{
  \begin{minipage}{0.32\linewidth}
  \centering
  \includegraphics[width=1.0\linewidth]{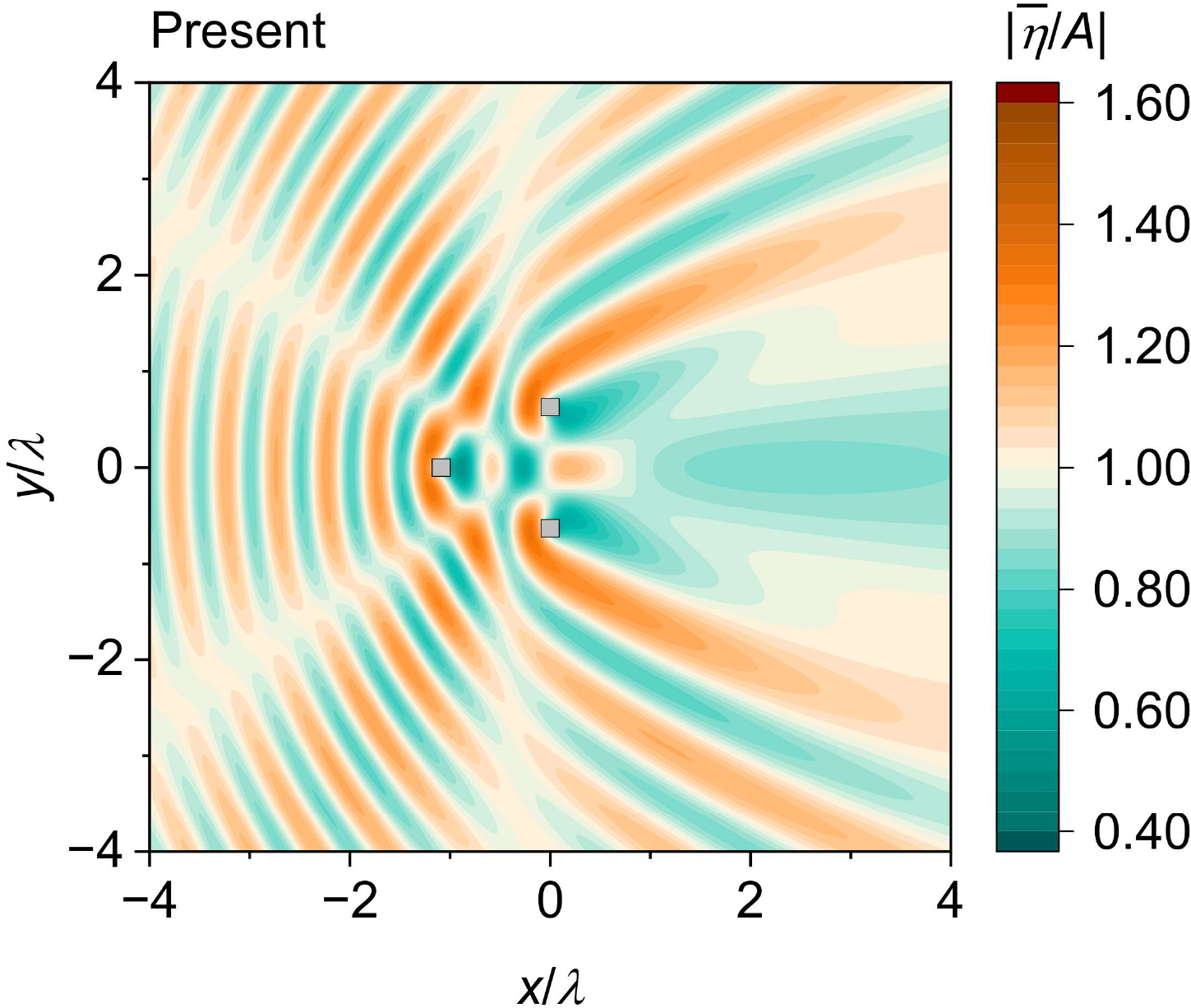}
  \subcaption{}
  \end{minipage}
}
{
  \begin{minipage}{0.32\linewidth}
  \centering
  \includegraphics[width=1.0\linewidth]{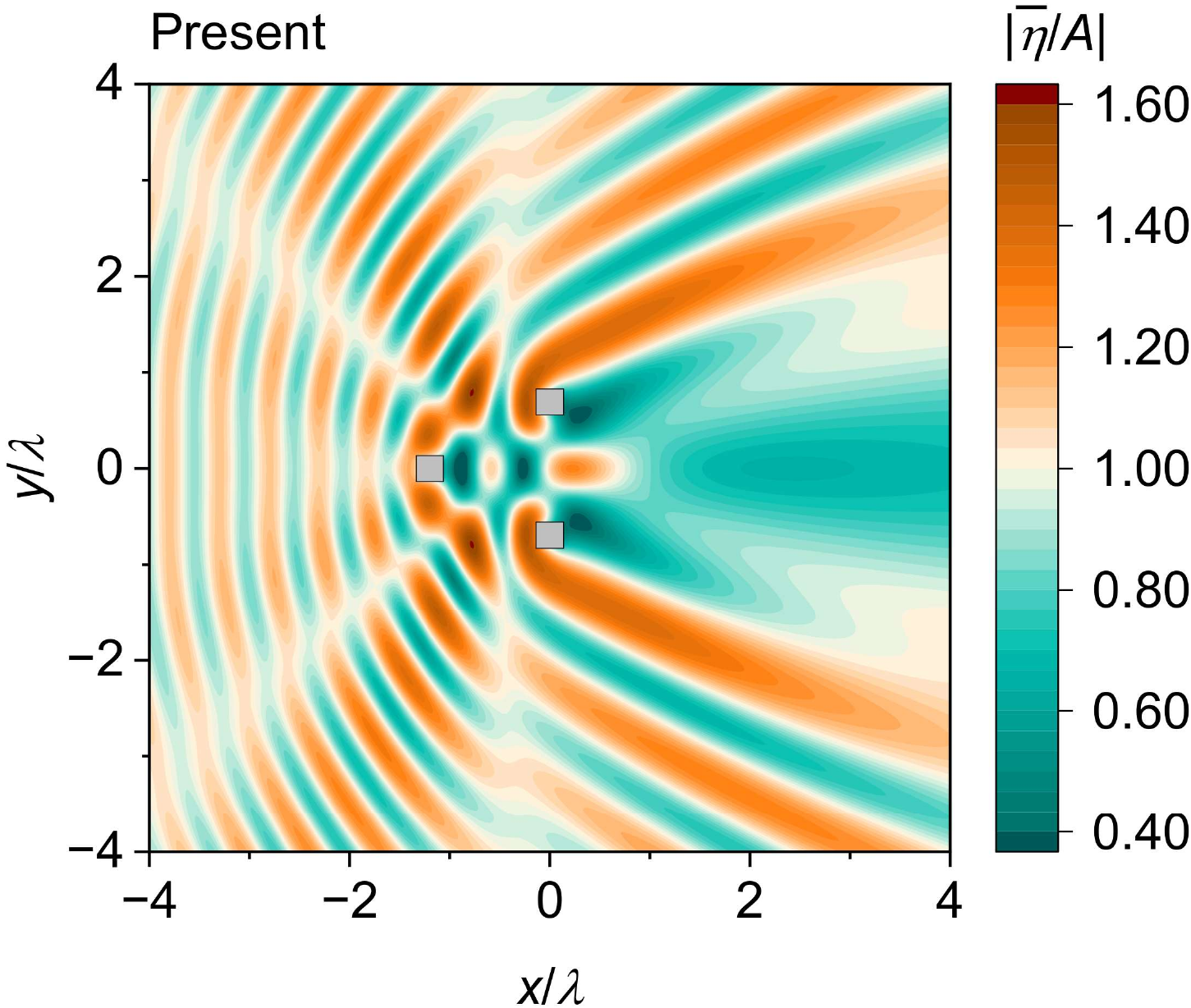}
  \subcaption{}
  \end{minipage}
}
\caption{
Amplitude distribution of wave field around three identical ice floes of Model-II with size $R_1=R_2=R_3=$ (a) $1.5R$, (b) $2.0R$, and (c) $3.0R$ in long waves.}
\label{fig:size_effect_3D}
\end{figure}

The effects of ice-floe size on the wave field around three identical ice floes are finally examined.
The size of each ice floe ranges from $R_1 = R_2 = R_3 = 1.0R$ to $3.0R$.
Fig. \ref{fig:size_effect_3D} shows the amplitude distribution of the wave field around three identical ice floes of Model-II in long waves.
Three sizes of the ice floes are considered, and the circumcircle distance is $d = 1.0\lambda$.
The general patterns of these amplitude distributions are similar. 
However, as the size of each ice floe increases, the difference between the maximum and minimum wave amplitudes is strengthened.
Fig. \ref{fig:size_error} shows the distribution of relative error along $y=0$ and $x=0$.
It still holds that the relative error of wave amplitude predicted by the EI method is always less than $5\%$ at a location outside the circumcircle of each ice floe.
\begin{figure}[!htp]
\centering
{
  \begin{minipage}{0.35\linewidth}
  \centering
  \includegraphics[width=1.0\linewidth]{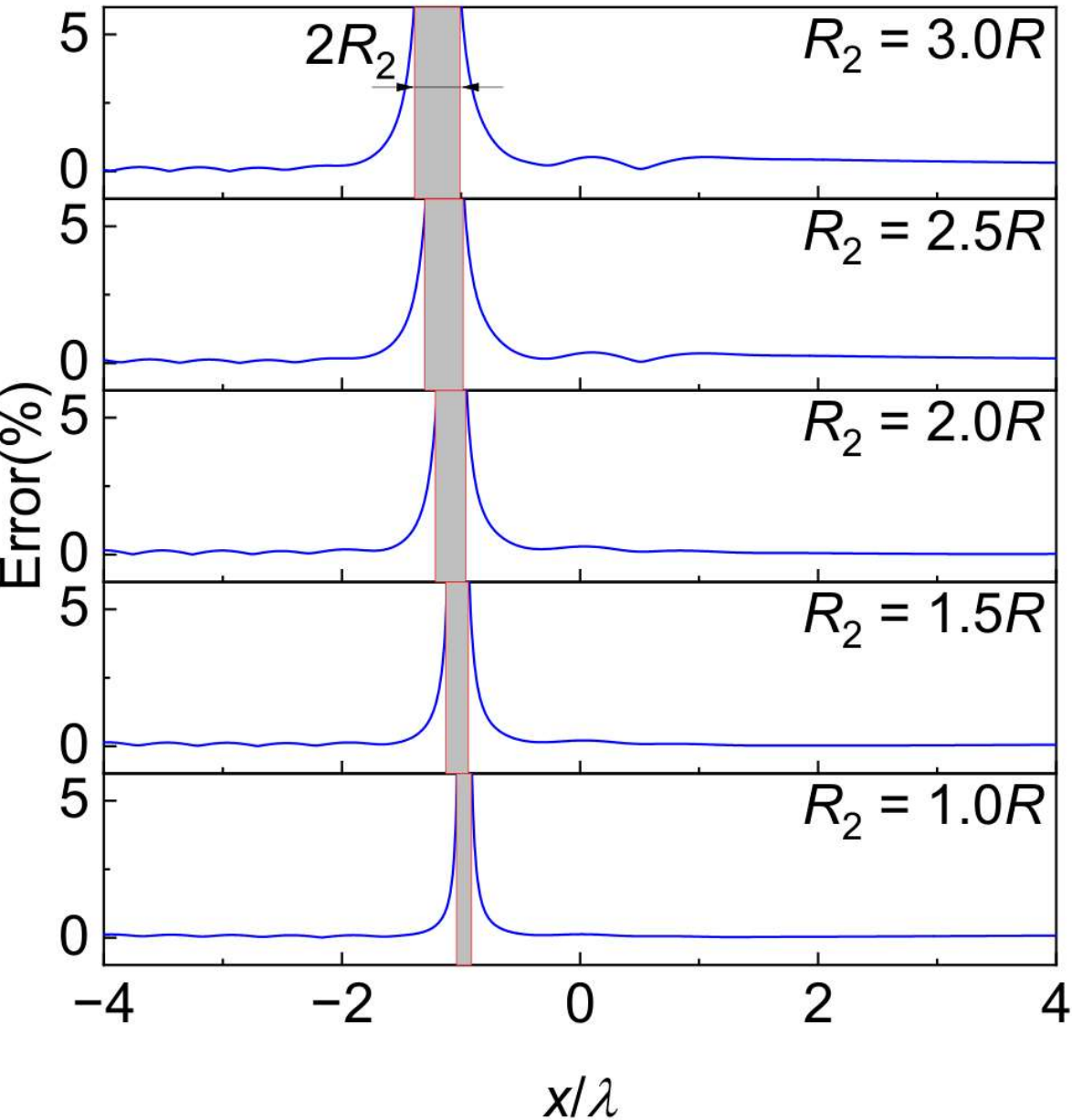}
  \subcaption{}
  \end{minipage}
}
{
  \begin{minipage}{0.35\linewidth}
  \centering
  \includegraphics[width=1.0\linewidth]{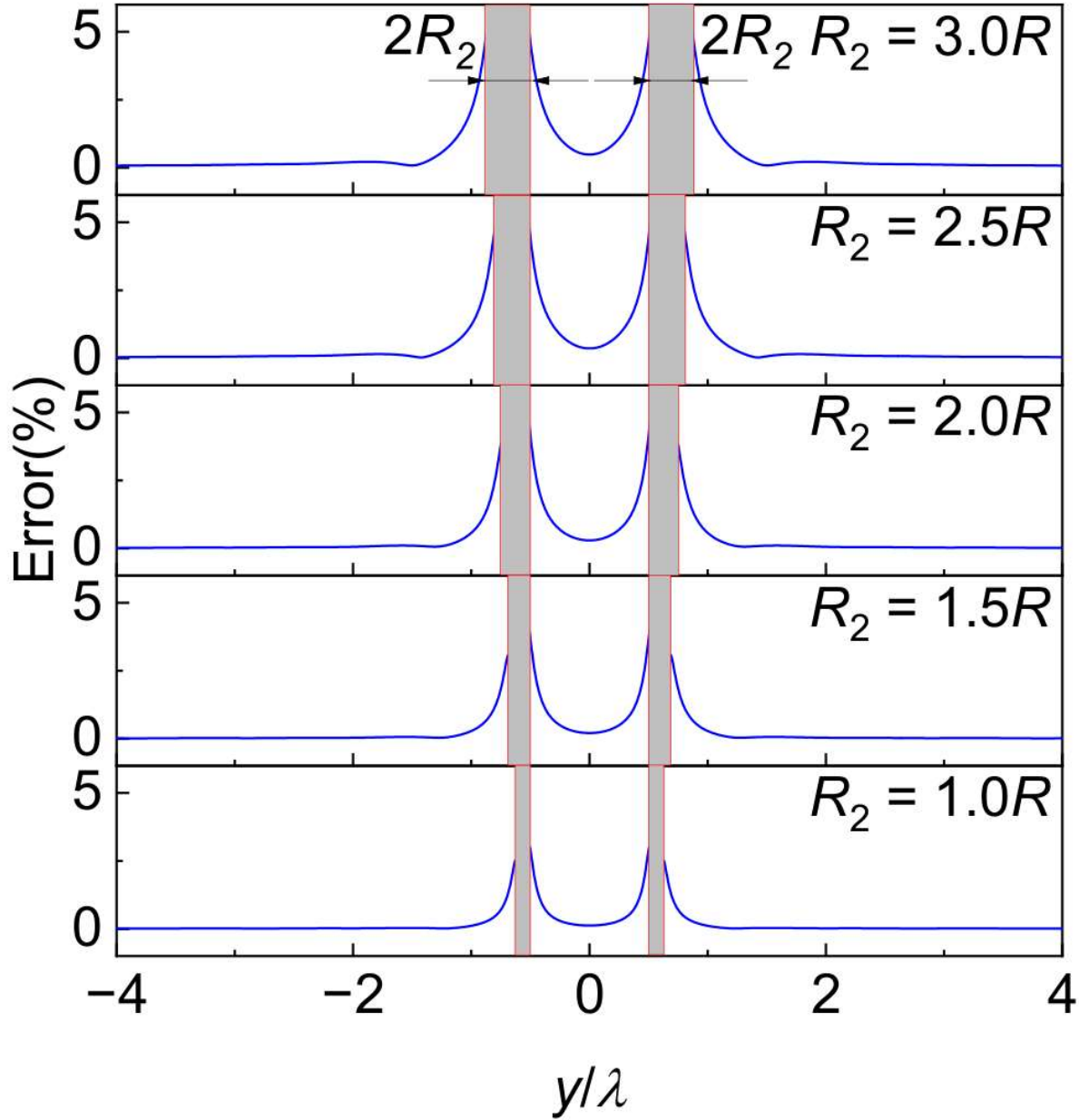}
  \subcaption{}
  \end{minipage}
}
\caption{
Distribution of relative error along (a) $y=0$ and (b) $x=0$, for three identical ices floe of Model-II with size $R_1=R_2=R_3$ ranging from $1.0R$ to $3.0R$ in long waves.
}
\label{fig:size_error}
\end{figure}

\subsection{Efficiency analysis of wave field influenced by ultra-large group of ice floes}

\begin{figure}[!htp]
\centering
{
  \begin{minipage}{0.7\linewidth}
  \centering
  \includegraphics[width=1.0\linewidth]{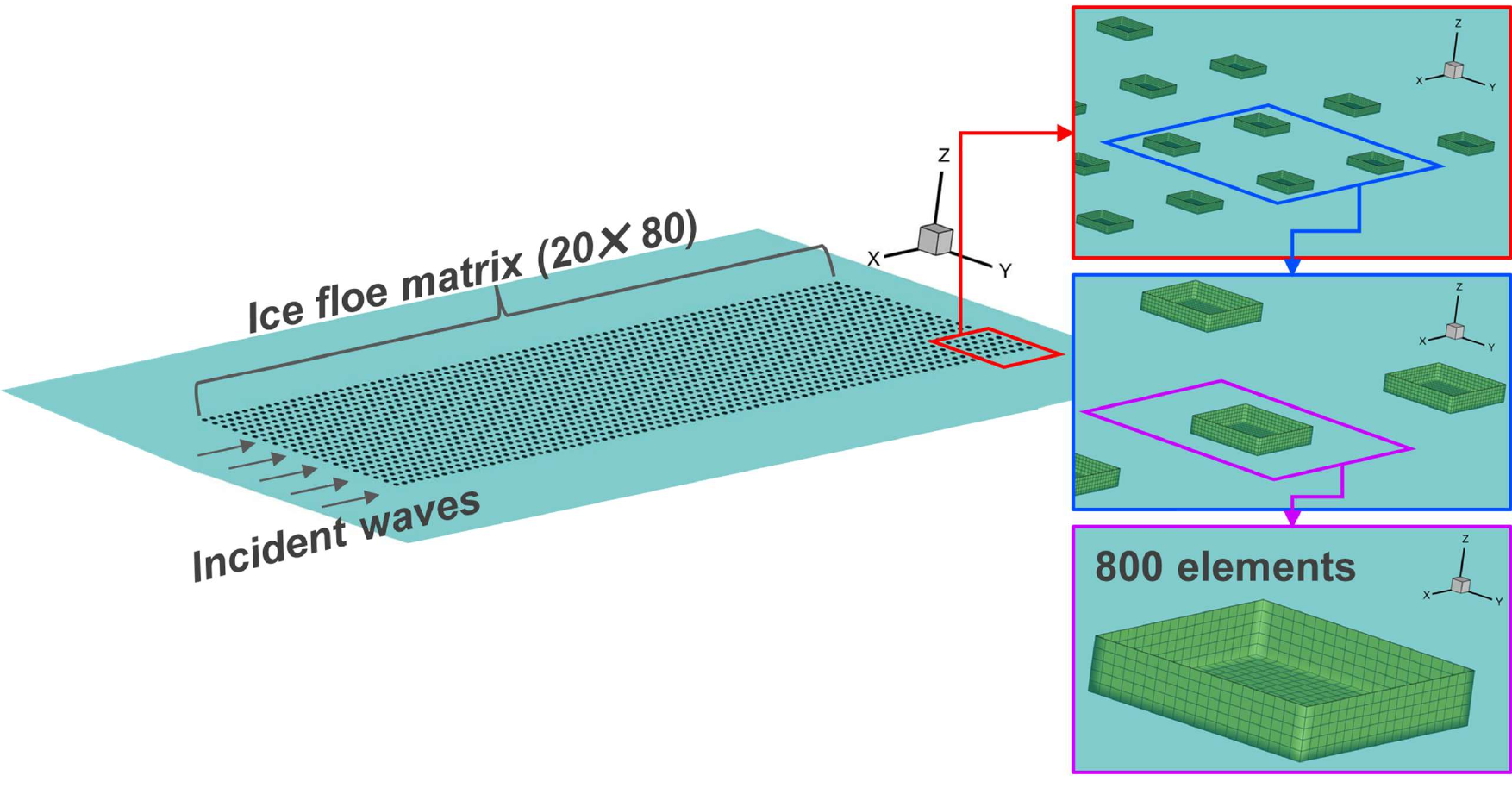}
  \end{minipage}
}
\caption{
Boundary elements on wet surface of $20\times 80$ ice floes of Model-II.}
\label{fig:efficiency_mesh}
\end{figure}

Subsection \ref{subsec:exam_EI} has tested the effectiveness of the present EI method in different multiple-ice-floe scenarios. 
This subsection further examines the efficiency of the EI method.
Here, an ultra-large group of ice floes is given as an example, as shown in Fig. \ref{fig:efficiency_mesh}.
A $20\times80$ matrix of ice floes of Model-II is formed. 
Each ice floe has 800 boundary elements on its mean wet surface.
For these 1600 ice floes, the total number of boundary elements for computation is 1,280,000.
In practice, such a computation load is beyond the capacity of any standard BEM solver on a personal computer.
Standard BEM solvers are usually limited to tens of thousands of boundary elements, because the system of linear equations derived from BEM has a fully-populated (non-sparse) coefficient matrix.
As the number of boundary elements increases, the computational costs of BEM grow in the order of the cube, and the memory grows in the order of the square.
However, computing the wave field around this ultra-large group of ice floes is not challenging for the present EI method.

\begin{figure}[!htp]
\centering
{
  \begin{minipage}{0.32\linewidth}
  \centering
  \includegraphics[width=1.0\linewidth]{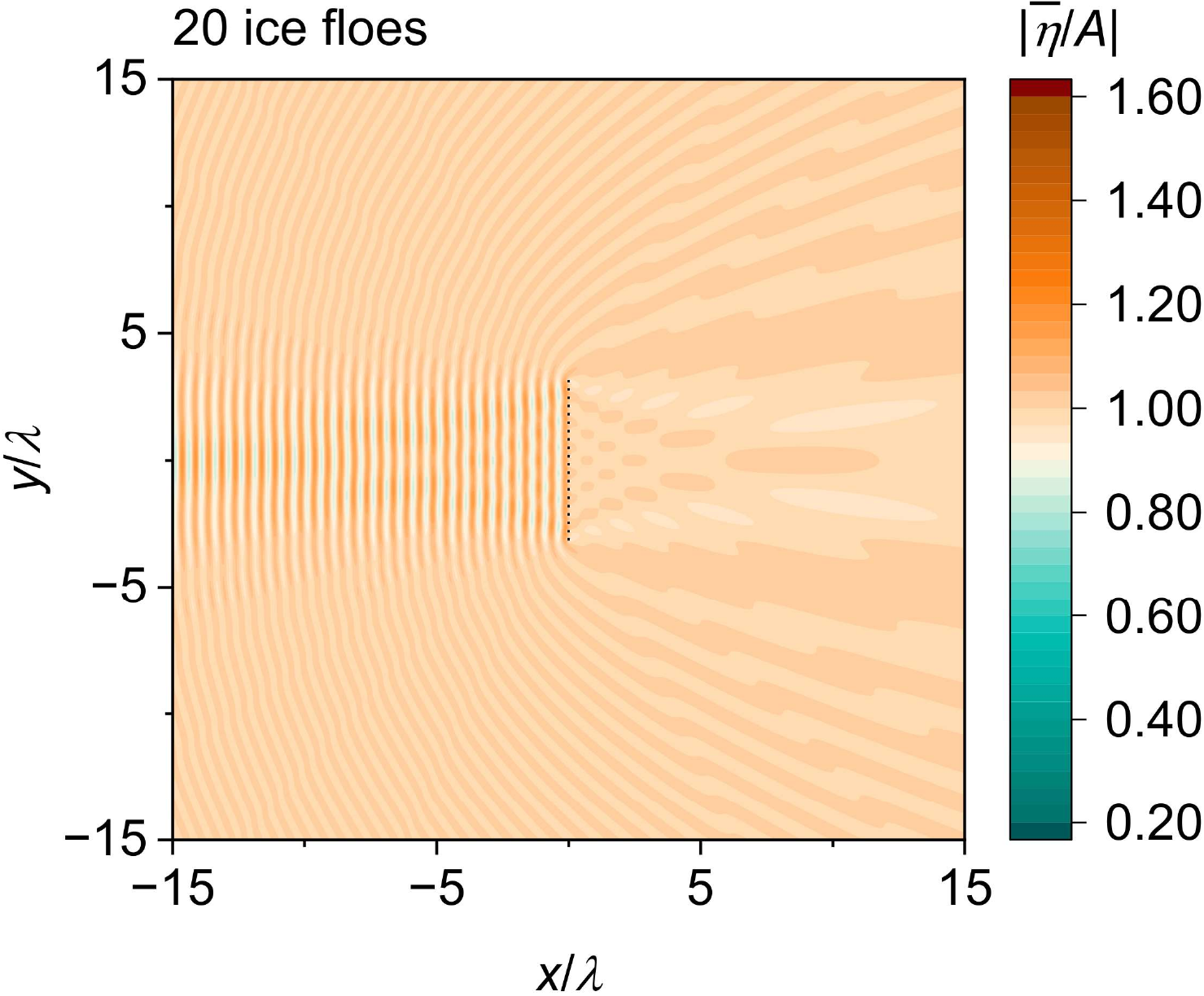}
  \subcaption{}
  \end{minipage}
}
{
  \begin{minipage}{0.32\linewidth}
  \centering
  \includegraphics[width=1.0\linewidth]{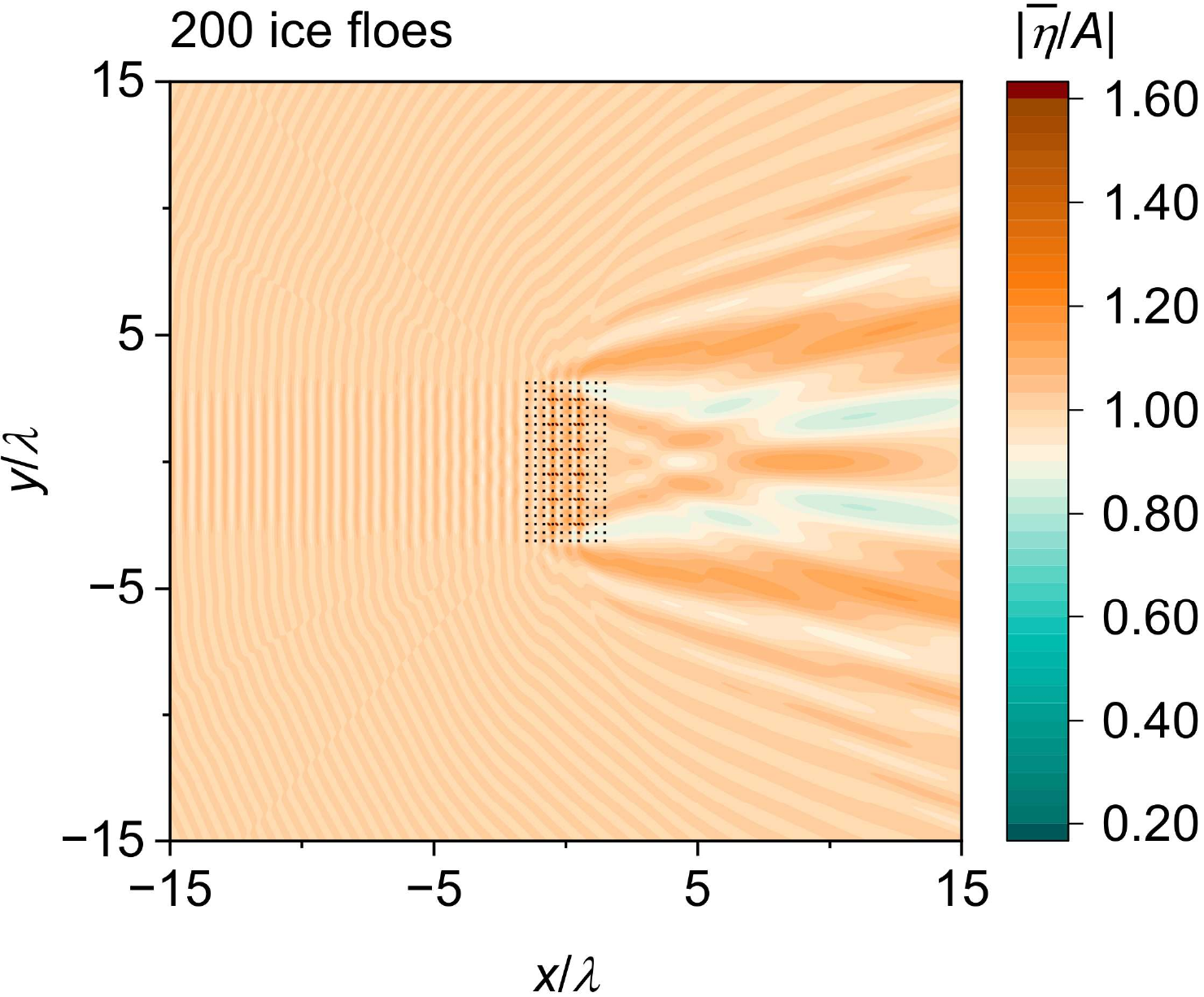}
  \subcaption{}
  \end{minipage}
}
{
  \begin{minipage}{0.32\linewidth}
  \centering
  \includegraphics[width=1.0\linewidth]{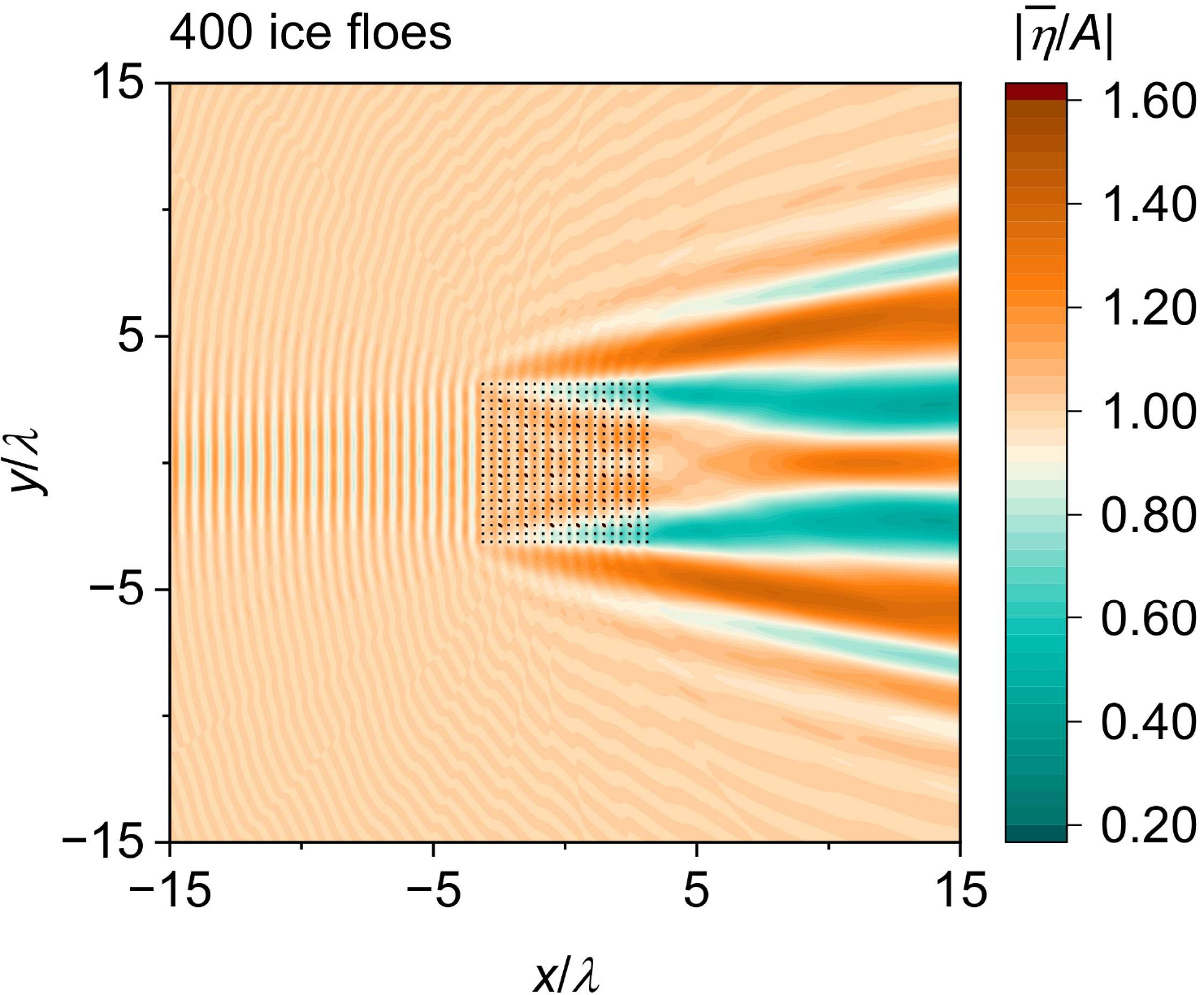}
  \subcaption{}
  \end{minipage}
}
{
  \begin{minipage}{0.32\linewidth}
  \centering
  \includegraphics[width=1.0\linewidth]{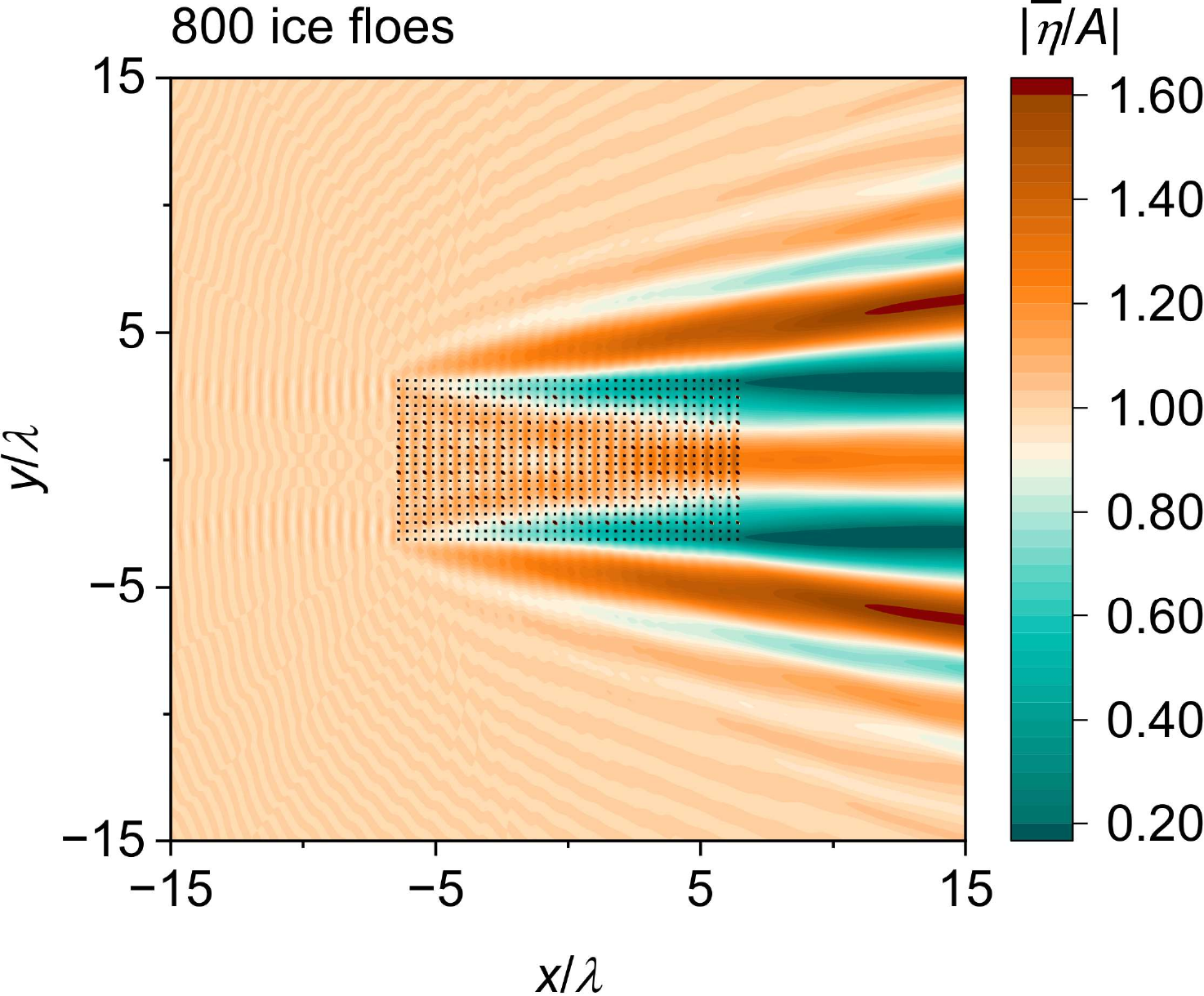}
  \subcaption{}
  \end{minipage}
}
{
  \begin{minipage}{0.32\linewidth}
  \centering
  \includegraphics[width=1.0\linewidth]{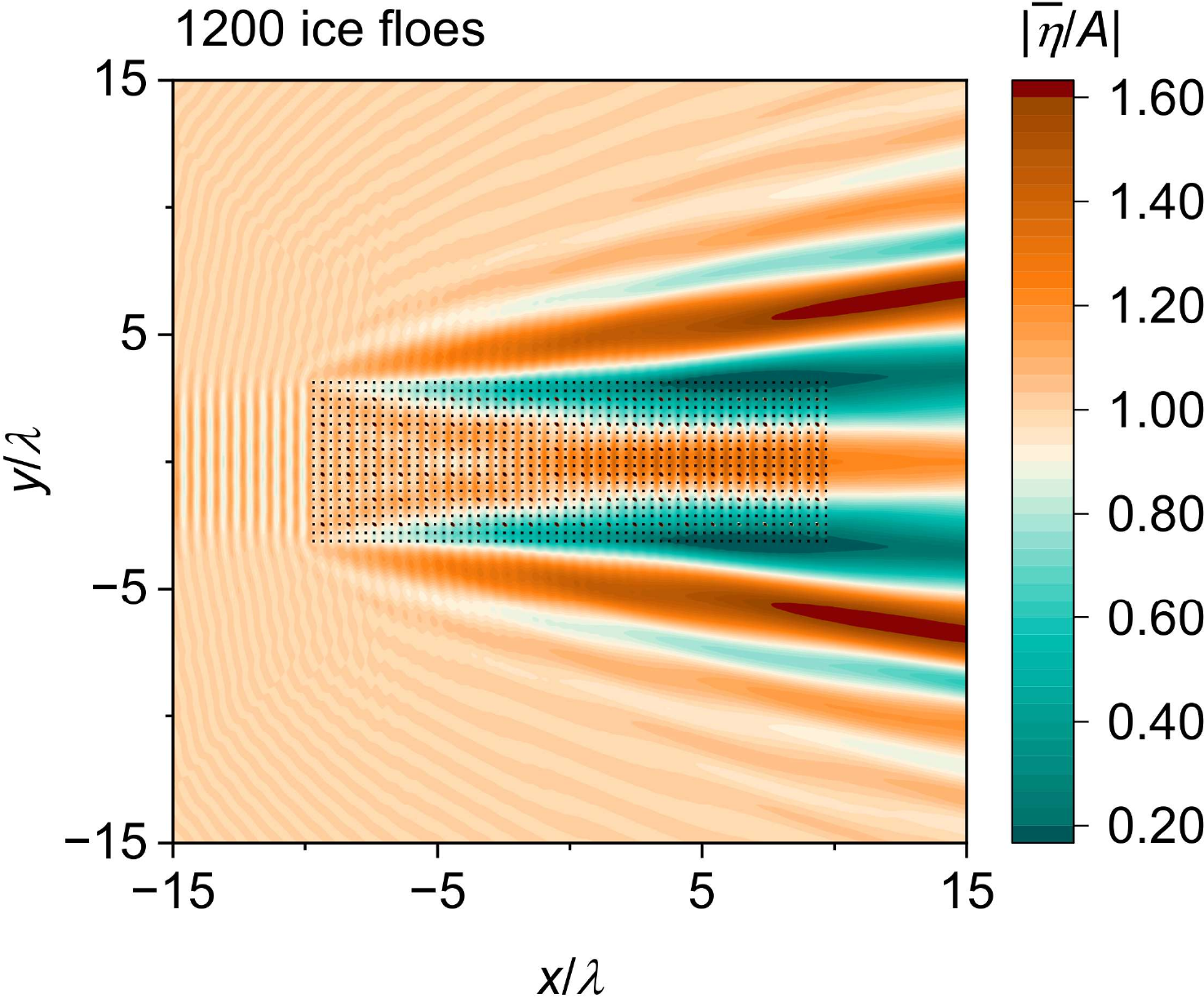}
  \subcaption{}
  \end{minipage}
}
{
  \begin{minipage}{0.32\linewidth}
  \centering
  \includegraphics[width=1.0\linewidth]{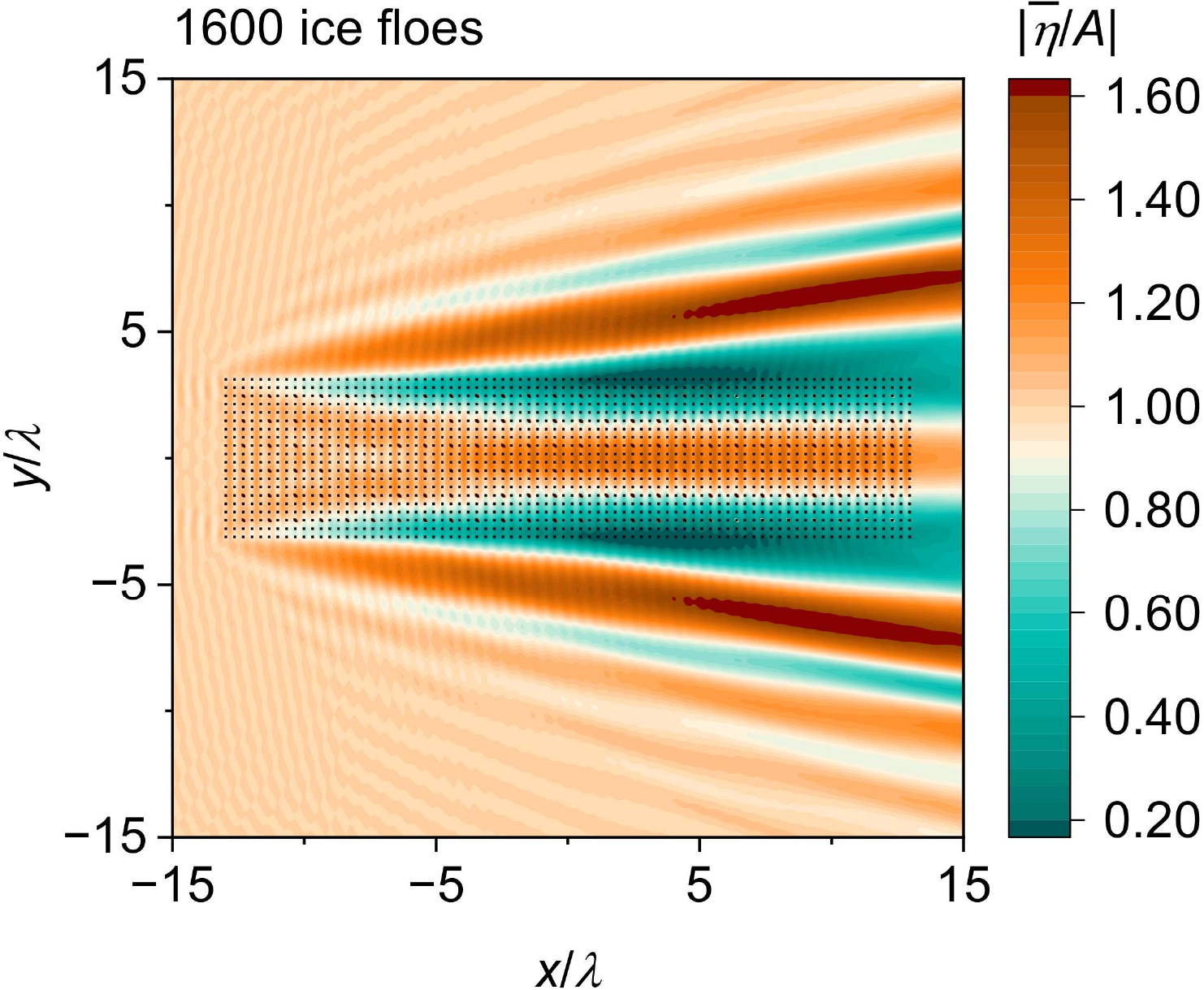}
  \subcaption{}
  \end{minipage}
}
\caption{
Distribution of wave amplitudes around identical ice floes of Model-II in a matrix of (a) $20\times 1$, (b) $20\times 10$, (c) $20\times 20$, (d) $20\times 40$, (e) $20\times 60$, and (f) $20\times 80$ in long waves.}
\label{fig:efficiency_field}
\end{figure}

Fig. \ref{fig:efficiency_field} shows the wave-amplitude distributions around identical Model-II ice floes in matrices of $20\times 1$, $20\times 10$, $20\times 20$, $20\times 40$, $20\times 60$, and $20\times 80$.
The direction of ambient incident waves meets $\beta = 0$, and the long-wave condition is considered.
The circumcircle distance between adjacent ice floes is $d = 0.2\lambda$ in both the $x$ and $y$ directions.
The wave amplitudes at $400\times400$ locations in the wave field are calculated to generate the contour of the wave-amplitude distribution.
As the number of columns in the ice-floe matrix increases from 1 to 80, the wake behind the ice floes becomes evident.
The maximum wave amplitude can be found at ``shoulder regions'' of the ice-floe matrix.
The wave amplitude distributions in these cases are graphically reasonable and have sufficient details.

\begin{figure}[!htp]
\centering
{
  \begin{minipage}{0.5\linewidth}
  \centering
  \includegraphics[width=1.0\linewidth]{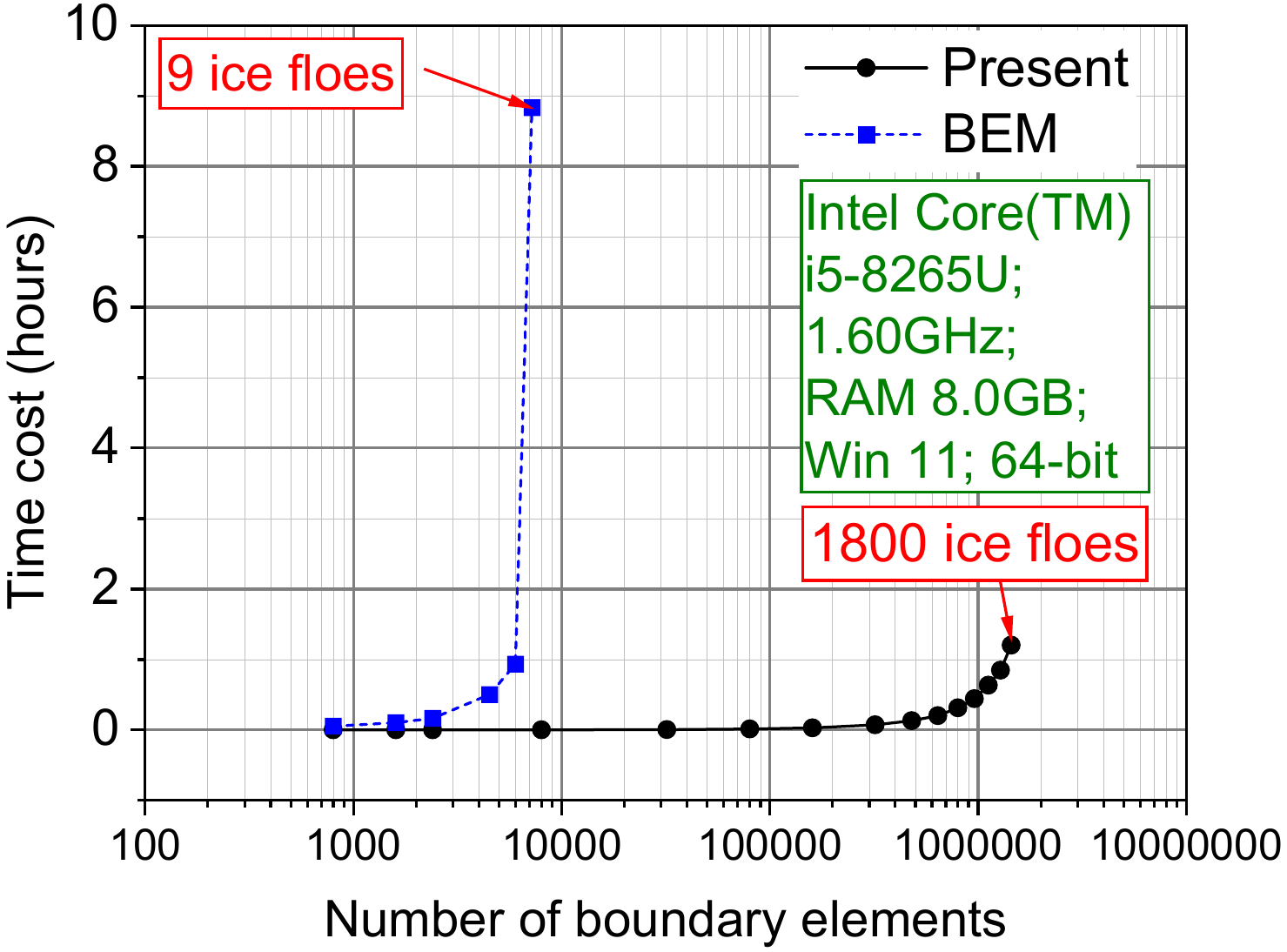}
  \end{minipage}
}
\caption{
Efficiency comparison between present enhanced interaction method and BEM.}
\label{fig:efficiency}
\end{figure}

Fig. \ref{fig:efficiency} compares the time costs of the EI method and BEM for different numbers of boundary elements.
The calculations are done on an ordinary personal computer with an Intel Core (TM) i5-8265U processor (released in 2017) and 8.0 GB of RAM.
The operating system is 64-bit Windows 11.
The time cost refers to the computational time needed to calculate the wave amplitudes at 160,000 locations in the wave field.
An 8-thread OpenMP parallel acceleration procedure is implemented in both the BEM and the present EI methods.
Even with OpenMP acceleration enabled, the BEM takes nine hours to predict the wave field around 9 ice floes (7200 elements).
The time and memory costs of BEM increase sharply with the number of boundary elements, and further increasing the boundary elements will exceed the memory capacity of current personal computers.
Meanwhile, with the present EI method, it takes less than 1.5 hours to handle 1800 ice floes that have 1,440,000 boundary elements.
The super-high efficiency of the present EI method for the wave field around an ultra-large group of ice floes is evident.

\subsection{Optimized path planning for navigation through wave field of ice floes}

\begin{figure}[!htp]
\centering
{
  \begin{minipage}{0.5\linewidth}
  \centering
  \includegraphics[width=1.0\linewidth]{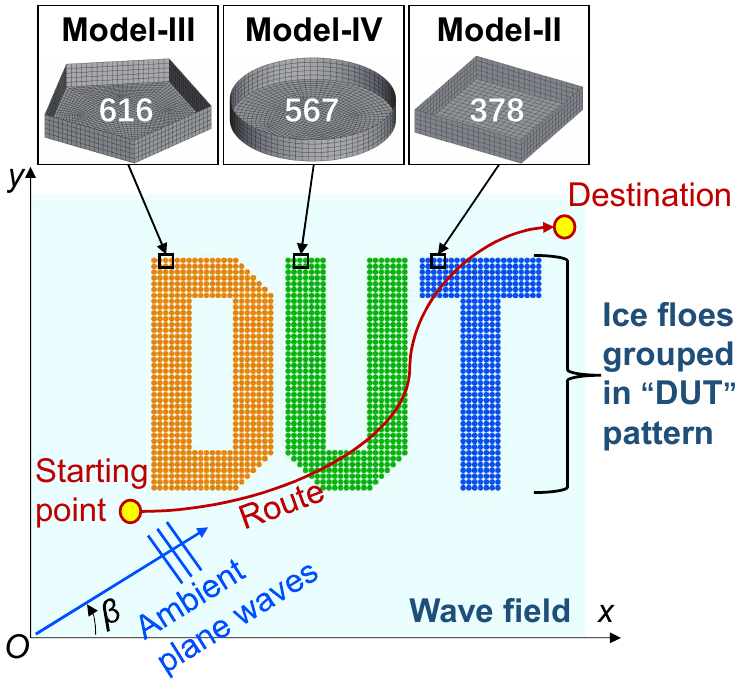}
  \end{minipage}
}
\caption{
Deployment of 1561 ice floes of Model-II, III and IV in ``DUT'' pattern}
\label{fig:DUT_shape}
\end{figure}

Using the present EI method, the wave field around a large group of arbitrary-shaped ice floes can be obtained.
Based on the knowledge of the wave-amplitude distribution map, various path-planning strategies can be used to recommend an optimized route for ship sailing.
This subsection demonstrates an optimized path planning practice for navigation routes through ice floes, echoing the background of the Arctic route planning system in Fig. \ref{fig:arctic}.
Fig. \ref{fig:DUT_shape} shows an example of an ice-floe field.
There are 1561 ice floes of mixed Model-II, III, and IV, grouped in a “DUT” pattern.
Specifically, 616 ice floes of Model-III form the letter ``D'', 567 ice floes of Model-IV form the letter ``U'', and 378 Model - II ice floes form the letter ``T''.
For each ``letter'', the circumcircle distance between adjacent ice floes in the $x$ or $y$ direction is $2R$.
Different incidence angles of the ambient plane waves are considered.
The ship starts near the lower left corner of the letter ``D''. 
Its destination is set near the upper right corner of the letter ``T''.
The purpose of path planning is to find an optimized route for the ship. 
On this route, the ship experiences the minimum average wave amplitude.

\begin{figure}[!htp]
\centering
{
  \begin{minipage}{0.32\linewidth}
  \centering
  \includegraphics[width=1.0\linewidth]{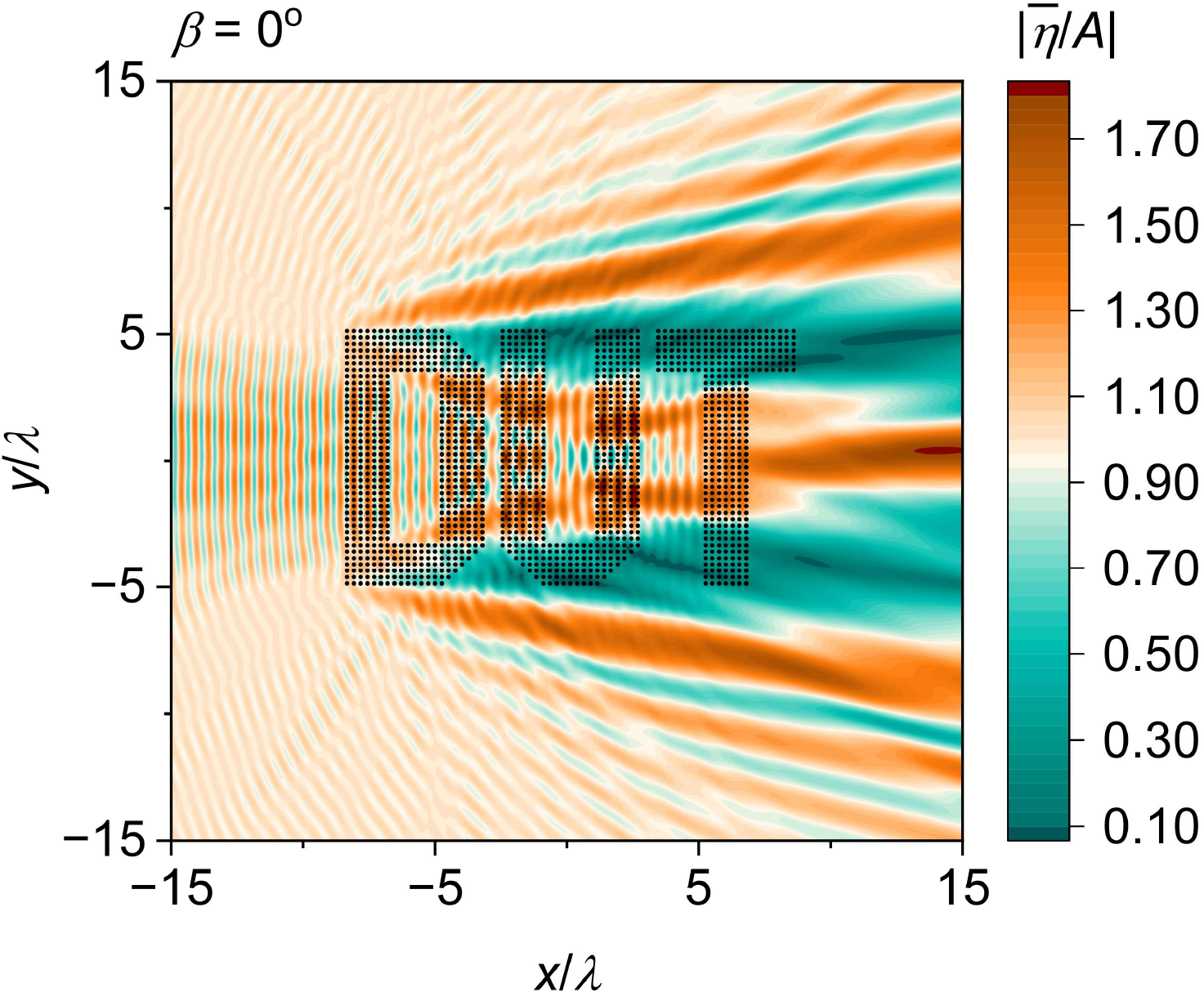}
  \subcaption{}
  \end{minipage}
}
{
  \begin{minipage}{0.32\linewidth}
  \centering
  \includegraphics[width=1.0\linewidth]{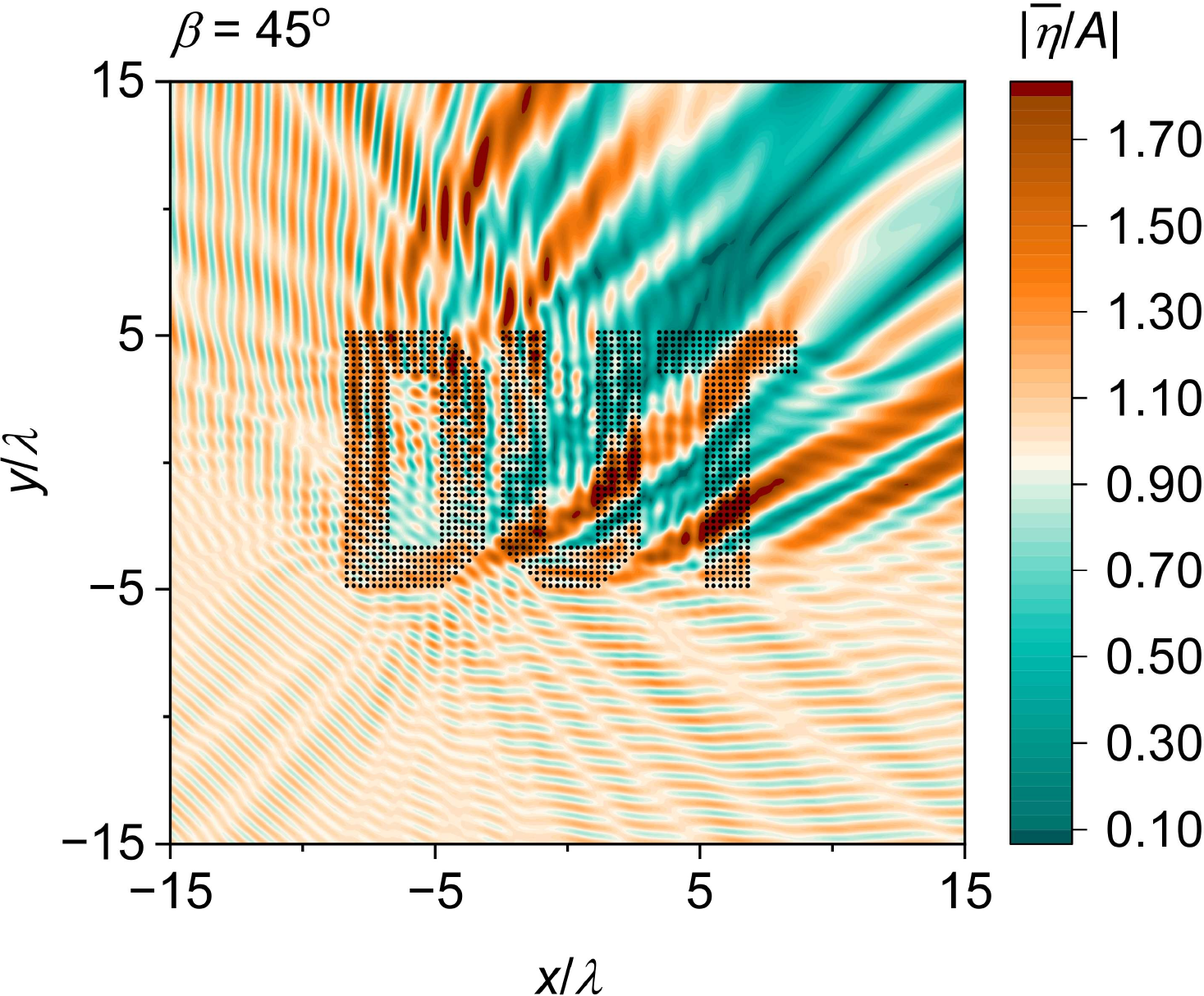}
  \subcaption{}
  \end{minipage}
}
{
  \begin{minipage}{0.32\linewidth}
  \centering
  \includegraphics[width=1.0\linewidth]{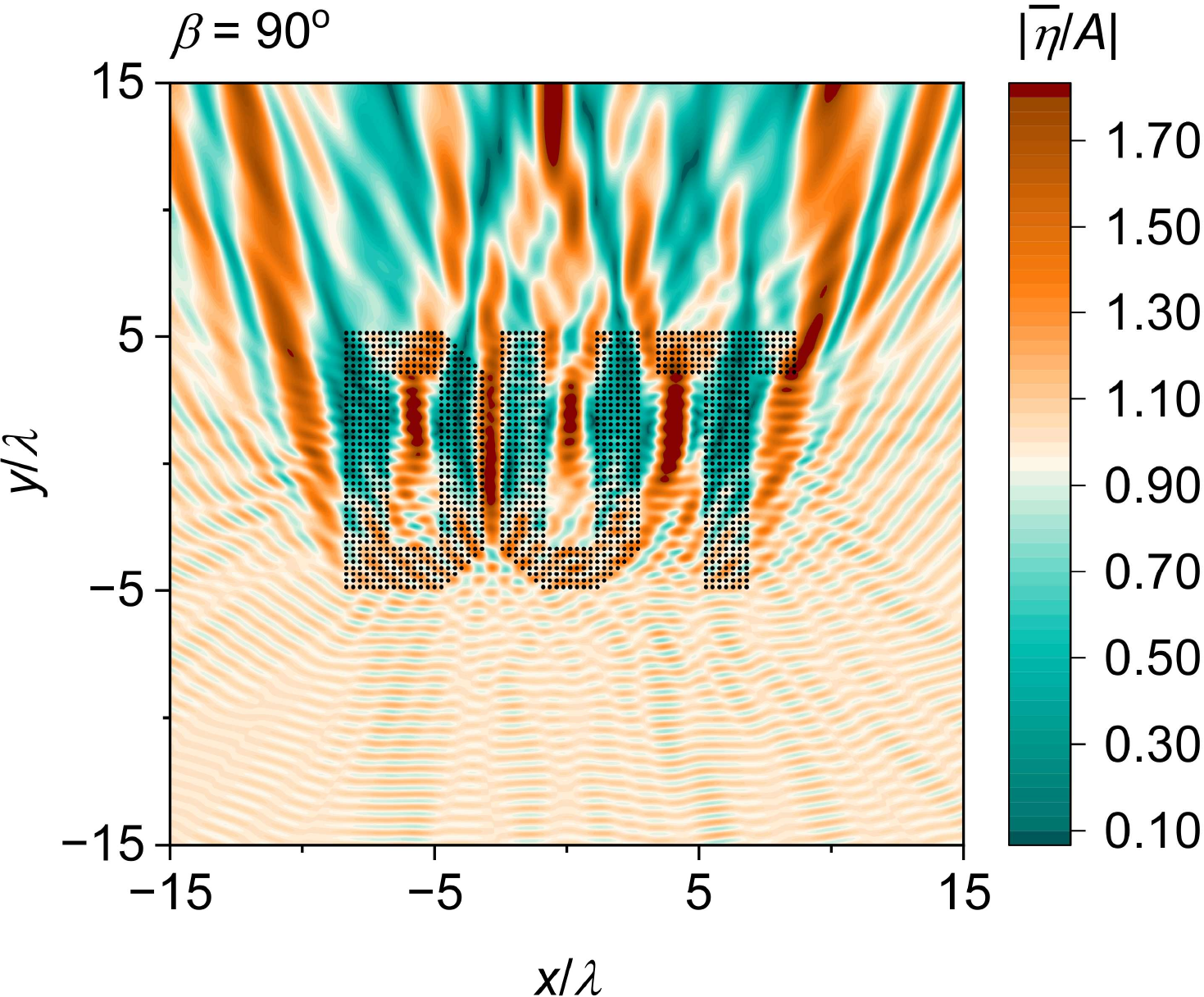}
  \subcaption{}
  \end{minipage}
}
\caption{
Distribution of wave amplitudes around ice floes grouped in ``DUT'' pattern in long waves with incidence angle of (a) $\beta=0^{\rm{o}}$, (b) $\beta=45^{\rm{o}}$, and (c) $\beta=90^{\rm{o}}$.}
\label{fig:DUT_direction}
\end{figure}

Fig. \ref{fig:DUT_direction} shows the distribution of wave amplitudes around ice floes grouped in the ``DUT'' pattern.
The ambient plane waves are long waves with the incidence angles $\beta = 0^{\circ}$, $45^{\circ}$, and $90^{\circ}$.
The amplitude of the upstream wave is evidently disturbed by the scattering waves of the ice floes.
In the downstream direction of the ice floes, several large areas of the free surface have a sufficiently small wave amplitude. 
This is due to the wave-sheltering effect of the ice floes.
It seems possible to make the route mostly cross the small-amplitude area.

Here, a path planning strategy based on dynamic programming is used.
Figs. \ref{fig:DUT_direction} (a), (b), and (c) are three wave-amplitude maps.
A rectangular zone on each wave-amplitude map is formed by the starting point and destination of the ship route.
The rectangular zone is discretized into $m \times n$ square units.
The square unit in the $i$th row and $j$th column is indexed as $(i,j)$.
As a route constraint, at each step, the ship must move either one unit to the right or one unit up.
Therefore, the ship needs to move $L = m + n - 1$ steps from the starting point to the destination.
An $m \times n$ matrix $[W]$ is defined.
The element in the $i$-th row and $j$-th column of $[W]$, denoted by $w_{i,j}$, is the average wave amplitude in the square unit $(i,j)$.
An additional $m \times n$ state matrix $[P]$ is defined. 
The $(i,j)$ element of $[P]$ is the minimum cumulative wave amplitude from the starting point to the unit $(i,j)$.
The first element of $[P]$ is initialized as $p_{1,1}=w_{1,1}$.
Elements in the first row and column of $[P]$ are defined as
\begin{align}
p_{1,j} = p_{1,j-1} + w_{1,j}, \textrm{ for } j=2,3,\cdots n,\\
p_{i,1} = p_{i-1,1} + w_{i,1}, \textrm{ for } i=2,3,\cdots m.
\end{align}
For non-boundary units, it is defined as 
\begin{equation} 
	p_{i,j} = w_{i,j} + \textrm{min}\left(p_{i-1,j}, p_{i,j-1}\right), \textrm{ for } 2\leqslant i \leqslant m \textrm{ and } 2\leqslant j \leqslant n.
\end{equation}
The state matrix $[P]$ is filled with elements $p_{i,j}$, as the index $i$ increases from 1 to $m$, and $j$ increases from 1 to $n$.
Thus, the minimum cumulative wave amplitude at the destination is $p_{m,n}$.
The minimum averaged wave amplitude along the optimal route is $p_{m,n}/L$.
Based on the state matrix $[P]$, the optimal route can be plotted in segments by backtracking from the destination $(m,n)$ to the starting point $(1,1)$.
The rules for backtracking are as follows:
(1) If $p_{i - 1,j}=p_{i,j}-w_{i,j}$, the optimal route links the adjacent units of $(i - 1,j)$ and $(i,j)$; 
and (2) If $p_{i,j - 1}=p_{i,j}-w_{i,j}$, the optimal route links the adjacent units of $(i,j - 1)$ and $(i,j)$.

\begin{figure}[!htp]
\centering
{
  \begin{minipage}{0.32\linewidth}
  \centering
  \includegraphics[width=1.0\linewidth]{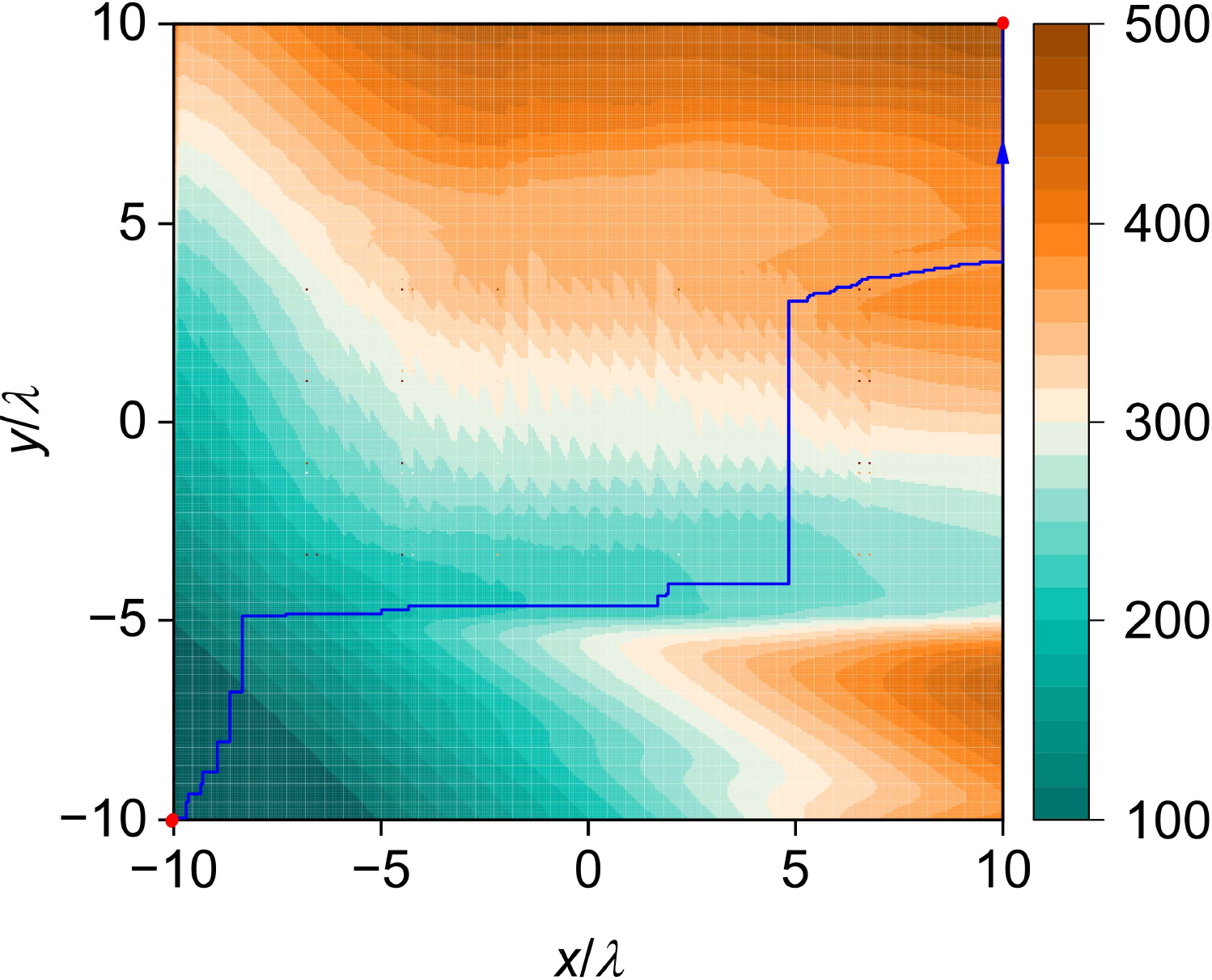}
  \subcaption{}
  \end{minipage}
}
{
  \begin{minipage}{0.32\linewidth}
  \centering
  \includegraphics[width=1.0\linewidth]{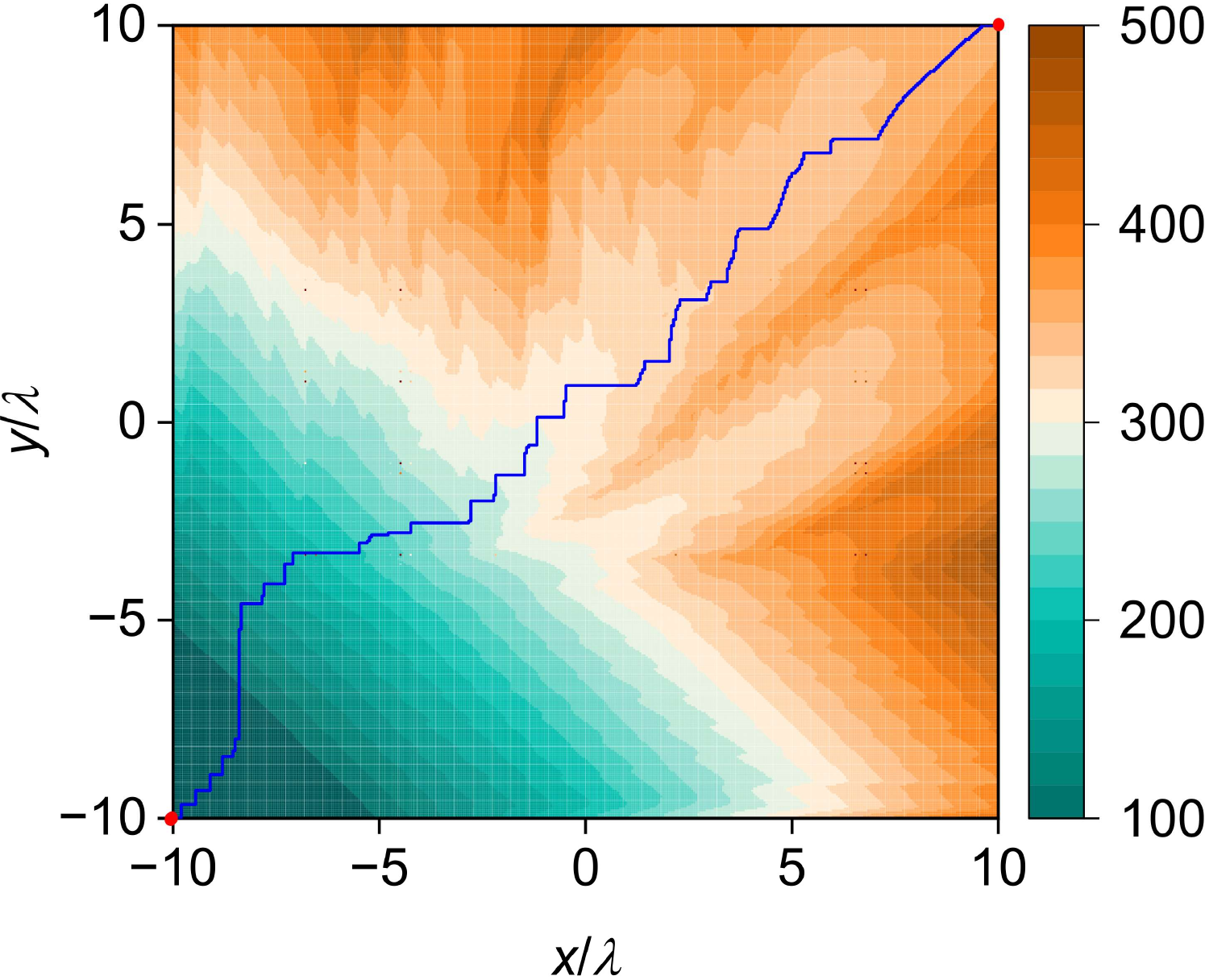}
  \subcaption{}
  \end{minipage}
}
{
  \begin{minipage}{0.32\linewidth}
  \centering
  \includegraphics[width=1.0\linewidth]{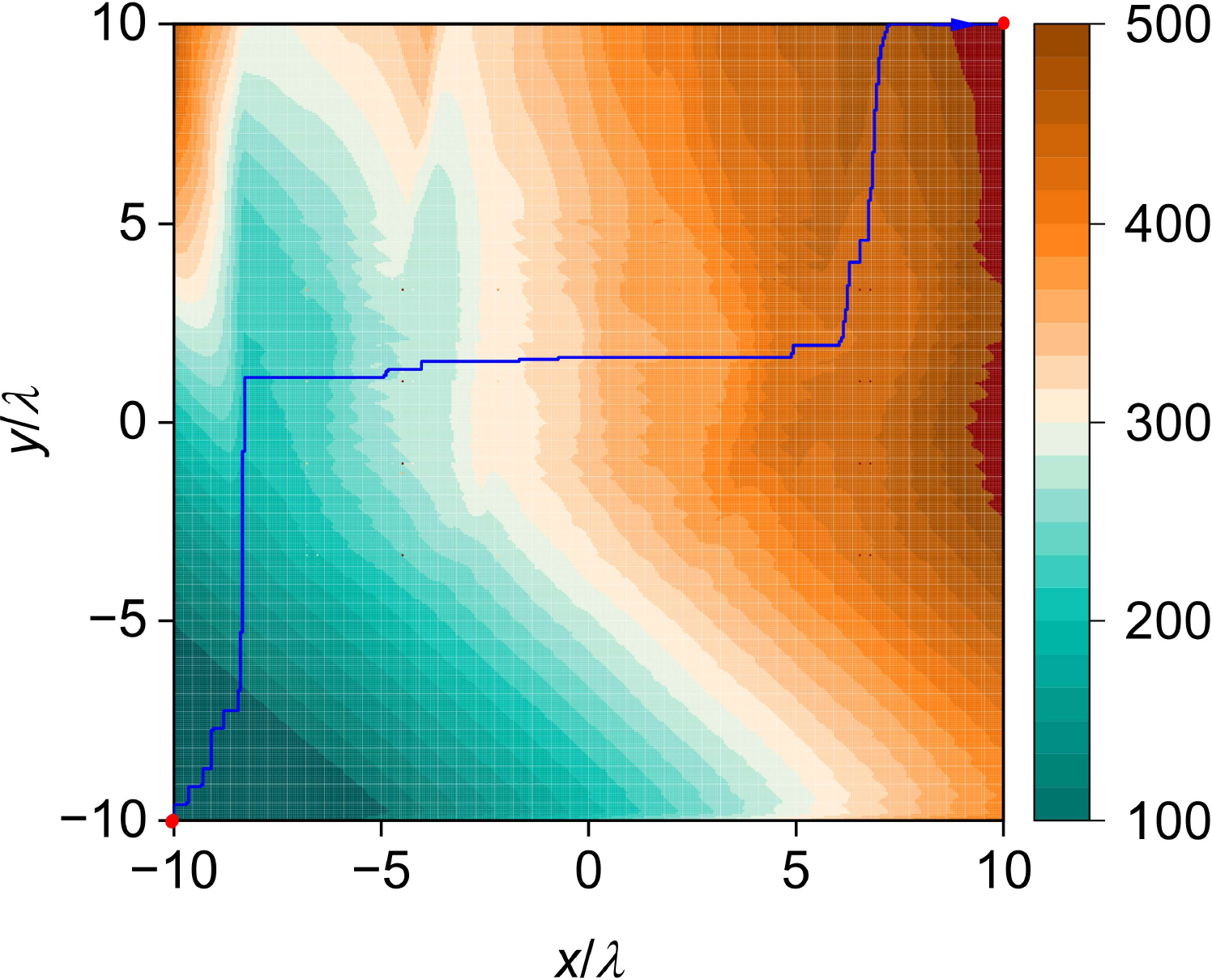}
  \subcaption{}
  \end{minipage}
}
\caption{
State matrices for optimal route planning in wave field around ice floes grouped in ``DUT'' pattern in long waves with incidence angle of (a) $\beta=0^{\rm{o}}$, (b) $\beta=45^{\rm{o}}$, and (c) $\beta=90^{\rm{o}}$.}
\label{fig:state_mat}
\end{figure}

\begin{figure}[!htp]
\centering
{
  \begin{minipage}{0.32\linewidth}
  \centering
  \includegraphics[width=1.0\linewidth]{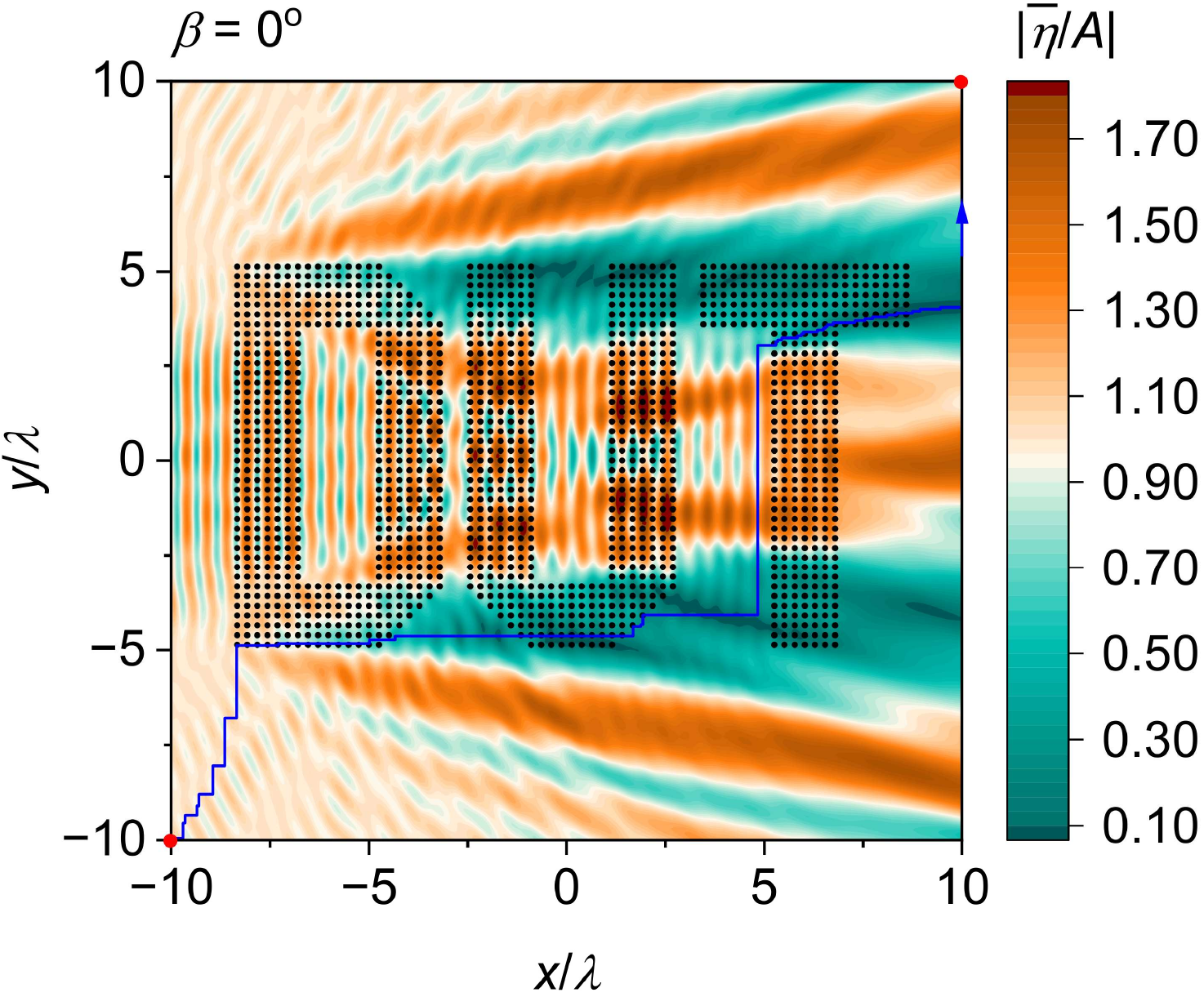}
  \subcaption{}
  \end{minipage}
}
{
  \begin{minipage}{0.32\linewidth}
  \centering
  \includegraphics[width=1.0\linewidth]{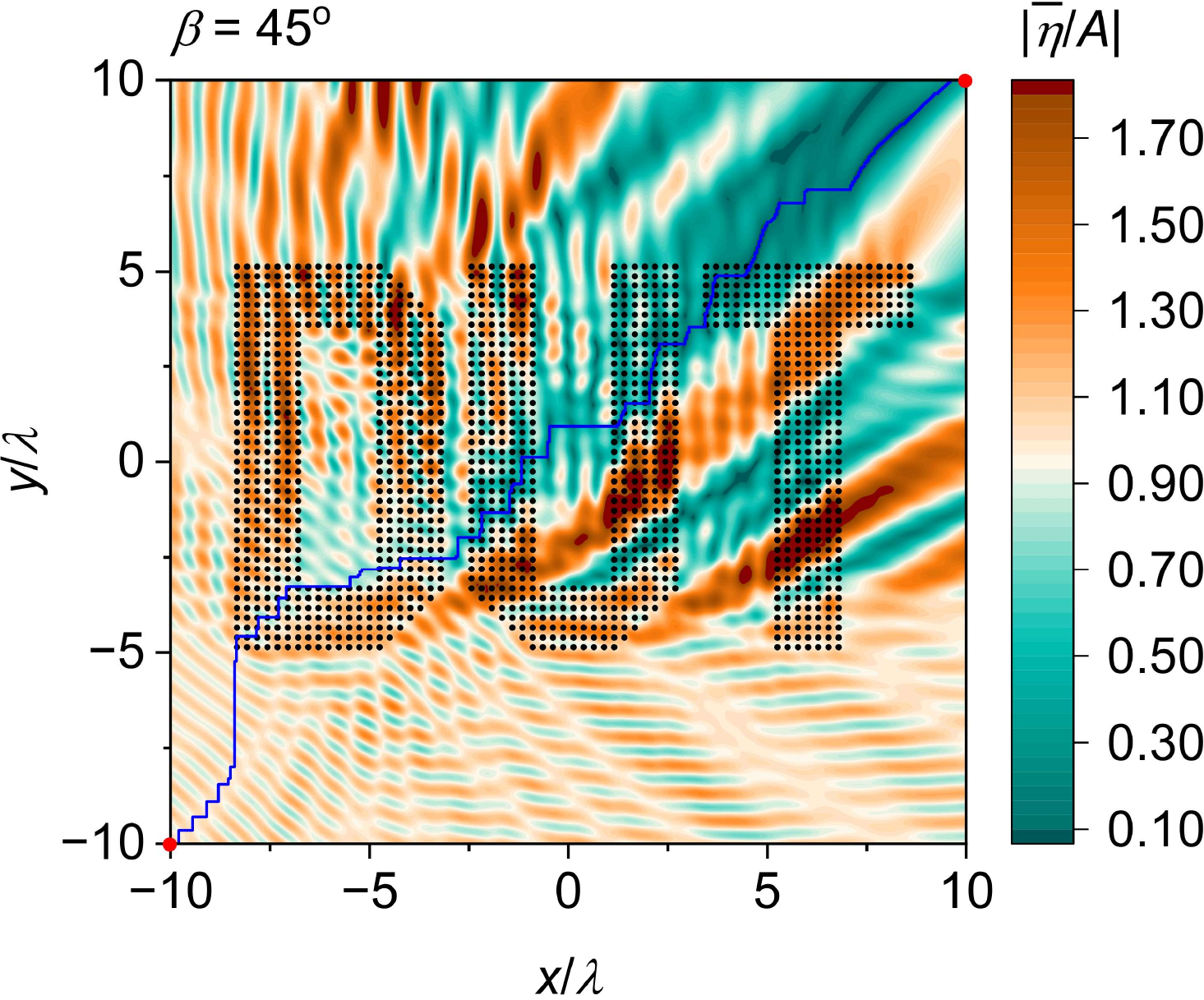}
  \subcaption{}
  \end{minipage}
}
{
  \begin{minipage}{0.32\linewidth}
  \centering
  \includegraphics[width=1.0\linewidth]{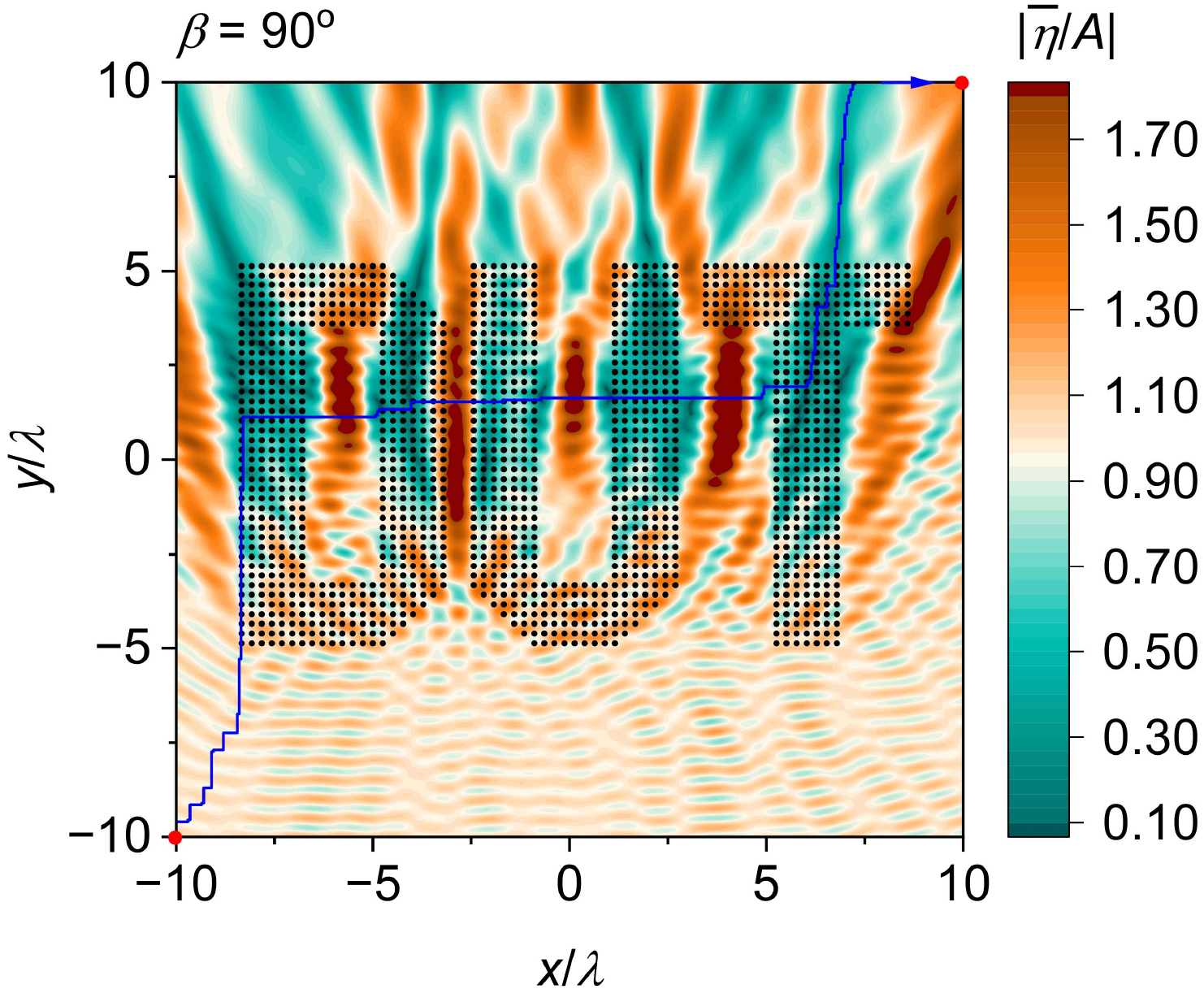}
  \subcaption{}
  \end{minipage}
}
\caption{
Optimal route planning in wave field around ice floes grouped in ``DUT'' pattern in long waves with incidence angle of (a) $\beta=0^{\rm{o}}$, (b) $\beta=45^{\rm{o}}$, and (c) $\beta=90^{\rm{o}}$.}
\label{fig:efficiency_route}
\end{figure}

Fig. \ref{fig:state_mat} shows elements of the state matrix $[P]$ for the wave fields in Fig. \ref{fig:DUT_direction}.
The starting point is set at $(-10\lambda,-10\lambda)$ and the destination is at $(10\lambda,10\lambda)$.
The rectangular zone is discretized into $400\times400$ square units.
The value of $p_{i,j}$ in each square unit is denoted by color.
The determined optimal routes across the wave fields of “DUT”-patterned ice floes are plotted.
Intuitively, the optimal routes are in the “valleys” of the contour maps of the state matrices.
Along each optimal route, the averaged wave amplitude $\bar{\eta}_{\textrm{ave}}$ is reduced to $0.49A$, $0.53A$, and $0.52A$ for the wave fields with incidence angles of $\beta = 0^{\circ}$, $45^{\circ}$, and $90^{\circ}$ respectively.
Figs. \ref{fig:efficiency_route} (a), (b) and (c) show the determined optimal routes across the wave-amplitude contours of Fig. \ref{fig:DUT_direction}.
It is reasonable to believe that with optimal Arctic route planning, the average sea state of the large group of ice floes the ship will encounter can be effectively reduced.

\color{black}
\section{Further discussion}\label{sec:discuss}

\subsection{Comparison of ``white-box'' and ``black-box'' models}

\begin{figure}[!htp]
\centering
{
  \begin{minipage}{0.9\linewidth}
  \centering
  \includegraphics[width=1.0\linewidth]{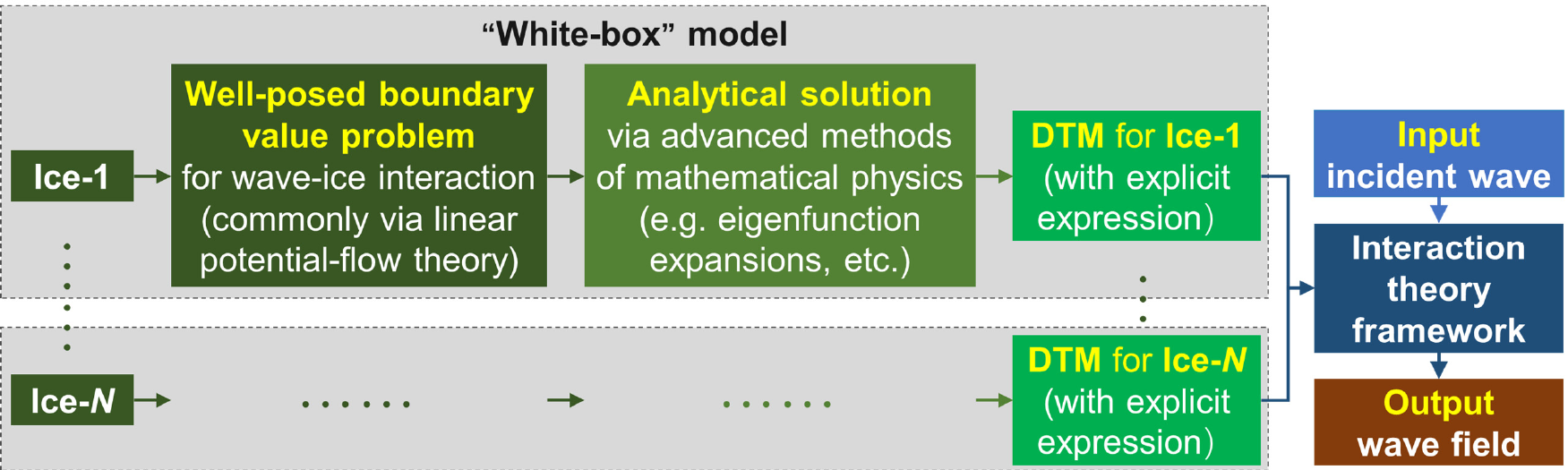}
  \subcaption{}
  \end{minipage}
}
{
  \begin{minipage}{0.8\linewidth}
  \centering
  \includegraphics[width=1.0\linewidth]{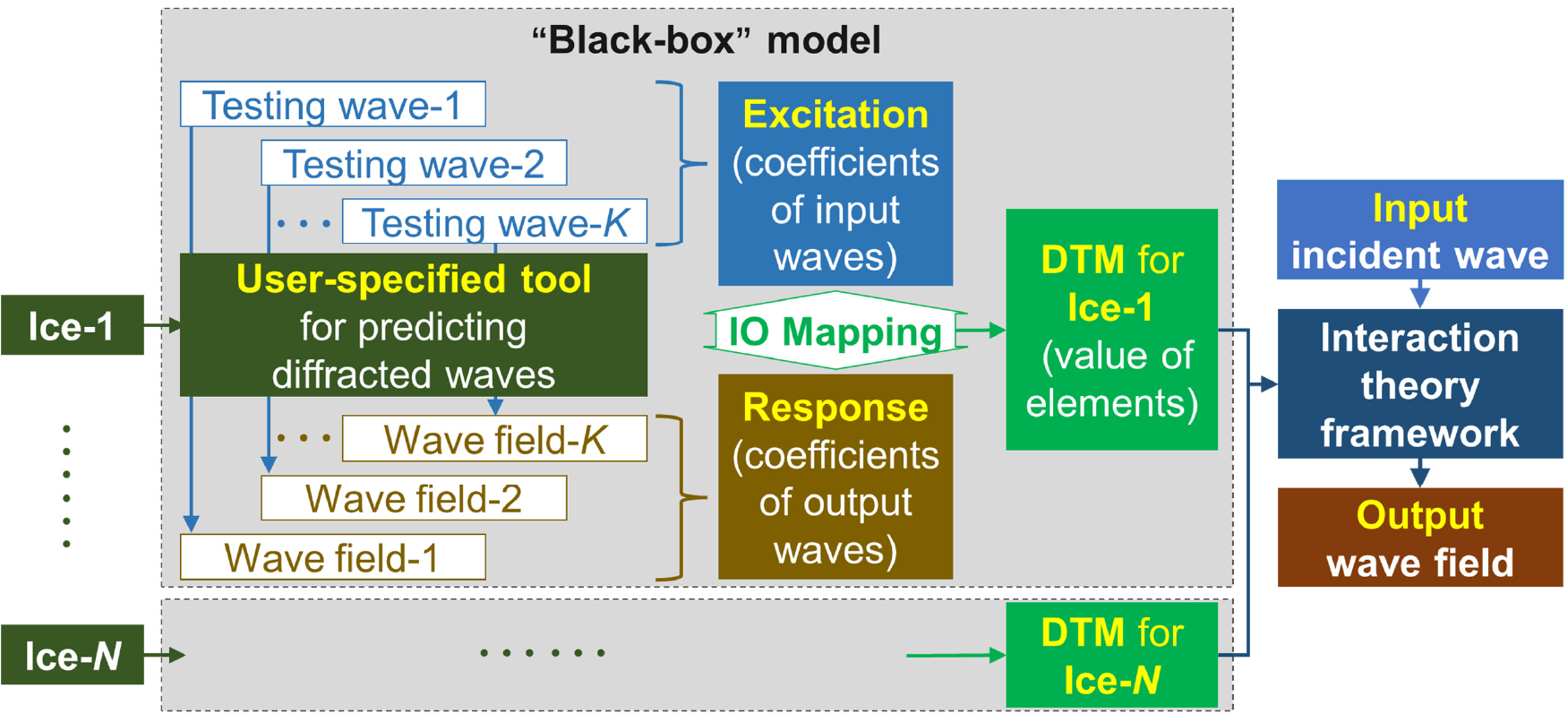}
  \subcaption{}
  \end{minipage}
}
\caption{Comparison of (a) conventional ``white-box'' model and (b) present ``black-box'' model for constructing ice floe DTMs in the application of interaction theory.}
\label{fig:black_box_process}
\end{figure}

This study is highlighted by the proposal of a novel ``black-box'' model for constructing the DTM of three-dimensional ice floes with arbitrarily complex geometry.
To better distinguish the present ``black-box'' model from the conventional ``white-box'' model, Fig. \ref{fig:black_box_process} compares their application logic when embedded within the interaction theory framework.
It is evident that both models are used to derive the DTM elements corresponding to each isolated ice floe.

Using the conventional ``white-box'' model, a well-posed boundary value problem should first be established for wave-ice interactions of each isolated ice floe, typically within the framework of linear potential-flow theory; analytical or semi-analytical solutions can be obtained by solving the problem using advanced methods of mathematical physics.
The elements of the DTM can be explicitly expressed based on the obtained analytical solutions.
Establishing and solving a precise and complete boundary value problem for ice floes is not always straightforward for users lacking solid mathematical foundations.

Unlike the ``white-box'' model, the present ``black-box'' model does not require accurate mathematical descriptions for wave-ice interaction problems involving individual ice floes.
The internal structure of the black box is detected by transmitting a sufficient number of predefined testing waves to each ice floe and analyzing the corresponding response wave fields.
The response wave fields are generated using efficient user-specified tools, including appropriate numerical solvers, data-driven predictions derived from experimental databases, analytical solution generators, and other relevant methods.
As an illustrative example of a numerical solver, the boundary element method is employed in Sec. \ref{sec:resul} to generate the response wave field, though its use is not compulsory.
Characteristic coefficients of the response wave field around each isolated ice floe are derived using the wave component detection (WCD) method--a surface-fitting technique based on measured wave elevations at a series of ``virtual wave gauges''.
Through input–output mapping of the characteristic parameter arrays for all excitation–response pairs, the element values of the DTM can be approximated.
Therefore, the present ``black-box'' model represents a more readily implemented and universally applicable approach for obtaining the DTM of three-dimensional ice floes with arbitrarily complex geometry.

Specifically for the WCD approach, a series of virtual wave gauges are arranged around the floe to sample the free-surface elevation of the scattered wave field.
The measured wave elevations are subsequently decomposed into a series of outgoing cylindrical wave modes, from which the modal scattering coefficients corresponding to different angular orders are derived via a least-squares fitting procedure.
These characteristic coefficients are employed to construct the DTM.
The DTM yields a compact representation of the scattering properties of the floe by correlating the incident and scattered cylindrical wave components.
Once the DTM is acquired, the interaction among multiple floes can be analytically described using Graf’s addition theorem to translate cylindrical wave fields between different local coordinate systems.
In this way, the complex multiple-scattering problem is reduced to solving a linear system of equations for the modal amplitudes, thus enabling efficient simulations of large floe arrays.

\subsection{Capability demonstration for ice floes with complex geometry}

\begin{figure}[!htp]
\centering
{
  \begin{minipage}{0.5\linewidth}
  \centering
  \includegraphics[width=1.0\linewidth]{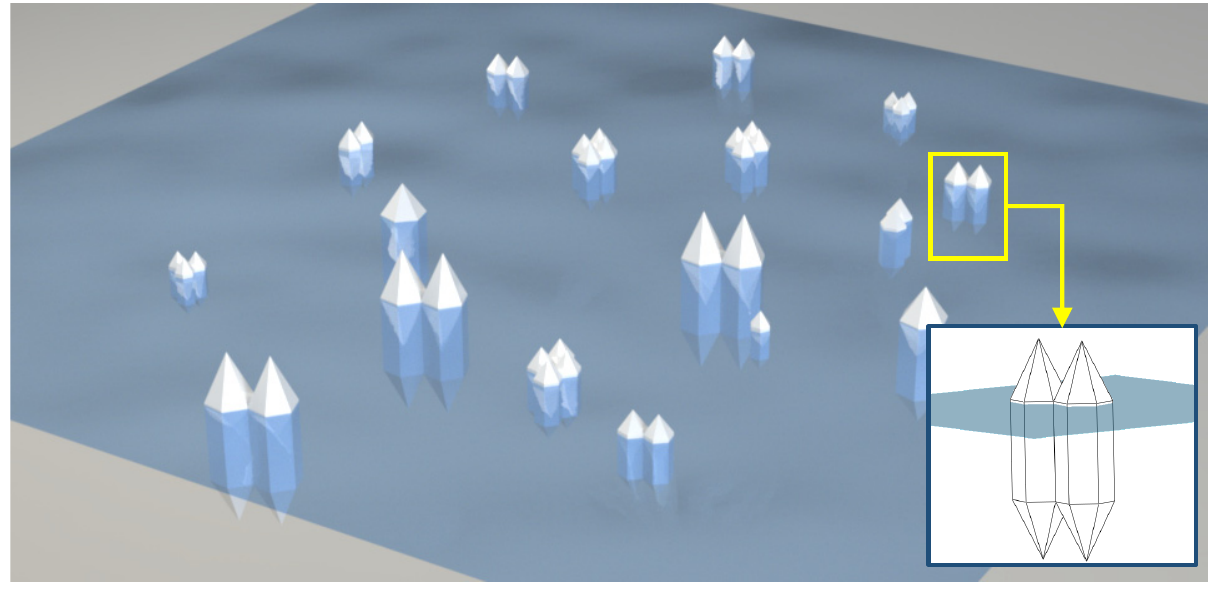}
  \end{minipage}
}
\caption{Scenario of a large group of ice floes with complex geometries for capability demonstration.}
\label{fig:DUT_shape_2}
\end{figure}

\begin{figure}[!htp]
\centering
{
  \begin{minipage}{0.22\linewidth}
  \centering
  \includegraphics[width=1.0\linewidth]{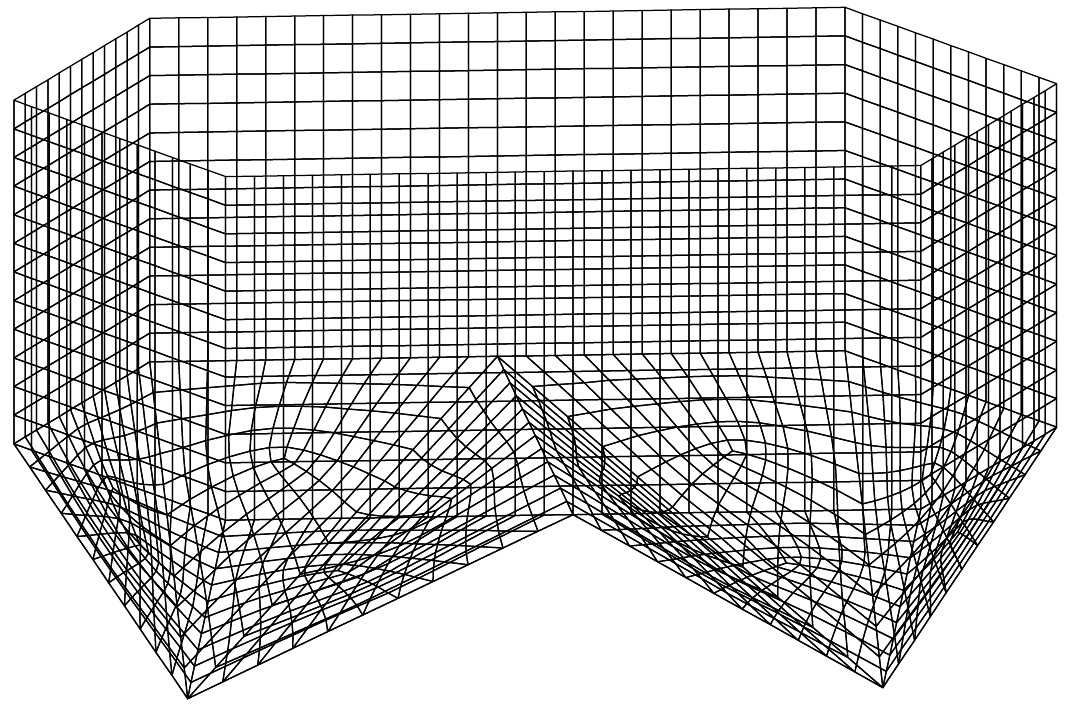}
  \subcaption{}
  \end{minipage}
}
{
  \begin{minipage}{0.22\linewidth}
  \centering
  \includegraphics[width=1.0\linewidth]{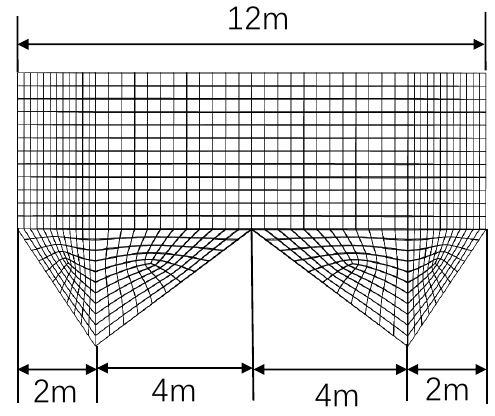}
  \subcaption{}
  \end{minipage}
}
{
  \begin{minipage}{0.22\linewidth}
  \centering
  \includegraphics[width=1.0\linewidth]{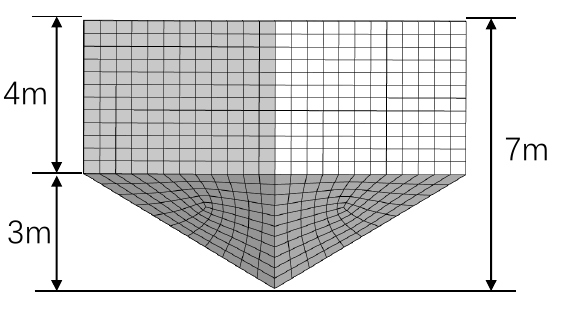}
  \subcaption{}
  \end{minipage}
}
{
  \begin{minipage}{0.22\linewidth}
  \centering
  \includegraphics[width=1.0\linewidth]{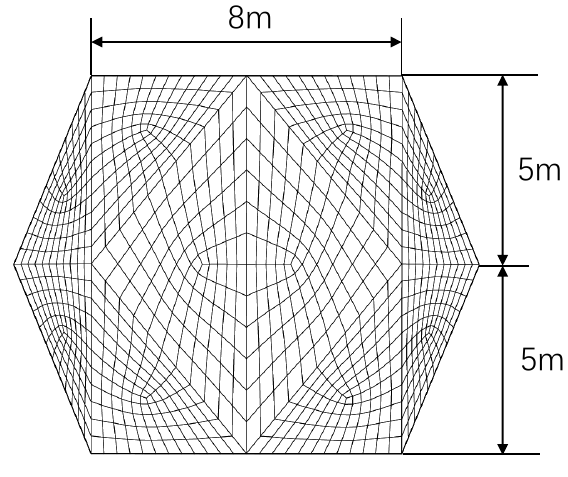}
  \subcaption{}
  \end{minipage}
}
\caption{Geometrical parameters of a typical ice floe: (a) boundary elements on the mean submerged body surface; (b) front view; (c) side view; and (d) top view.
}
\label{fig36}
\end{figure}

\begin{figure}[!htp]
\centering
{
  \begin{minipage}{0.32\linewidth}
  \centering
  \includegraphics[width=1.0\linewidth]{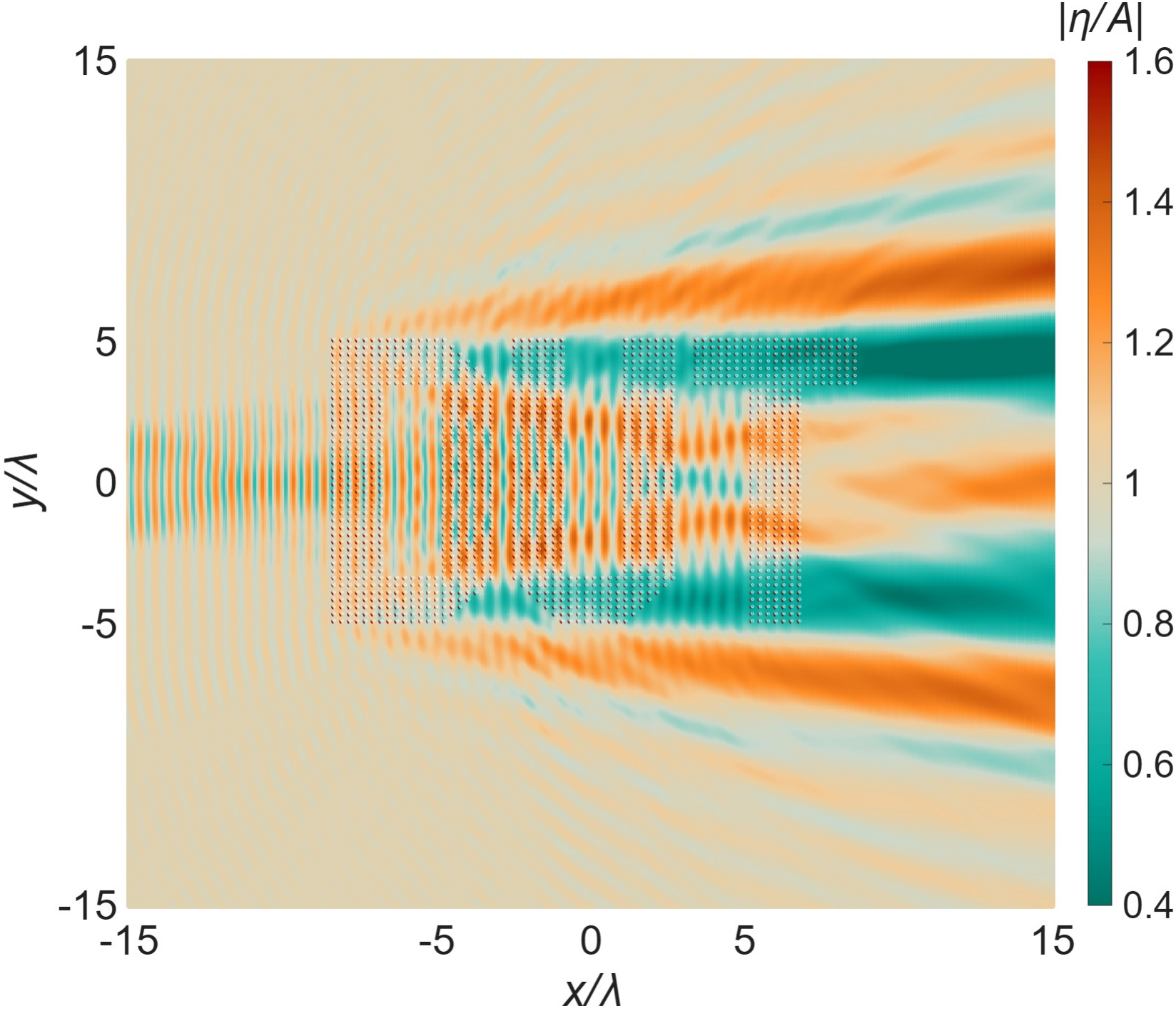}
  \subcaption{}
  \end{minipage}
}
{
  \begin{minipage}{0.32\linewidth}
  \centering
  \includegraphics[width=1.0\linewidth]{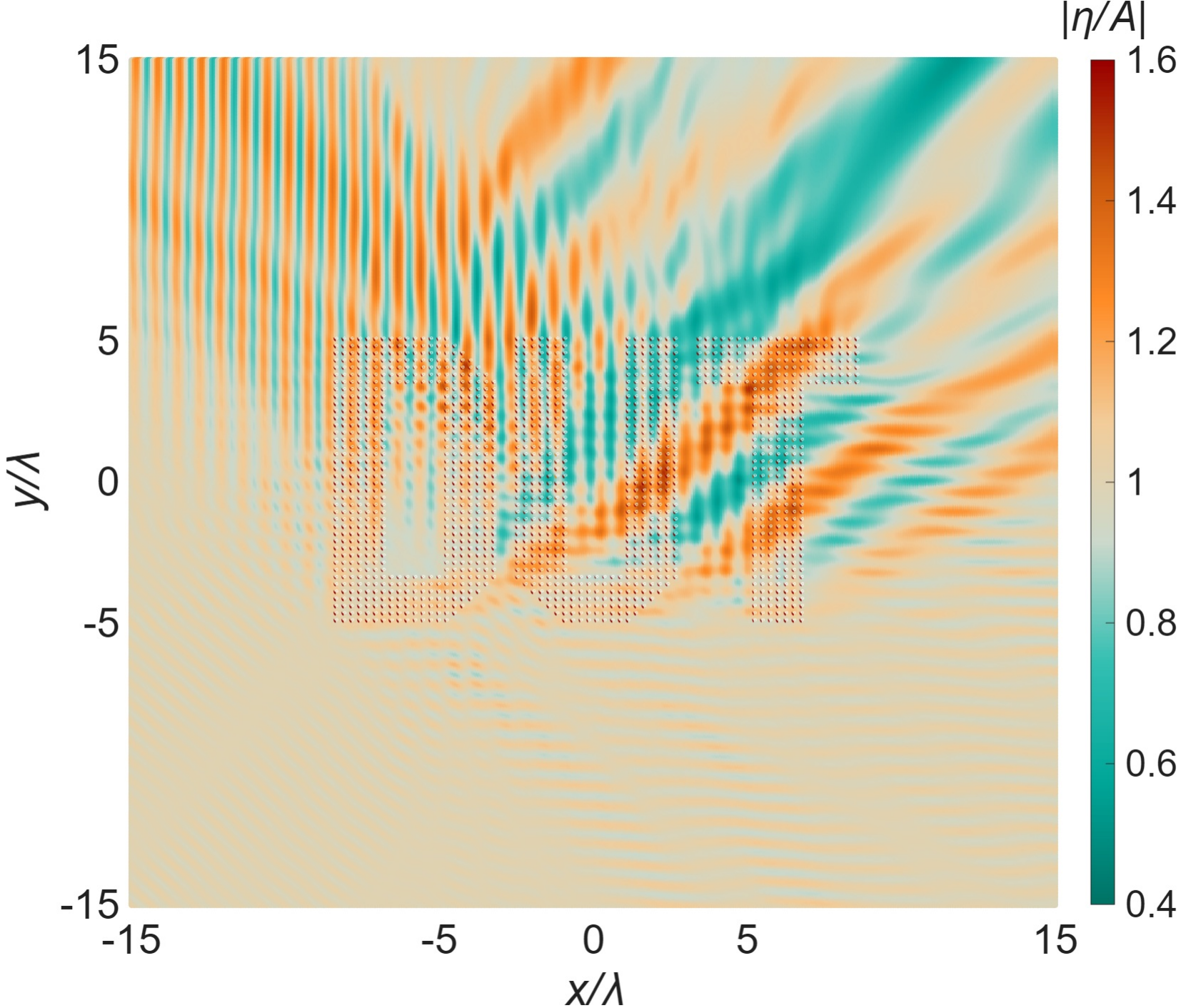}
  \subcaption{}
  \end{minipage}
}
{
  \begin{minipage}{0.32\linewidth}
  \centering
  \includegraphics[width=1.0\linewidth]{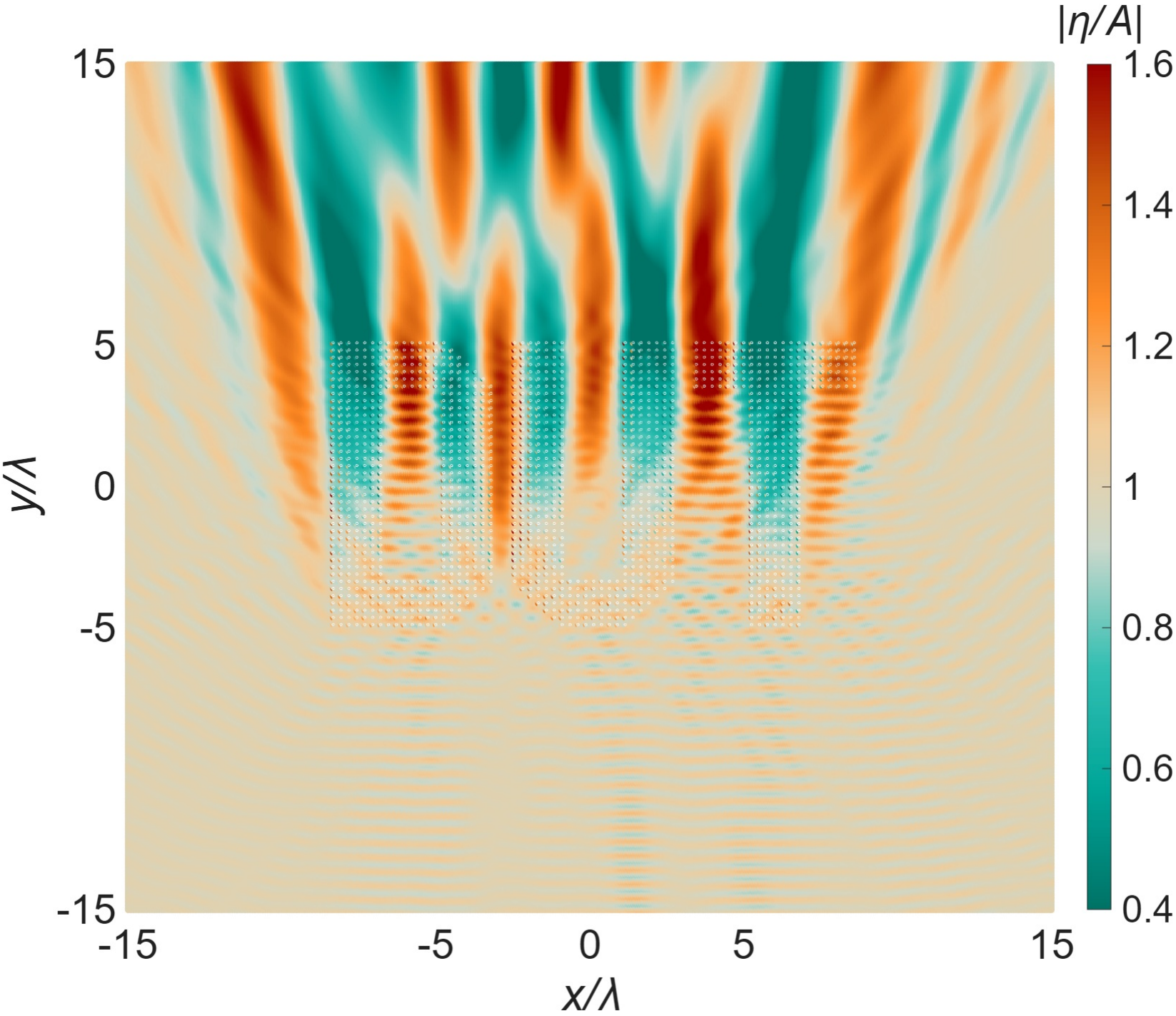}
  \subcaption{}
  \end{minipage}
}
\caption{Distribution of wave amplitudes around ice floes with complex geometry grouped in ``DUT'' pattern in long waves with incidence angle of (a) $\beta=0^{\rm{o}}$, (b) $\beta=45^{\rm{o}}$, and (c) $\beta=90^{\rm{o}}$.}
\label{fig37}
\end{figure}

Although only three regular geometries are considered in the calculation as shown in Fig. \ref{fig:DUT_shape}, the present black-box-model-enhanced interaction method is applicable to three-dimensional ice floes with arbitrarily complex geometries.
To demonstrate the capability of the proposed method, a scenario involving a large group of ice floes with complex geometries is presented, as illustrated in Fig. \ref{fig:DUT_shape_2}.
Fig. \ref{fig36} presents the geometrical details of a typical ice floe with three-dimensional characteristics.
The 1561 ice floes arranged in the ``DUT'' pattern as shown in Fig. \ref{fig:DUT_shape} are replaced by ice floes with considerably more complex geometries, including the configuration illustrated in Fig. \ref{fig36}.
Fig. \ref{fig37} shows the distribution of wave amplitudes around these ice floes grouped in the ``DUT'' pattern.
The ambient plane waves are long waves with the incidence angles $\beta = 0^{\circ}$, $45^{\circ}$, and $90^{\circ}$.
The distribution of wave amplitudes around this large group of ice floes with complex geometry is similar to that shown in Fig. \ref{fig:DUT_direction}.
The computational efficiency is maintained in these large-scale wave field simulations.
Thus, the capability of the proposed method for predicting wave-scattering wave fields by a large group of ice floes with complex geometries is confirmed.

\subsection{Limitations of black-box-model-enhanced interaction method}

Despite its advantages, several limitations remain in the black-box-model-enhanced interaction method.
First, the accuracy of the reconstructed scattering coefficients rely on the spatial distribution and number of virtual wave gauges employed for sampling the scattered wave field. 
Insufficient angular resolution or improper wave gauge placement may introduce fitting errors in the cylindrical wave decomposition.
Second, the present formulation employs linear wave theory and disregards the hydroelastic and motion responses of ice floes; consequently, nonlinear wave–structure interactions are not involved.
Third, the method relies on the assumption that each floe can be represented as an isolated scatterer whose scattering characteristics remain unchanged when embedded in an array. 
In reality, near-field hydrodynamic interactions between closely spaced floes may slightly alter the local scattering behaviour.
Finally, effects such as viscous dissipation, wave breaking, and ice fracture are not incorporated into the current model, despite potentially playing an important role in wave attenuation within the marginal ice zone.

Considering more realistic Arctic conditions, future research should focus on addressing these limitations. 
These include exploring improvements to the wave-field sampling strategy, incorporating higher-order hydroelastic effects, and extending the present framework to consider dissipative processes and irregular floe geometries.

\color{black}

\section{Conclusion}\label{sec:concl}

\color{black}
This study is motivated by the practical maritime safety requirements of the Chinese research icebreaker Xue Long. 
An intelligent route planning system is designed for ship navigation in Arctic waters with icebergs and large ice floes.
An enhanced interaction (EI) method is developed for predicting the water-wave field within a large group of arbitrarily shaped ice floes.
A novel black-box model, referred to as the wave component detection (WCD) method, is proposed for constructing the diffraction transfer matrix (DTM) of the interaction theory framework.
The DTM, which is mathematically intractable for three-dimensional ice floes with arbitrarily complex geometry, can now be solved using an easily implemented and universally applicable approach.

\color{black}
Without loss of generality, four ice-floe shapes are taken as example models to demonstrate the capability of the EI method.
Boundary element method solutions are used to validate the wave-reconstruction effectiveness of the EI method.
The parameters of virtual wave gauge layout in the WCD approach are examined.
Three operation rules are recommended for implementing the WCD approach.
The reliability of the EI method is confirmed in scenarios with multiple ice floes of different sizes and distances.
The error-range statements can be conservatively made as follows:
(i) at a location more than 1/5 of the circumradius of each ice floe away from its circumcircle, the relative error in the reconstructed wave amplitude is always less than $5\%$; 
and (ii) at any location more than half the wavelength away from the circumcircle, the relative error is always less than $0.5\%$.

The super-high efficiency of the EI method is demonstrated in cases involving an ultra-large group of ice floes.
As example cases, up to 1800 ice floes with 1,440,000 boundary elements are considered. 
It takes less than 1.5 hours to calculate wave amplitudes at 160,000 locations in the wave field with 1800 ice floes on an ordinary personal computer with an Intel Core (TM) i5-8265U processor (released in 2017).

Based on the knowledge of the wave-amplitude distribution map, various path-planning strategies can be used to recommend an optimized route for ship sailing.
As a demonstration, the dynamic programming strategy is used to recommend optimized navigation routes through 1561 ice floes of mixed shapes.
In wave field with incidence angles of $0^{\circ}$, $45^{\circ}$, and $90^{\circ}$, along each optimal route, the average wave amplitude the ship encounters is reduced to 0.49, 0.53, and 0.52 times the incident wave amplitude.
Based on the wave field predicted by the EI method, users can take advantage of the wave-sheltering effect of the ice floes for route optimization.

\color{black}
It should be noted that the present EI method still relies on linear potential-flow theory.
The applicability of the EI method to large-amplitude waves requires further investigation.
Meanwhile, how to efficiently incorporate the hydroelastic properties of the ice floes into the WCD approach deserves further study.
In the near future, physical experiments on wave scattering by a large group of arbitrarily shaped ice floes will be performed to validate the proposed black-box-model-enhanced interaction method.
\color{black}

\section*{Acknowledgements}
This work is supported by the National Natural Science Foundation of China (Grant No. 52471271). 

\section*{Conflict of Interest}
The authors have no conflicts to disclose.

\printcredits

\section*{Data Availability}
The data that support the findings of this study are available from the corresponding author, upon reasonable request.

\section*{Declaration of generative AI and AI-assisted technologies in the writing process}
During the preparation of this work, generative AI and AI-assisted technologies were not used.

\bibliographystyle{cas-model2-names}

\bibliography{refs}

@book{Abramowitz1964,
author = {Abramowitz, M.  and Stegun, I. A. },
title = {Handbook of Mathematical Functions},
year = {1964},
publisher = {Dover, Washington, D.C.}
}

@article{bennetts2009wave,
	title={Wave scattering by multiple rows of circular ice floes},
	author={Bennetts, LG and Squire, VA2558590},
	journal={J. Fluid Mech.},
	volume={639},
	pages={213--238},
	year={2009},
	publisher={Cambridge University Press}
}

@article{bennetts2010wave,
	title={Wave scattering by ice floes and polynyas of arbitrary shape},
	author={Bennetts, LG and Williams, TD2728216},
	journal={J. Fluid Mech.},
	volume={662},
	pages={5--35},
	year={2010},
	publisher={Cambridge University Press}
}

@article{bennetts2010three,
	title={A three-dimensional model of wave attenuation in the marginal ice zone},
	author={Bennetts, Luke G and Peter, Malte A and Squire, VA and Meylan, Michael H},
	journal={J. Geophys. Res. Oceans},
	volume={115},
	number={C12},
	year={2010},
	pages={C12043},
	publisher={Wiley Online Library}
}

@article{fox1990reflection,
	title={Reflection and transmission characteristics at the edge of shore fast sea ice},
	author={Fox, Colin and Squire, Vernon A},
	journal={J. Geophys. Res. Oceans},
	volume={95},
	number={C7},
	pages={11629--11639},
	year={1990},
	publisher={Wiley Online Library}
}

@article{HUANG2020102817,
title = {Ship resistance when operating in floating ice floes: A combined {CFD DEM} approach},
journal = {Mar. Struct.},
volume = {74},
pages = {102817},
year = {2020},
issn = {0951-8339},
author = {Luofeng Huang and Jukka Tuhkuri and Bojan Igrec and Minghao Li and Dimitris Stagonas and Alessandro Toffoli and Philip Cardiff and Giles Thomas}
}

@article{kagemoto1986interactions,
	title={Interactions among multiple three-dimensional bodies in water waves: an exact algebraic method},
	author={Kagemoto, Hiroshi and Yue, Dick KP},
	journal={J. Fluid Mech.},
	volume={166},
	pages={189--209},
	year={1986},
	publisher={Cambridge University Press}
}

@article{Keller1998,
author = {Keller, Joseph B.},
title = {Gravity waves on ice-covered water},
journal = {J. Geophys. Res. Oceans},
volume = {103},
number = {C4},
pages = {7663-7669},
year = {1998}
}

@article{linton2003reflection,
	title={Reflection and transmission at the ocean/sea-ice boundary},
	author={Linton, CM and Chung, Hyuck},
	journal={Wave Motion},
	volume={38},
	number={1},
	pages={43--52},
	year={2003},
	publisher={Elsevier}
}

@article{meylan1993finite,
	title={Finite-floe wave reflection and transmission coefficients from a semi-infinite model},
	author={Meylan, Michael and Squire, Vernon A},
	journal={J. Geophys. Res. Oceans},
	volume={98},
	number={C7},
	pages={12537--12542},
	year={1993},
	publisher={Wiley Online Library}
}

@article{meylan1996response,
	title={Response of a circular ice floe to ocean waves},
	author={Meylan, Michael H and Squire, Vernon A},
	journal={J. Geophys. Res. Oceans},
	volume={101},
	number={C4},
	pages={8869--8884},
	year={1996},
	publisher={Wiley Online Library}
}

@article{meylan2015surge,
	title={Surge motion of an ice floe in waves: comparison of a theoretical and an experimental model},
	author={Meylan, Michael H and Yiew, Lucas J and Bennetts, Luke G and French, Benjamin J and Thomas, Giles A},
	journal={Ann. Glaciol.},
	volume={56},
	number={69},
	pages={155--159},
	year={2015},
	publisher={Cambridge University Press}
}

@article{montiel2016attenuation,
	title={Attenuation and directional spreading of ocean wave spectra in the marginal ice zone},
	author={Montiel, Fabien and Squire, VA and Bennetts, LG},
	journal={J. Fluid Mech.},
	volume={790},
	pages={492--522},
	year={2016},
	publisher={Cambridge University Press}
}

@article{montiel2015reflection,
	title={Reflection and transmission of ocean wave spectra by a band of randomly distributed ice floes},
	author={Montiel, Fabien and Squire, Vernon A and Bennetts, Luke G},
	journal={Ann. Glaciol.},
	volume={56},
	number={69},
	pages={315--322},
	year={2015},
	publisher={Cambridge University Press}
}

@article{mosig2019transport,
	title={A transport equation for flexural-gravity wave propagation under a sea ice cover of variable thickness},
	author={Mosig, JEM and Montiel, F and Squire, VA},
	journal={Wave Motion},
	volume={88},
	pages={153--166},
	year={2019},
	publisher={Elsevier}
}

@article{orzech2018coupled,
	title={A Coupled System for Investigating the Physics of Wave--Ice Interactions},
	author={Orzech, Mark D and Shi, Fengyan and Veeramony, Jayaram and Bateman, Samuel and Calantoni, Joseph and Kirby, James T},
	journal={J. Atmos. Oceanic Technol.},
	volume={35},
	number={7},
	pages={1471--1485},
	year={2018}
}

@article{peter2004infinite,
	title={Infinite-depth interaction theory for arbitrary floating bodies applied to wave forcing of ice floes},
	author={Peter, Malte A and Meylan, Michael H},
	journal={J. Fluid Mech.},
	volume={500},
	pages={145--167},
	year={2004},
	publisher={Cambridge University Press}
}

@article{peter2010water,
	title={Water-wave scattering by vast fields of bodies},
	author={Peter, Malte A and Meylan, Michael H},
	journal={SIAM J. Appl. Math.},
	volume={70},
	number={5},
	pages={1567--1586},
	year={2010},
	publisher={SIAM}
}

@article{porter2004approximations,
	title={Approximations to wave scattering by an ice sheet of variable thickness over undulating bed topography},
	author={Porter, D and Porter, R},
	journal={J. Fluid Mech.},
	volume={509},
	pages={145--179},
	year={2004},
	publisher={Cambridge University Press}
}

@article{porter2019coupling,
	title={The coupling between ocean waves and rectangular ice sheets},
	author={Porter, Richard},
	journal={J. Fluid. Struct.},
	volume={84},
	pages={171--181},
	year={2019},
	publisher={Elsevier}
}

@article{sahoo2001scattering,
	title={Scattering of surface waves by a semi-infinite floating elastic plate},
	author={Sahoo, T and Yip, Tsz Leung and Chwang, Allen T},
	journal={Phys. Fluids},
	volume={13},
	number={11},
	pages={3215--3222},
	year={2001},
	publisher={American Institute of Physics}
}

@article{SCHOYEN2011977,
title = {The Northern Sea Route versus the {Suez Canal}: cases from bulk shipping},
journal = {J. Transport Geogr.},
volume = {19},
number = {4},
pages = {977-983},
year = {2011},
author = {Halvor Schoyen and Svein Brathen}
}

@article{Shen2022,
author = {Shen, Hayley H. },
title = {Wave-in-ice: theoretical bases and field observations},
journal = {Philos. Trans. A Math. Phys. Eng. Sci.},
volume = {380},
number = {2235},
pages = {20210254},
year = {2022}
}

@article{Squire2018,
author = {Squire, Vernon A. },
title = {A fresh look at how ocean waves and sea ice interact},
journal = {Philos. Trans. A Math. Phys. Eng. Sci.},
volume = {376},
number = {2129},
pages = {20170342},
year = {2018}
}

@article{Thomson2024,
author = {Thomson, Jim and Rogers, W. Erick},
title = {Swell and sea in the emerging Arctic Ocean},
journal = {Geophys. Res. Lett.},
volume = {41},
number = {9},
pages = {3136-3140},
year = {2014}
}

@article{wang2023numerical,
	title={Numerical Study on Wave--Ice Floe Interaction in Regular Waves},
	author={Wang, Chunhui and Wang, Jiaan and Wang, Chao and Wang, Zeping and Zhang, Yuan},
	journal={J. Mar. Sci. Eng.},
	volume={11},
	number={12},
	pages={2235},
	year={2023},
	publisher={MDPI}
}

@article{WANG201090,
title = {Experimental study on surface wave propagating through a grease–pancake ice mixture},
journal = {Cold Reg. Sci. Technol.},
volume = {61},
number = {2},
pages = {90-96},
year = {2010},
author = {Ruixue Wang and Hayley H. Shen}
}

@article{Waseda2018,
title = {Correlated Increase of High Ocean Waves and Winds in the Ice-Free Waters of the Arctic Ocean},
journal = {Sci. Rep.-UK},
volume = {8},
pages = {4489},
year = {2018},
author = {
Waseda, Takuji and Webb, Adrean and Sato, Kazutoshi and Inoue, Jun and Kohout,  Alison and  Penrose, Bill and Penrose, Scott}
}

@article{Yoshida1990ANM,
  title={A numerical method for huge semisubmersible responses in waves},
  author={K. Yoshida and Goo},
  journal={Trans. Soc. Nav. Archit. Mar. Eng.},
  year={1990},
  volume={98},
  pages={365-387},
}

@article{zhang2023resonance,
  author = {Zhang, Chongwei and Wang, Pengfei and Huang, Luofeng and Zhang, Mengke and Wu, Haitao and Ning, Dezhi},
  title = {Resonance mechanism of hydroelastic response of multipatch floating photovoltaic structure in water waves over stepped seabed},
  journal = {Phys. Fluids},
  volume = {35},
  pages = {107137},
  year = {2023}
}

@article{Meylan2021floe,
  author = {Meylan, Michael H. and Horvat, Christopher and Bitz, Cecilia M. and Bennetts, Luke G.},
  title = {A floe size dependent scattering model in two-and three-dimensions for wave attenuation by ice floes},
  journal = {Ocean Modell.},
  volume = {161},
  pages = {101779},
  year = {2021}
}

@article{mohapatra2025three,
  author = {Mohapatra, Sarat Chandra and Soares, C. Guedes and Meylan, Michael H.},
  title = {Three-Dimensional and Oblique Wave-Current Interaction with a Floating Elastic Plate Based on an Analytical Approach},
  journal = {Symmetry},
  volume = {17},
  number = {6},
  pages = {831},
  year = {2025}
}

@article{yu2022new,
  author = {Yu, Jie and Rogers, W. Erick and Wang, David W.},
  title = {A new method for parameterization of wave dissipation by sea ice},
  journal = {Cold Reg. Sci. Technol.},
  volume = {199},
  pages = {103582},
  year = {2022}
}

@article{montiel2024kernel,
  author = {Montiel, F. and Meylan, M. H. and Hawkins, S. C.},
  title = {Scattering kernel of an array of floating ice floes: application to water wave transport in the marginal ice zone},
  journal = {Phil. Trans. R. Soc. A},
  volume = {480},
  pages = {20230633},
  year = {2024}
}

@article{Squire2007,
  author = {Squire, V. A.},
  title = {Of ocean waves and sea-ice revisited},
  journal = {Cold Reg. Sci. Technol.},
  volume = {49},
  number = {2},
  pages = {110--133},
  year = {2007}
}

@article{Bennetts2012,
  author = {Bennetts, Luke G. and Squire, Vernon A.},
  title = {On the calculation of an attenuation coefficient for transects of ice-covered ocean},
  journal = {Phil. Trans. R. Soc. A},
  volume = {468},
  number = {2137},
  pages = {136--162},
  year = {2012}
}

@article{Kohout2008,
  author = {Kohout, A. L. and Meylan, M. H.},
  title = {An elastic plate model for wave attenuation and ice floe breaking in the marginal ice zone},
  journal = {J. Geophys. Res.: Oceans},
  volume = {113},
  pages = {C09016},
  year = {2008}
}

@article{Bennetts2010,
  author = {Bennetts, L. G. and Peter, M. A. and Squire, V. A. and Meylan, M. H.},
  title = {A three-dimensional model of wave attenuation in the marginal ice zone},
  journal = {J. Geophys. Res.},
  volume = {115},
  pages = {C12043},
  year = {2010}
}

@article{Meylan2018,
  author = {Meylan, M. H. and Bennetts, L. G.},
  title = {Three-dimensional time-domain scattering of waves in the marginal ice zone},
  journal = {Phil. Trans. R. Soc. A},
  volume = {376},
  number = {2129},
  pages = {20170334},
  year = {2018}
}

\end{document}